\pdfoutput=1
\documentclass[aps, prb, amsfonts, amsmath, twocolumn, superscriptaddress, 10pt, floatfix]{revtex4-1}
\usepackage{natbib}

\usepackage[utf8]{inputenc}
\usepackage[english]{babel}
\usepackage{amsmath}
\usepackage{amsfonts}
\usepackage{amssymb}
\usepackage{graphicx}
\usepackage{xcolor}

\begin{document}

\title{Coherent quantum dynamics launched by incoherent relaxation in a quantum circuit simulator of a light-harvesting complex.}

\author{A. W. Chin}
\affiliation{Laboratoire Chimie Physique (LCP)-CNRS,Universit\'e Paris Saclay, Univ. Paris Sud,  F-91405 Orsay, France}
\affiliation{Institut des NanoSciences de Paris, Sorbonne Universit\'e, 4 place Jussieu, boite courrier 840, 75252 PARIS Cedex 05, France}

\author{E. Mangaud}
\affiliation{Institut des NanoSciences de Paris, Sorbonne Universit\'e, 4 place Jussieu, boite courrier 840, 75252 PARIS Cedex 05, France}
\affiliation{Laboratoire Collisions Agr\'egats R\'eactivi\'e (IRSAMC), Universi\'e Toulouse III Paul Sabatier, UMR 5589, F-31062 Toulouse Cedex 09, France}	

\author{O. Atabek}
\affiliation{Institut des Sciences Mol\'eculaires d'Orsay (ISMO) UMR CNRS 8214, Universit\'e Paris Saclay, Univ. Paris Sud,  F-91405 Orsay, France}

\author{M. Desouter-Lecomte}
\affiliation{Laboratoire Chimie Physique (LCP)-CNRS,Universit\'e Paris Saclay, Univ. Paris Sud,  F-91405 Orsay, France}
\affiliation{D\'epartement de Chimie, Universit\'e de Li\`ege, Sart Tilman, B6, B-4000 Li\`ege, Belgium}

\begin{abstract}
Engineering and harnessing coherent excitonic transport in organic nanostructures has recently been suggested as a promising way towards improving man-made light harvesting materials. However, realising and testing the dissipative system-environment models underlying these proposals is presently very challenging in supramolecular materials. A promising alternative is to use simpler and highly tunable `quantum simulators' built from programmable qubits, as recently achieved in a superconducting circuit by Poto{\v{c}}nik \textit{et al}. Nature Communications, 9, 904 (2018) \cite{potovcnik2018studying}. In this article, we simulate the real-time dynamics of an exciton coupled to a quantum bath as it moves through a network based on the quantum circuit of \cite{potovcnik2018studying}. Using the numerically exact hierarchical equations of motion to capture the open quantum system dynamics, we find that an ultrafast but completely \emph{incoherent} relaxation from a high-lying `bright' exciton into a doublet of closely spaced 'dark'  excitons can spontaneously generate electronic coherences and oscillatory real-space motion across the network (quantum beats). Importantly, we show that this behaviour also survives when the environmental noise is classically stochastic (effectively high temperature), as in present experiments. These predictions highlight the possibilities of designing matched electronic and spectral noise structures for robust coherence generation that doesn't require coherent excitation or cold environments.
 \end{abstract}

\maketitle

\section{INTRODUCTION}
Creating and sustaining `long-lived' electronic coherences in complex, multi-component supramolecular systems has recently been highlighted as an exciting route towards advanced molecular nanodevices with applications ranging from energy harvesting to optomechanics and sensing \cite{bredas2017photovoltaic,Scholes2017}. In this context, `long-lived' refers to decoherence times of comparable duration to the `functional' timescales of the system, which might, for example, correspond to energy transport times or charge generation, in the case of photovoltaic light-harvesting structures \cite{Castro2014,chin2013role,Collini2010,fuller2014vibronic,Kreisbeck2012,Lambert_2013,Lee1462,Panitchayangkoon2010,romero2014quantum}. However, many - if not all - reliable molecular functions are driven in a thermodynamic direction by noisy interactions between electronic degrees of freedom and their thermal environments, so complete suppression of environmental couplings - as is desirable for quantum computation - is not a fruitful strategy for the multitude of applications discussed in Refs\cite{bredas2017photovoltaic,Scholes2017}.

Instead, there has been an emerging interdisciplinary focus on understanding how it may be possible to exploit the non-perturbative and non-Markovian dynamics of structured system-environment interactions in nanostructured systems \cite{Akihito09,Castro2014,Chen2015,Chin2012,Chin2013,Dijkstra_2010,elinor1,Iles_Smith_2015,killoran2015,Lee1462,Mal__2016,Qin2017,Santamore2013,Stones2016}, with mounting theoretical evidence that a transient and correlated interplay of dissipative and coherent dynamics may be advantageous for a wide range of ultrafast optoelectronic processes. Indeed, although this essential idea has an origin in studies of photosynthetic pigment-protein complexes (PPCs), it is in rationally designed, organic functional materials, such as DNA origami, polymer-fullerene heterojunctions, carbon nano tubes and molecular dimer systems, that the existence and potentially beneficial impacts of electronic coherence and 'noise-assisted' dynamics on light-harvesting processes have been most cleanly and recently demonstrated \cite{bakulin2016real,falke2014coherent,lim2015vibronic,novelli2015vibronic,boulais2018programmed,gelinas2014ultrafast,hemmig2016programming}. Examples of theoretically proposed `noise-asisted' quantum phenomena and their potential applications are reviewed in Refs \cite{Lambert_2013,bredas2017photovoltaic,Scholes2017}.

\begin{figure*}
\centering
\includegraphics[width=\textwidth]{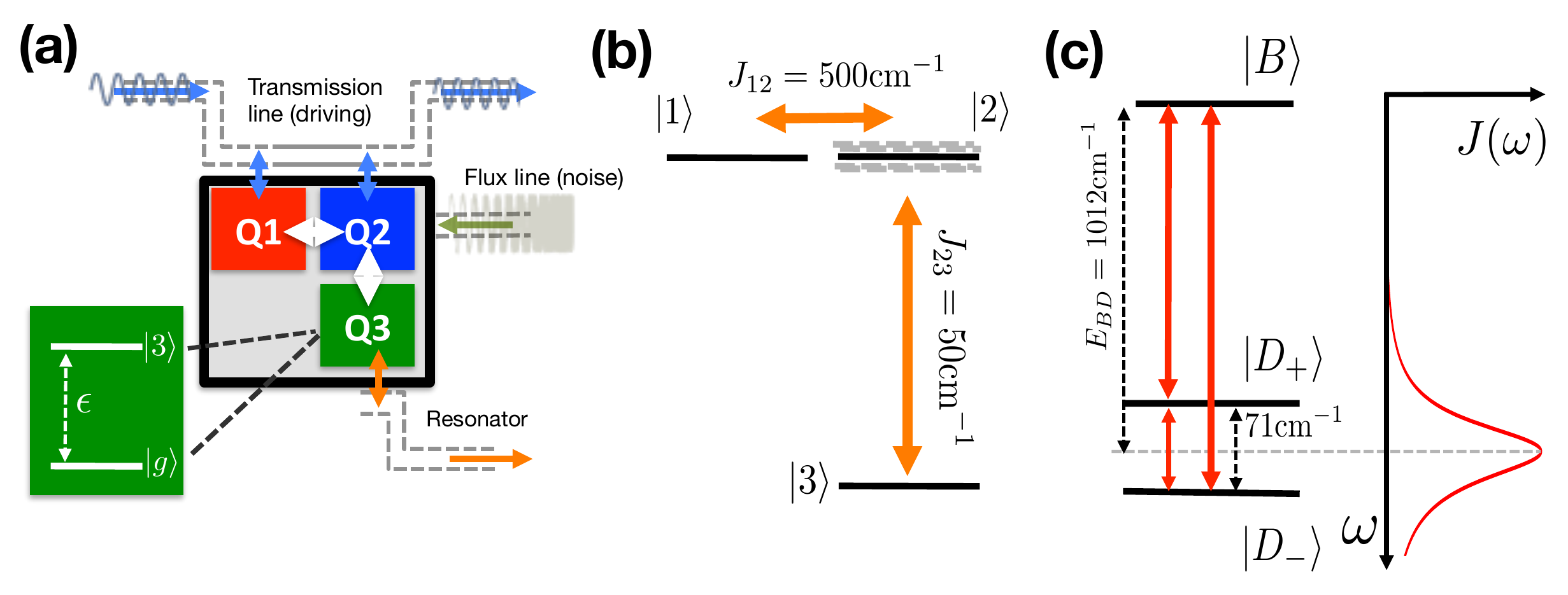}	
\caption{(a) a schematic representation of the superconducting quantum circuit used in Ref. \cite{potovcnik2018studying} to simulate energy transport in a photosynthetic light-harvesting array. Here, three qubits ($Q1-Q3$) act as chromophores with a tunable excitation energy $\epsilon_{i}$ and are coupled together by nearest-neighbour capacitive interactions (white arrows). Qubits $1$ and $2$ are coupled identically to a transmission line which carries the excitation/pump fields, while only emission in the resonator line is uniquely sensitive to the excitation of qubit $3$. The flux lines are used to tune $\epsilon_{i}$, allowing the application of stochastic signals to mimic an arbitrary classical dephasing noise on the qubit (chromophore) system. (b) The rescaled electronic couplings and detunings of the indivudal qubits/sites used in \cite{potovcnik2018studying} and in this paper. (c) The resulting spectrum of the bright $|B\rangle$ and dark $|D_{\pm}\rangle$ eigenstates and the structured (peaked) spectral noise density investigated in the experiment and in this article.}
\label{fig1}
\end{figure*}

Regardless of whether Nature got there first, or at all \cite{duan2017nature}, these latter studies underscore the new possibilities arising from exploiting emerging nanofabrication techniques to tune both the properties of photo-excited states (delocalisation, dipoles, energy spectrum) and their environments to obtain novel optoelectronic materials based on taylored system-environment interactions. Recently, Poto{\v{c}}nik \textit{et al.} have demonstrated the first experimental `quantum simulator' of an open quantum light harvesting model built from transmon qubits in a superconducting circuit (Fig.\ref{fig1}) \cite{potovcnik2018studying}. Using three individually tunable qubits coupled to a transmission line (for photoexcitation) and a resonator (to detect emission), Poto{\v{c}}nik \textit{et al.} demonstrated the formation of robust, delocalised photoexcited states with optical properties analogous to the Davydov-split (Frenkel) excitonic states found in PPCs or J-aggregates \cite{may2008charge,renger2001ultrafast}. These states are engineered, as in photosynthetic antennae complexes, so that energy absorbed by the highest energy state is spatially directed by dissipation towards the lowest energy state \cite{renger2001ultrafast,scholes2011lessons}, which is proximate to a `reaction centre' that transduces this incoming energy (here, the resonator).  Although impossible in any real supramolecular structure, this set up also allows controllable application of environmental dephasing noise of arbitrary strength and spectral properties, which in the basis of delocalised states (\textit{vide infra}) leads to controllable incoherent transitions between the single particle excited states of the network. This feature of the experiment makes it a near-ideal platform for testing theories of open dynamics, and by varying the noise coupling strength, it was demonstrated that the energy transfered across the network was maximised at an optimal value of the dephasing rate, precisely as predicted by recent theories of `noise-assisted transport' (also known as `ENAQT') \cite{RebentrostEtAlNJP2009,Caruso09,Plenio08,Plenio10}.  In further agreement, this optimal dephasing noise strength was found to be of similar magnitude to the smallest coherent coupling between the qubits, occuring at the `strong-to-weak coupling' transition point where the lowest energy delocalised eigenstates begin to collapse into effectively localised on-site excitations with sequential hopping transport \cite{potovcnik2018studying}.

Additionally, they also confirmed that energy transfer is considerably more efficient when the spectral function of the environment is strongly peaked around the energy differences between the excitonic excited states. Such structured environmental spectral functions are characterisitic of molecular vibrations, and have become intensively studied in open quantum system theory due to their multi-scale linear response functions (memory effects) and often non-perturbative coupling to the excited states at specific frequencies. These properties allow potentially qualitative and non-stationary modifications of excited state dynamics, c.f. simple heat baths, that have been connected to (transient) phenomena such as violation of detailed balance, extension of electronic coherence times and vibronic mixing of electronic states \cite{Castro2014,Chen2016,Chin2013,elinor1,Huelga2013,killoran2015,Kreisbeck2012,Mal__2016,novelli2015vibronic,Santamore2013,schulze2015explicit,Stones2016}. However, to describe the dissipative dynamics of systems coupled to such environments requires advanced numerical and analytical techniques, and approaches ranging from many-body methods to advanced master equation formulations have recently been applied or developed for this aim \cite{Akihito09,deVega2017,Fruchtman2016,Hughes_2009,chin2010exact,prior2010efficient,Iles_Smith_2015,Makri_1998,manthe2008multilayer,Martinazzo_2011,Peter11,prior2013quantum,schroder2017multi,Strasberg_2016,Tanimura_1989,Thoss_2001}. 

In this article, we explore the real-time dynamics of the three-qubit model implemented by Poto{\v{c}}nik \textit{et al}, using the numerically exact hierarchical equations of motion (HEOM) technique \cite{Tanimura_1989, Ishizaki_2005, Tanimura_2006, Xu_2007, Schulten_2012, Ishizaki_2009, Shi_2009} to address a number of theoretical questions that could be verified in a future time-resolved version of the experiment. Specifically we shall show that the set up of Ref.\cite{potovcnik2018studying} is an ideal platform to demonstrate the generation of coherence by \emph{incoherent} processes, in this case caused by the noise-induced relaxation of a high energy state into a closely spaced doublet of lower energy states. This is a timely topic, as most observations of coherent optoelectronic phenomena only appear under coherent excitation, whereas functional light harvesting devices are likely to operate under incoherent illumination, i.e. sunlight \cite{brumer2016,han2013nature}. Without access to excitation sources that can generate electronic coherencs, i.e. laser pulses, future coherent devices must rely on transient internal or inter-component dynamics to induce wave-like phenomena. The present work offers some insight into how this might be obtained from an engineering of electronic eigenstates to match a structured bath spectral density, highlighting the importance of different energy scales in the problem. We also note that another type of quantum simulator for light-harvesting using trapped ions has also recently been demonstrated \cite{gorman2018engineering}. 

Specifically, we use HEOM to prove that although dissipative `population-to-coherence' processes are non-secular in nature, these often neglected transitions in the perturbative Redfield theory can, following decay of the high energy state, generate long-lasting coherence between the lower energy eigenstates. We demonstrate that these quantum beats are a manifestation of real-space coherent motions that could be detectable in a superconducting circuit experiment. To make further connections to general experimental conditions, we also show that interactions with both quantum and classically stochastic fluctuating environments can generate these coherent dynamics, and find that there is an optimal coupling for coherence generation that lies in an intermediate coupling regime. The possible role of non-Markovianity in these phenomena is also studied, as the structured spectral density we consider has a longer correlation time than the incoherent decay dynamics. However,  while we explicitly demonstrate that the use of HEOM is essential to account for the strong environmental memory effects (especially for classical noise), the correlation betwen the generation of coherence and the formal measure of non-Markovianity we use appears to be weak. 
 
The paper is organized as follows. Section II describes the three-state model and the system-bath interaction. Qualitative predictions and concepts emerging from Redfield approach are given in section III with a presentation of the operational HEOM equations. Section IV presents our numerical results for a quantum or classical noise and Section V provides some discussion and perspectives for future investigations.

\section{Models and parameters}
\subsection{Electronic system}
A virtue of the model `excitonic' Hamiltonian implemented in Ref. \cite{potovcnik2018studying} is that its dynamics, with typical timescales of $\mu$s in superconducting circuits, remains unchanged when all energetic parameters are rescaled to optical frequencies (factor $\approx 10^5$). In light of this, and in the interest of understanding coherence generation in molecular systems, we will retain the relative energy level and coupling structure of the qubits in Ref.\cite{potovcnik2018studying}, but work at the time/frequency scales of molecular optics and replace qubits with `sites' representating chromophores. 
   
The "electronic" system consists of three chromophoric sites (two-level systems) but the active Hilbert space is confined to the single excitation sector and therefore is described by three states $|n\rangle$ ($n=1,2,3$) corresponding to a single localised excitation on site $n$. Following the qubit topology of Fig. \ref{fig1}, the model Hamiltonian for the chromophore system is given by
		\begin{equation}
		H_S = \sum_{n = 1}^{3} \varepsilon _n \vert n \rangle \langle n \vert  + \sum_{n=1}^{3}\sum_{m\neq n =1}^{3}  J_{nm} \vert n \rangle \langle m \vert 
		\end{equation}
As in the experiment, the first two states ($\vert 1 \rangle $ and $ \vert 2 \rangle $) are tuned to degeneracy ($\epsilon_{1}=\epsilon_{2}=0$) and strongly coupled by a coherent coupling ${J_{12}}$ while state $\vert 2 \rangle $ is weakly coupled to a third state $\vert 3 \rangle $, through ${J_{23}} = {J_{12}}/10 $.  $J_{13}=0$, which is an approximation very close to the experimental realisation (qubit 1 and qubit 3 are not physically close to each other - see Fig.\ref{fig1}a. The energy gap between the degenerate levels and the lower state energy is equal to the first coupling $\epsilon_{3}  = -J_{12}$. Diagonalising this simple Hamiltonian then leads to the eigenstate spectrum shown in Fig. \ref{fig1}c, which is characterised by a single high energy state and a low-lying doublet of states with an energy splitting $\approx 2J_{12}$, approximately ten times smaller than the mean doublet-to-high-energy state gap. Due to this tuning of states, the eigenstates are highly delocalised over the sites. The high energy `bright' state is approximately given by $|B\rangle=\frac{1}{\sqrt{2}}(|1\rangle+|2\rangle)$, whereas the lower-energy `dark' states are given by $|D_{\pm}\rangle= \frac{1}{2}(|1\rangle-|2\rangle)\pm\frac{1}{\sqrt{2}}|3\rangle$. 

We remark that, here, `bright' and `dark' refers to the coupling of these states to the transmission line. As seen in Fig.\ref{fig1}a, sites $1\&2$ are close to the waveguide and both couple to the excitation field with the same strength. Consequently, the transition dipoles of these sites interfere constructively in the $|B\rangle$ eigenstate, making this `bright', while destructive interference decouples the two dark states $|D_{\pm}\rangle$ from the excitation fields. Experimentally, this configuration is very useful, as it allows optical population of a single, well-defined state from which transport then proceeds, while the non-emissive nature of the dark states prevents radiative losses and noise which might obscure the signatures of energy flow across the site network. Beyond practical considerations, it has also been proposed that using such dark states to reduce radiative losses could boost the efficiency of 'bio-inspired' organic photovolatics devices \cite{Chen2016,Creatore_2013,Dorfman_2013,guzik,killoran2015,Newman_2016,Scully2010,Zhang_2015}. Finally, the resonator is only coupled to state $|3\rangle$, so only $|D_{\pm}\rangle$ will emit into this channel. Experimentally, it is this resonator emission that is used to quantify the energy transfer from the $|B\rangle$ state. For clarity of discussion, we will not explicitly model the coupling of the system to the excitation and read-out fields, but instead consider dynamics beginning with a population prepared in the $|B\rangle$ state.         

\subsection{System-bath interaction: quantum and classically stochastic environments}
In the quantum simulator, noise is generated by applying a stochastic signal along the flux lines that are used to tune the excited state energies of the individual qubits. This effectively introduces site-selective, stochastic (gaussian) noise that is diagonal in the basis of the localised qubit excitations (pure dephasing noise), but this noise is essentially classical in nature (\textit{vide infra} and see section \ref{classical}). Nevertheless, this noise is generated by a signal generator that can produce almost arbitrary stochastic power spectra (see section \ref{sec:redfield}), providing a versatile tool for probing dissipative quantum transport. In order to connect with molecular systems, we will consider \emph{both} classical and quantum noise within a common framework in which the environment is treated as a continuum of quantum harmonic oscillators coupled linearly to the electronic system. Following Ref. \cite{potovcnik2018studying}, we will consider the case of noise applied to only one site of the network, site $2$, so that the quantum system-bath coupling is given by      
\begin{equation}		
		{H_{SB}} = SX = - \vert 2 \rangle \langle 2 \vert \frac{1}{\sqrt{2}} \sum\limits_k  g_k \left( {{a_k} + a_k^\dag } \right)
\end{equation}
where $S = \vert 2 \rangle \langle 2 \vert $ is the system operator and the displacement enviromental  operator is defined by $X =  - \frac{1}{\sqrt{2}}  \sum_{k} g_k \left( {{a_k} + a_k^\dag } \right)$ where  ${a_{k}^{\dagger},a_k}$ are the bosonic creation and annihilation operators, respectively, of an oscillator of frequency $\omega_{k}$ that is coupled to state $|2\rangle$ with amplitude $g_{k}$. The Hamiltonian of the oscillator bath is $H_{B}=\sum_{k}\omega_{k}a^{\dagger}_{k}a_{k}$ (with $\hbar  = 1$). The system bath coupling also leads to an energy shift $\lambda= -1/2 \sum_{k}g_{k}^{2} / \omega_{k}$ of state $|2\rangle$ (the reorganisation energy) that is added to the system Hamiltonian to define an effective system Hamiltonian $H_{S,eff}= H_S + \vert 2 \rangle \langle 2 \vert \lambda$. The total system-environment Hamiltonian becomes $H=H_{S,eff}+H_{SB}+H_{B}$. By diagonalising $H_{S,eff}$, all the eigen states are coupled through the environment by off diagonal terms of the system coupling operator in the eigenbasis set 
\begin{equation}
 V_{\lambda} = U^{-1}_{\lambda} S U_{\lambda}.
\label{V} 
\end{equation}
The reorganisation energy is an indicator of the coupling strength since it modifies the system coupling operator and the energy gap among the eigenstates. We shall use a dimensionless parameter 
\begin{equation}
\eta = \lambda / E_{BD_+}
\label{eta} 
\end{equation} 
giving the ratio between the renormalisation energy and the dissipation free $E_{BD_+}$ energy gap \cite{Thoss_2001}.

\begin{figure}
 \centering
	\includegraphics[width =1.\columnwidth]{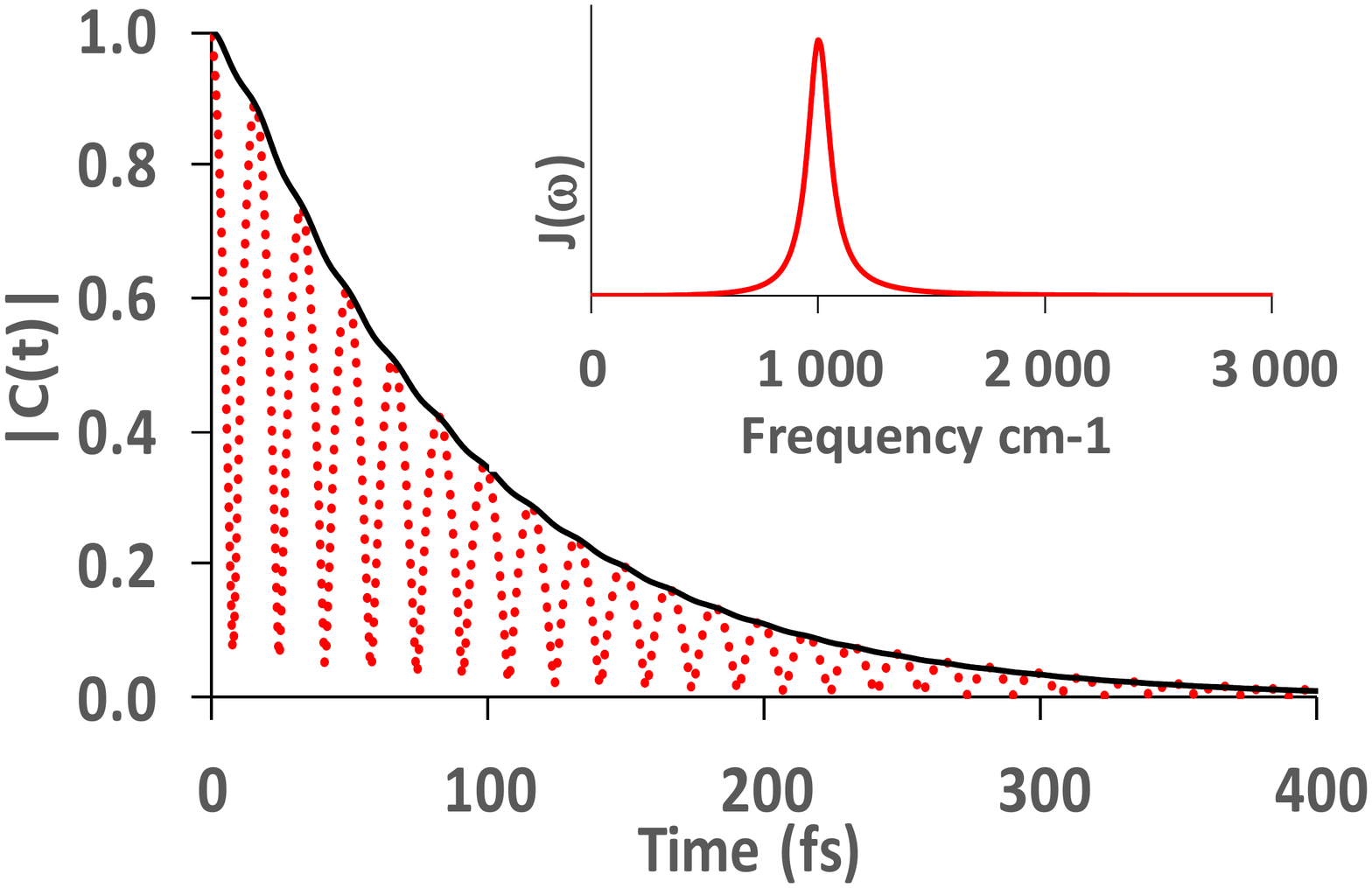}	
\caption{Modulus of the normalized two-time bath correlation function $C(t - \tau )$ in a.u. with $\tau $ = 0 at room temperature (full line) and in the high temperature limit used to simulate classical noise where the correlation function becomes a real oscillatory function (dots). Inset : corresponding spectral density in arbitrary units. The system-bath coupling strength and therefore the renormalisation energy or the $\eta$ parameter are scanned by varying the $p$ factor in Eq.\ref{J}. }
\label{fig:corre}
\end{figure}

Prior to excitation of state $|B\rangle$, we will always assume the environment osillators to be in thermal equilibrium w.r.t their free Hamiltonian $H_{B}$. With this common asumption, the behaviour of the reduced density matrix of the electronic subsystem is completely characterised by the environment's spectral function $J(\omega ) = \sum_k (g_k^2 / \omega _k ) \delta (\omega  - {\omega _k})$ and its temperature through the Bose function $n(\omega)=(e^{\beta\omega}-1)^{-1}$ where $\beta =1/{k_B} T$ and $k_B$ is the Boltzmann constant. Both appear in the thermal two-time correlation function of the oscillator displacements, which ultimately determines the dissipative physics of the system (see below) \cite{Breuer2002}. The correlation function is given by $C(t-\tau)=\mathrm{Tr}_{B}[\rho_{B}X(t)X(\tau)]$, where $\rho_{B}$ is the equilibirum density matrix of the environment oscillators and the time-dependent operators are in the Heisenberg picture w.r.t. the Hamiltonian $H_{B}$ of the environmental displacement operator $X$. This leads to    
\begin{equation}
		C\left( {t - \tau } \right) = \frac{1}{\pi }\int\limits_{ - \infty }^{ + \infty } {d\omega J\left( \omega  \right) n\left( \omega \right) e^{i\omega \left( t - \tau \right)}} .	
		\label{ct}
		\end{equation}
In the context of open system theory, the difference between quantum and classical stochastic noise is most clearly seen in $C(t)$; the two-time correlation function for non-commuting operators is complex-valued, while for a classical scalar variable it is real. From Eq. \ref{ct} and the fact that $J(-\omega)=-J(\omega)$, it can be seen that $C(t)$ becomes real in the limit of $\beta\rightarrow 0$ (high temperature limit) and we shall later exploit this to extract results about classical noise from our HEOM simulations. This is further illustrated in Fig. \ref{fig:corre} for the structured spectral density that we will consider in our numerical results.  The spectral density is here a superohmic Lorentzian function 
\begin{equation}
J\left( \omega  \right)=\frac{p{{\omega }^{3}}}{{{\Lambda }_{1}}({{\Omega }_{1}},{{\Gamma }_{1}}){{\Lambda }_{2}}({{\Omega }_{2}},{{\Gamma }_{2}})}.        
\label{J}
\end{equation}
where ${{\Lambda }_{k}}=\left[ {{\left( \omega +{{\Omega }_{k}} \right)}^{2}}+\Gamma _{k}^{2} \right]\left[ {{\left( \omega -{{\Omega }_{k}} \right)}^{2}}+\Gamma _{k}^{2} \right]. $

The parameters are chosen to create a sharp spectral density peaked at the energy gap $E_{BD}$, as shown in the inset of Fig. \ref{fig:corre}. The numerical values of the parameterisation used are given in Appendix \ref{appendix:spectraldensity}, with the $p$ parameter being used as a scaling factor that allows us to vary the reorganisation energy of the bath.

\section{Simulation techniques for obtaining and characterising reduced density matrices}
\subsection{Redfield Equations}\label{sec:redfield}
For open quantum systems, the principal object of interest is the reduced density matrix of the system $\rho_{S}(t)=\mathrm{Tr}_{B}[\rho(t)]$, where $\rho(t)$ is the joint density matrix of the system-environment state. In general, determining $\rho(t)$ in order to obtain $\rho_{S}(t)$ is highly demanding, however it is possible in many cases to find approximations that greatly simplify this task, and which also provide very useful concepts and intuition for discussing more complex open physics.

For the case of weak coupling (second order perturbation theory w.r.t. system-bath coupling) leading to a broad spectral function characterised by a cutoff frequency $\omega_{c}$ that is much larger than any energy difference in the system Hamiltonian $H_{S,eff}$, the dynamics of $\rho_{S}(t)$ are often well-described by the Bloch-Redfield theory. Detailed derivations of the Bloch-Redfield master equation can be found in Refs. \cite{blum2013density, Breuer2002}, here we shall simply state the results of immediate consequence for our results and discussion. Following the Born-Markov approximation, the equation of motion for the reduced density matrix in the interaction picture $\tilde{\rho}_{S}(t) =e^{-iH_{S,eff}t}\rho_{S}(t)e^{+iH_{S,eff}t}$ is given by the time-local master equation

\begin{eqnarray}
\dot{\tilde{\rho}}_{nm}(t)=\sum_{j,k}R_{nmjk}\tilde{\rho}_{jk}(t)e^{i(E_{nm}-E_{jk})t}, \label{BR}
\end{eqnarray}
where $\rho_{nm}=\langle E_{n}|\rho|E_{m}\rangle$ and $|E_{m}\rangle$ is the $m-th$ eigenstate of $H_{S,eff}$ with energy $E_{m}$, which we number in order of ascending energy. This form, containing explicit time-dependence, is often referred to the Non-Secular Bloch Redfield equation, distinguishing it from the Secular Bloch Redfield equations which are obtained from Eq.(\ref{BR}) by only retaining terms in the RHS summation for which $E_{nm}-E_{jk}=0$, where $E_{nm}=E_{n}-E_{m}$. This last approximation is normally justifed when the energy differences between different transition energies $E_{nm}-E_{jk}\gg R_{nnjk}^{-1}, \forall  n,m,i,j$, so that on the typical timescales on which the density matrix evolves under the action of the Redfield tensor $R_{nmjk}$, the highly oscillatory terms with $E_{nm}-E_{jk}\neq0$ average to zero. When this is indeed valid, the Secular Bloch Redfield equations have a particularly simple form, as population (diagonal) and coherence (off diagonal) terms of $\rho_{S}$ are completely decoupled. The populations then obey a Pauli (kinetic) master equation, while any coherences present in the intial condition independently and exponentially decay to zero. Crucially, there are no terms in the SBR equations that allow for the \textit{ex nihilo} generation of coherences.         

However, non-secular terms can create dynamical coupling between populations and coherences, as has been widely discussed in the context of ultrafast spectrosopies. The significance of these terms has also recently been highlighted in a number of papers, showing that their inclusion often provides a more accurate description when compared with more advanced numerical treatments of open quantum systems  \cite{jeske2015bloch}. Interestingly, well-known problems related to the potential lack of positivity of reduced density matrices under BR evolution have also been shown to arise from problems related to the failure of the Born-Markov asumption, rather than the structure of the master equations when non-secular terms are included \cite{eastham2016bath}.

\subsection{Non-secular generation of spontaneous coherence via incoherent decay}

Of particular relevance for our three-level model is the structure of the population-to-coherence elements of the Redfield tensor $R_{nmii}=R_{mnii}^{*}$, especially the term $R_{D_{+}D_{-}BB}$ which describes the generation of coherence by a population in the initially excited state $|B\rangle$. This is given by 

\begin{eqnarray}
R_{D_{+}D_{-}BB}&=&\pi V_{D_{+}B}^{*}V_{D_{-}B}J(E_{BD_{+}})(n(E_{BD_{+}})+1) \nonumber \\ 
&+&\pi V_{D_{+}B}^{*}V_{D_{-}B}J(E_{BD_{-}})(n(E_{BD_{-}})+1)
\label{RBD}
\end{eqnarray} 
Comparing this to the incoherent decay rate from state $|B\rangle$ to the lower energy doublet of states (population-to-population transfer) 
\begin{eqnarray}
R_{D_{+}D_{+}BB}&=&2\pi |V_{BD_{+}}|^{2}J(E_{BD_{+}})(n(E_{BD_{+}})+1))\\
R_{D_{-}D_{-}BB}&=&2\pi |V_{BD_{-}}|^{2}J(E_{BD_{-}})(n(E_{BD_{-}})+1)),
\label{RBBDD}
\end{eqnarray}   
we can see that, if $J(E_{D_{\pm}B}), n(E_{BD_{\pm}})$ and matrix elements $V_{D_{\pm}B}$ are equal, then $R_{D_{+}D_{+}BB}=R_{D_{-}D_{-}BB}=R_{D_{+}D_{-}BB}$. For this case, and in the absence of the time-dependence associated with the term $R_{D_{+}D_{-}BB}$ in Eq. \ref{BR}, the incoherent decay of state $B$ would create population and coherence in the lower doublet at an equal rate, i.e would tend to generate a coherent superposition of the lower doublet states. Indeed, if the decay of the $|B\rangle$ state is much faster than the oscillation period (set by $E_{D_{+}D_{-}}$), one would expect the relaxation to occur into a pure superposition state of $|\psi\rangle\approx|D_{+}\rangle\pm|D_{-}\rangle$ ( with phase depending on the sign of the matrix elements). If the decay is much slower than the oscillatory period, the coherence generation will be negligible, leading to an incoherent mixture of (approximately equal) populations in the doublets. 

Within the scope of BR theory, the eigenstate level scheme and spectral density that we consider enables the conditions for coherence generation to be almost perfectly met at least for small reorganisation energy when the model parameters remain close to those predicted by the eigenstates of $H_S$. Firstly, the matrix elements for transition between the $|B\rangle$ state and the doublets $|D_{\pm}\rangle$ induced by the spatially local operator $V_{\lambda}$ (\ref{V}) are almost equal, due to the real-space delocalisation of the states and the large energy gap between the $|B\rangle$ and $|D_{\pm}\rangle$ manifolds (their fractional difference is no more than $\approx J_{23}/J_{12}\ll1$). Secondly, by applying a spectral function that is symmetric and peaked at an energy $\frac{1}{2}(E_{BD_{+}}+E_{BD_{-}})$, we obtain equality of $J(E_{D_{\pm}B})$. Thirdly, by working at a temperature such that $\beta E_{BD_{\pm}}\gg1$, or $\beta E_{BD_{\pm}}\ll1$ all relevant elements of the Redfield tensor approach equality. This only leaves the condition that the transition rates for decay of the $|B\rangle$ state should be faster than the period of coherent oscillations in $|D_{\pm}\rangle$ states. This can be controlled by varying the coupling strength to the environment, although we note that obtaining this condition violates the standard application of the Markov approximation, additionally motivating our use of non-perturbative HEOM methods (see below).
  
Finally, we note that the same qualitative analysis can be applied to coherences generated between the $|B\rangle$ and $|D_{\pm}\rangle$ states due to decay of the $|B\rangle$ state. However, the relevant oscillatory time period to compare to the decay rate is now set by the bright-dark state splitting $E_{BD}$, which is ten times larger than $E_{D_{+}D_{-}}$. It could therefore be possible to find a parameter space in which large inter-doublet coherence is generated \emph{without} any significant coherence generation between the bright and dark states. This scenario defines what we mean by coherence generation arising from \emph{incoherent} relaxation. 

\subsection{HEOM}

We recall here the derivation already presented in previous works \cite{Tanimura_1989, Ishizaki_2005, Tanimura_2006, Xu_2007, Schulten_2012, Ishizaki_2009, Shi_2009} and give, for a purpose of completeness, the main equations.
For the efficiency of the HEOM algorithm, the spectral density is parametrized so that the two-time bath correlation function is expressed as a sum of complex exponential functions \cite{Shi_2014}: 
				\begin{equation}
				C\left( {t - \tau } \right) = \sum\limits_{k = 1}^{{n_{cor}}} {{\alpha _k}{e^{i{\gamma _k}\left( {t - \tau } \right)}}}. 
				\label{Corre}
				\end{equation}
Explicit expressions of the ${\alpha _k}$ and ${\gamma _k}$ by the analytical integration of Eq.(\ref{ct}) with the superohmic parametrization of the spectral density (\ref{J}) can be found in the Appendix of ref. \cite{Mangaud_2017}. ${n_{cor}}$ is the sum of the four terms coming from the four simple poles in the upper complex plane and, in principle, an infinite number of terms related to the poles (Matsubara frequencies) of the Bose function on the imaginary axis   $\nu_j$. In practice, we find that the number of Matsubara terms remains small (about 10) at and above room temperature for the model under study. The complex conjugate of the correlation function can be expressed by keeping the same coefficients ${\gamma _k}$ in the exponential functions with modified coefficients ${\tilde \alpha _k}$ according to :
				\begin{equation}
				{C^*}\left( {t - \tau } \right) = \sum\limits_{k = 1}^{{n_{cor}}} {{{\tilde \alpha }_k}} {e^{i{\gamma _k}\left( {t - \tau } \right)}}
				\label{Correconj}
				\end{equation}
				with	${\tilde \alpha _{1}} = \alpha _{2}^*$, ${\tilde \alpha _{2}} = \alpha _{1}^*$, ${\tilde \alpha _{3}} = \alpha _{4}^*$, ${\tilde \alpha _{4}} = \alpha _{3}^*$ and ${\tilde \alpha _{j,matsu}} = {\alpha _{j,matsu}}$ where the ${\alpha _{m}}$ with $m = 1,4$ are related to the four poles of the superohmic Lorentzian function and ${\alpha _{j,matsu}}$ refer to the  Matsubara terms \cite{Tannor_2010}.

By assuming an initial factorization of the total density matrix, the time evolution of the reduced density matrix, in interaction representation $\tilde{\rho}_{S}(t)$ is given by		
\begin{eqnarray}
\tilde{\rho}_{S}(t) = &T{r_B}\left[ {{e^{\int\limits_0^t {d\tau L\left( \tau  \right)} }}\rho _B^{eq}{\tilde{\rho}_{S}(0)}} \right] \nonumber \\
= & e^{\int\limits_0^t {d\tau \int\limits_0^\tau  {dt'T{r_B}\left[ {L\left( \tau  \right)L\left( {t'} \right)\rho _B^{eq}} \right]} } }\tilde{\rho}_{S}(0)
\end{eqnarray}
where $L\left( t \right)\centerdot =-\frac{i}{\hbar }\left[ S\left( t \right)X\left( t \right),\centerdot  \right]$  is the Liouvillian of the system-bath interaction with the system coupling operator in interaction representation $S\left( t \right) = {e^{i{{\hat H}_{S,eff}}t}}\hat S{e^{ - i{{\hat H}_{S,eff}}t}}$ and the bath operator as given above. Expressions (\ref{Corre}) and (\ref{Correconj}) for $C\left( {t - \tau } \right)$ and ${C^*}\left( {t - \tau } \right)$, correspond to the relaxation of $n_{cor}$ effective bath modes. Each set of the corresponding occupation numbers is represented by a collective index ${\bf{n}} = \left\{ {{n_1}, \cdots ,{n_{{n_{cor}}}}} \right\}$ and is associated to an auxiliary density matrix.  The master equation is then written as a time-local hierarchical system of coupled equations among the auxiliary operators. Each matrix can communicate only with the superior and inferior level in the hierarchy for which one occupation number is varied by one unity ${\bf{n}}_k^{\pm}  = \left\{ {{n_1}, \cdots ,{n_k} {\pm}, \ldots ,{n_{{n_{cor}}}}} \right\}$ :
\begin{eqnarray}
\mathop {{\rho _{\bf{n}}}}\limits^\bullet \left( t \right) &=& i\sum\limits_{k = 1}^{{n_{cor}}} {{n_k}{\gamma _k}{\rho _{\bf{n}}}\left( t \right)} 
 - i\left[ {S\left( t \right),\sum\limits_{k = 1}^{{n_{cor}}} {{\rho _{{\bf{n}}_k^ + }}\left( t \right)} } \right] \nonumber \\
  &-& i\sum\limits_{k = 1}^{{n_{cor}}} {{n_k}\left( {{\alpha _k}S\left( t \right){\rho _{{\bf{n}}_k^ - }} - {{\tilde \alpha }_k}{\rho _{{\bf{n}}_k^ - }}S\left( t \right)} \right)} 
  \label{heom}
\end{eqnarray}

In this hierarchy of auxiliary density matrices, the system density matrix is given by top row, i.e. for ${\bf{n}} = \left\{ {0, \cdots ,0} \right\}$ hence ${ \tilde \rho_S}\left( t \right) = {\rho _{\left\{ {0, \cdots ,0} \right\}}}\left( t \right)$. The level of the hierarchy is chosen until convergence is reached for the system density matrix. As previously stated, it can be seen that the equations of motion that determine the reduced density matrix of the system are completely determined by the expansion coefficients of the correlation functions $C(t)$ and $C^{*}(t)$ that appear in Eqs.\ref{Corre} and \ref{Correconj}, and which are ultimately determined by the environment spectral function and temperature.   

The HEOM equations are efficient to go beyond the second order perturbative regime of the Redfield equations even if the Markovian approximation could still be valid at higher orders. However, stronger couplings are often linked to non-Markovian dynamics.  Signature of non-Markovian behaviour for strong system-bath coupling is analyzed in appendix \ref{appendix:NM} where we also illustrate convergence of HEOM equations and compare numerically exact HEOM simulations with Redfield results.  

\section{RESULTS}
\subsection{QUANTUM NOISE}
\label{sec:quantumnoise}

\begin{figure}
  \centering
	\includegraphics[width =1.\columnwidth]{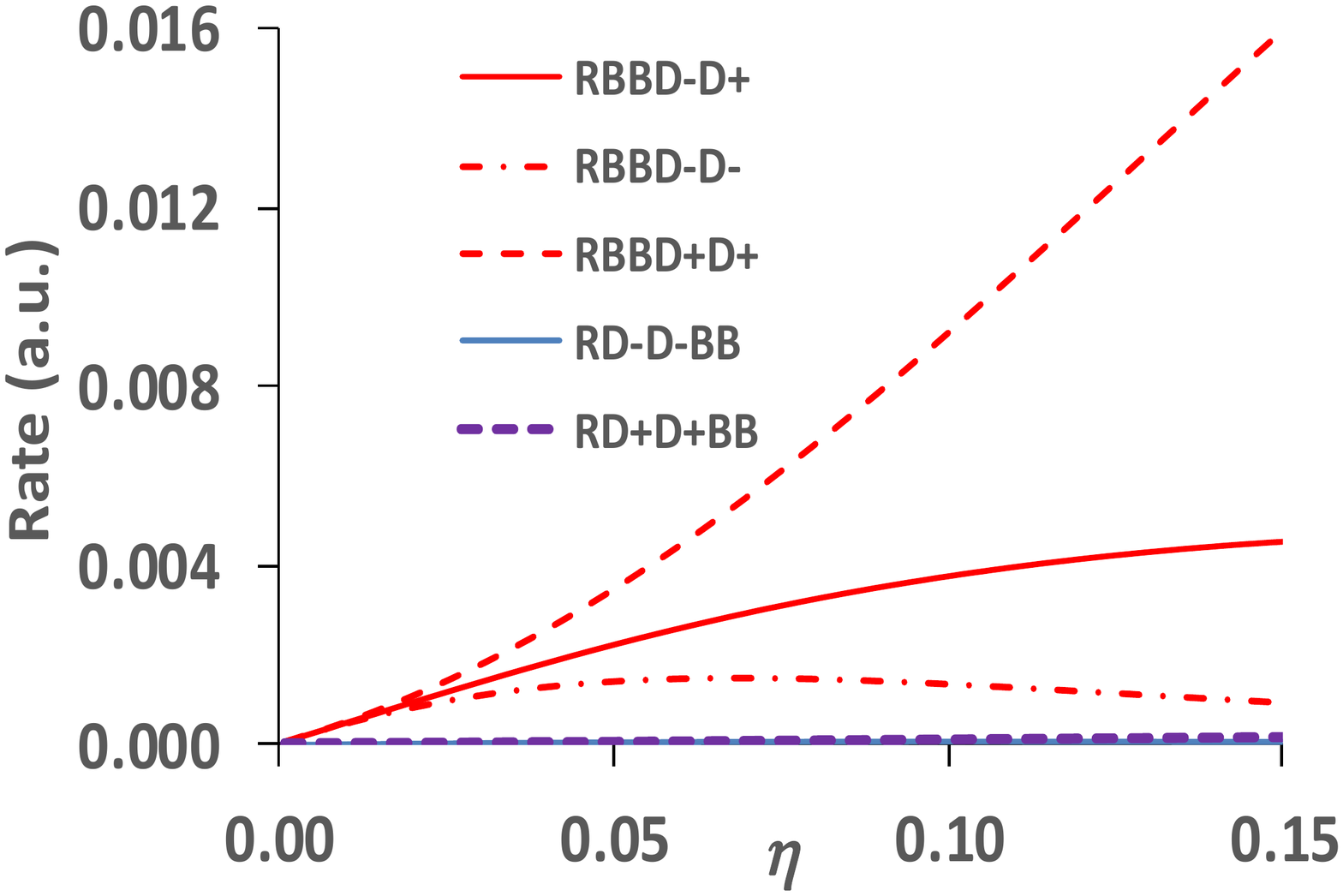}
\caption{ Main Redfield tensor elements for population to population or population to coherence transfer as a function of the coupling $\eta$ parameter Eq.\ref{eta}.}
\label{fig:rate298}
\end{figure}

In each simulation, the initial state is the bright eigenstate $\vert B \rangle $ and the bath is at room temperature $T$ = 298$K$ which corresponds to quantum noise for our spectral density, as the peak frequency $\Omega\approx 5k_{B}T$. We note, however, that this temperature give an energy scale $k_{B}T/E_{D_{+}D_{-}}\approx3$, which would be expected to drive equal (mixed) populations of the doublet population and rapid coherence loss. From the analysis based on the Redfield theory, we first examine the evolution of the main tensor elements to predict the best range of the $\eta$ parameter (Eq.\ref{eta}) to create the expected long lived superposed state in the doublet. The three main tensor elements of the `downhill' transitions at $T$ = 298$K$ $R_{D_{+}D_{+}BB}, R_{D_{-}D_{-}BB}, R_{D_{+}D_{-}BB}$ are displayed in Fig.~\ref{fig:rate298} as a function of the $\eta$ parameter by accounting for the variation of the coupling $V_{\lambda}$ (Eq.\ref{V}) and of the eigen energy gap induced by the renormalization energy. The best expected domain for the coherence creation appears to be around $\eta = 0.015$ which corresponds here to a renormalisation of about 15 cm$^{-1}$. This value provides the largest decay rate while maintaining equality of the three relevant Redfield tensor terms. The reverse rates $R_{BBD_{-}D_{-}}, R_{BBD_{+}D_{-}}$ remain negligible at room temperature due to detailed balance, potentially extending the lifetime of any decay-generated coherent states.  Simulation with HEOM will allow us to probe stronger couplings beyond the perturbative regime and examine the stability of the process. 

\begin{figure}
  \centering
	\includegraphics[width =1.\columnwidth]{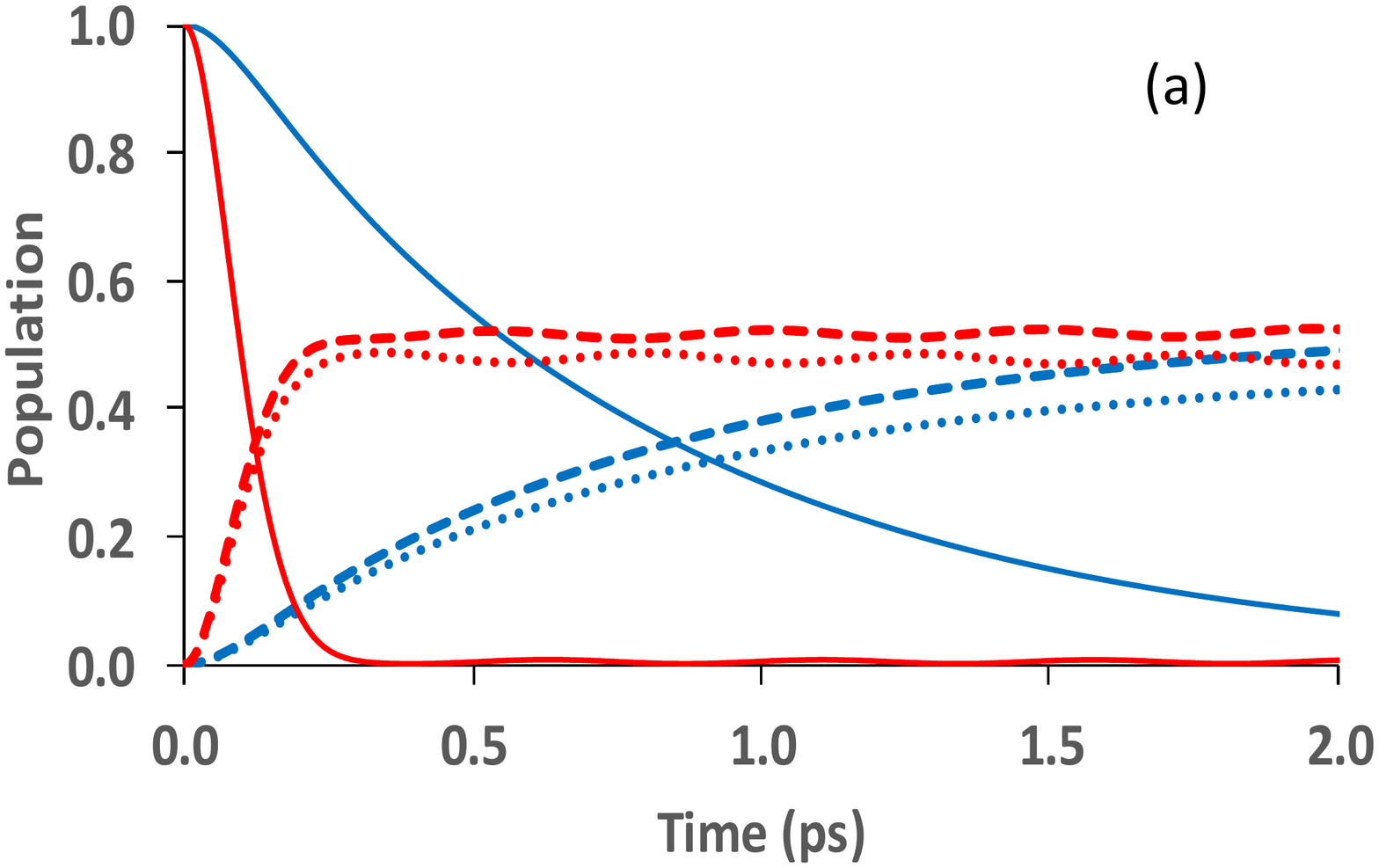}
	\includegraphics[width =1.\columnwidth]{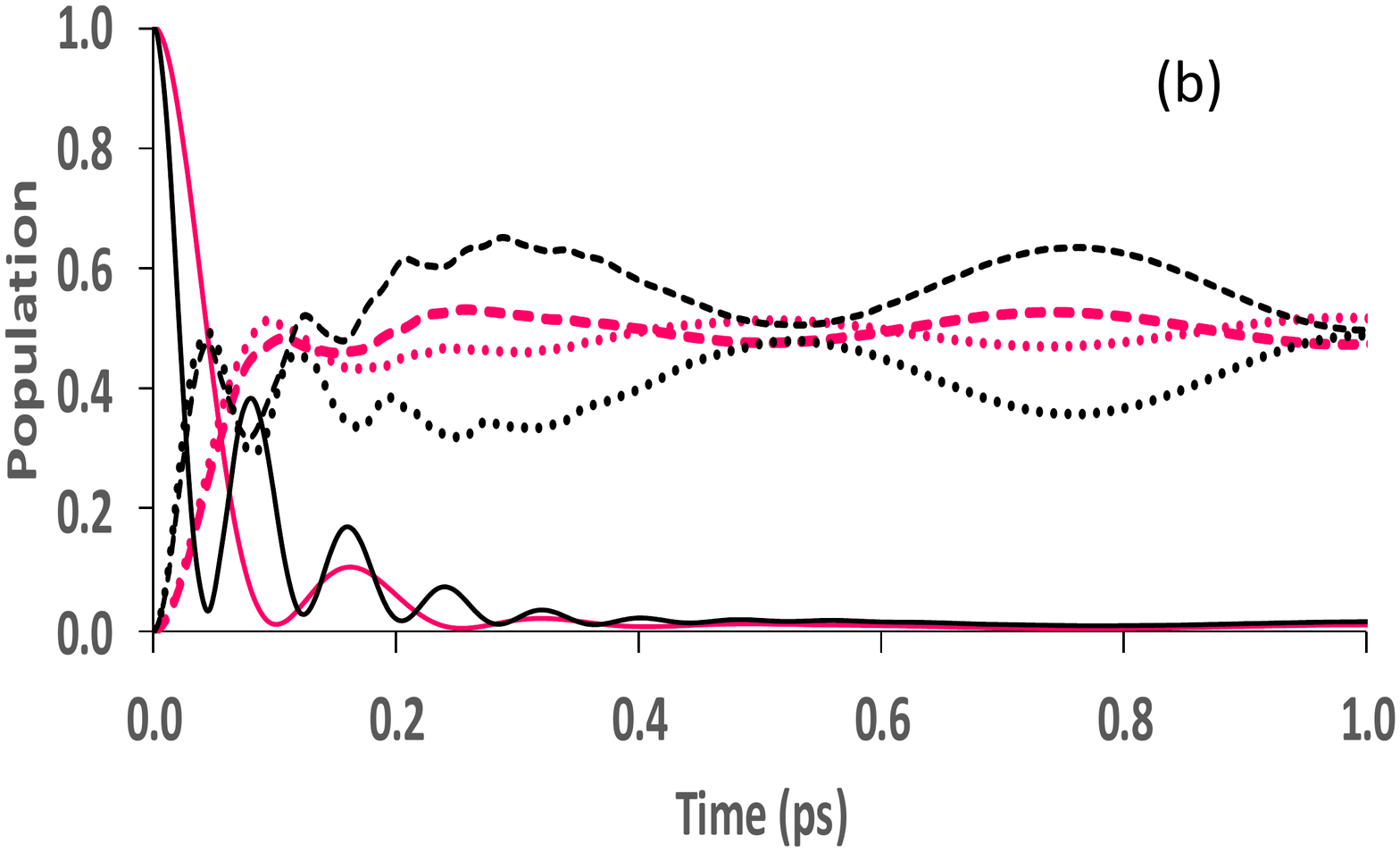}
\caption{ Population evolution of the bright state $\vert B \rangle $ (full lines) and of the dark doublet $\vert {{D_+}} \rangle $ (dots) and $\vert {{D_-}} \rangle $ (dashes) for different coupling strengths. Panel (a) weak system bath-coupling $\eta$ = 0.001 (blue) and $\eta$ = 0.01 (red); panel (b) strong coupling $\eta$ = 0.04 (pink) and $\eta$ = 0.16 (black).}
\label{fig:pop298}
\end{figure}

Figure ~\ref{fig:pop298} shows the population evolution (i.e. the diagonal elements of $\rho_S (t)$) in the $\vert B \rangle $ state (full lines) and in the doublet $\vert {{D_+}} \rangle $ (dashes) and $\vert {{D_-}} \rangle $ (dots) for different coupling regimes. In the perturbative regime (Fig.~\ref{fig:pop298}a), the decay is monotonous while for the strong coupling case (Fig.~\ref{fig:pop298}b), oscillations occur which are related to quasi-reversible energy exchange between the system and the environment, as the coupling strength exceeds the line width of the spectral function (strong-coupling cavity limit). These latter dynamics lead to features in the measure of non-Markovianity we present in Appendix \ref{appendix:NM}, as predicted in ref. \cite{Paternostro_2011}. The possible creation of a superposition state is suggested by the simultaneous growing of population in the $\vert {{D_{\pm}}} \rangle $ doublet states, although this could also arise without generating any coherence in the doublet. Close to the best expected coupling regime for $\eta $ = 0.01 (red curves), the populations rise monotonically and plateau to equilibrium values expected from the Boltzman distribution at this temperature. The populations of the two doublet states show oscillatory behaviour in the strong regime $\eta $ = 0.16 (black curves), with a period and duration much longer than the oscillations seen in the decay of the bright state. We will return to this non-perturbative effect in Section \ref{classical}.   

\begin{figure}
\includegraphics[width =1.\columnwidth]{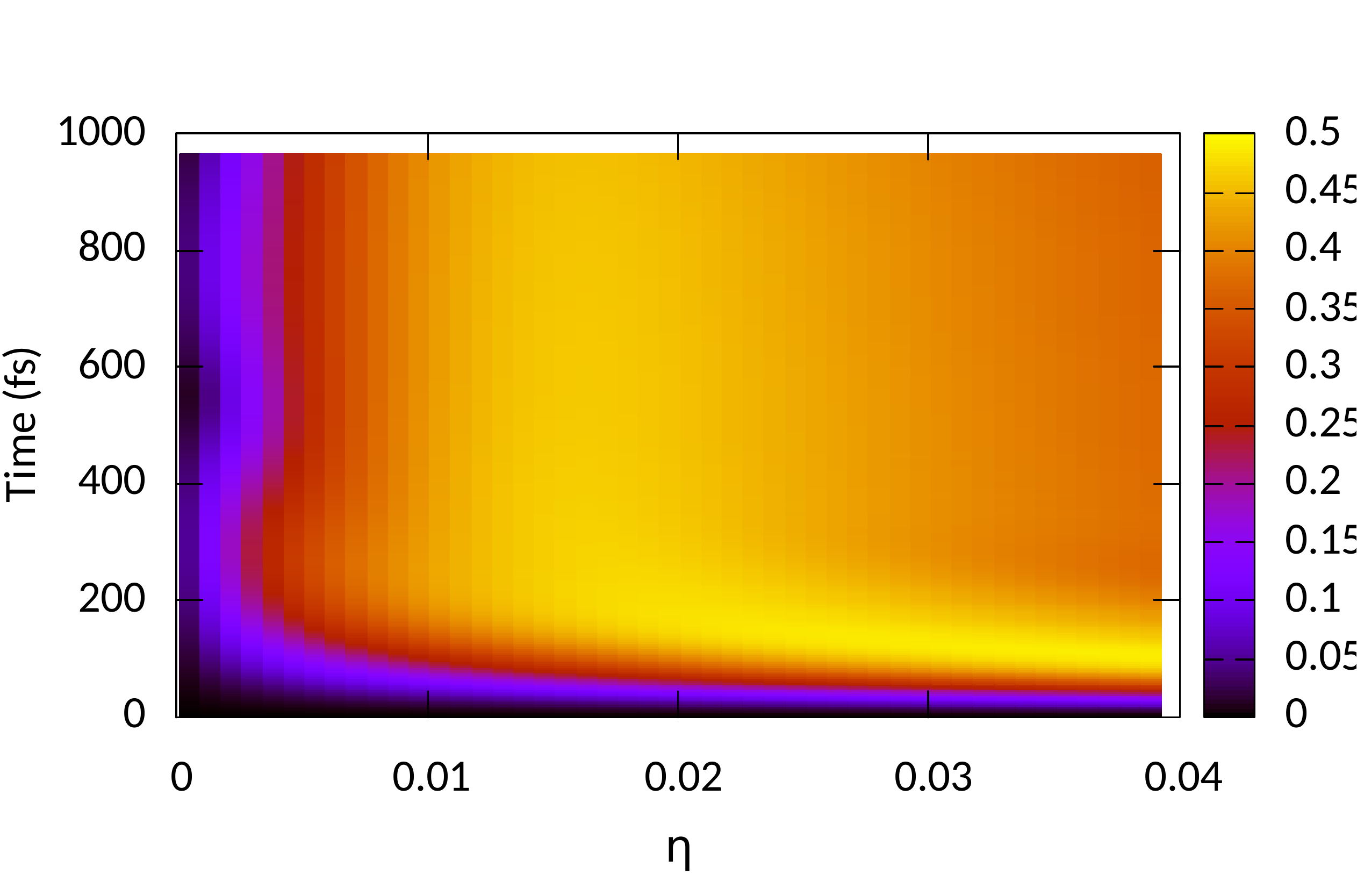}
\caption{Iso-value contours in the modulus of the coherence between the doublet states ${\rho _{{D_+}{D_-}}}(t)$ for different $\eta$ parameters (Eq. \ref{eta}) at $T =$298$K$. }
\label{fig:cohe}
\end{figure}

\begin{figure}
\includegraphics[width =1.\columnwidth]{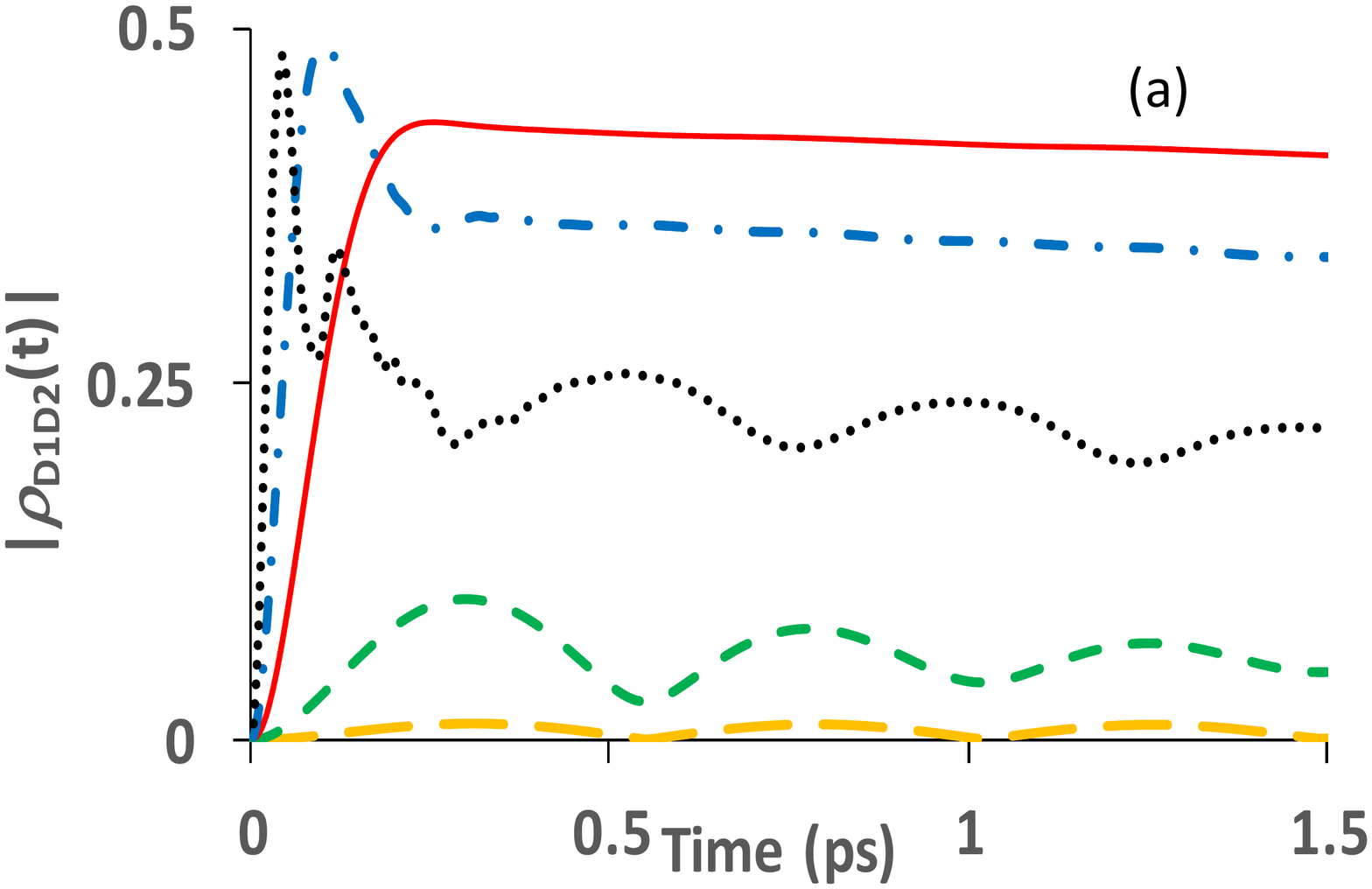}
\includegraphics[width =1.\columnwidth]{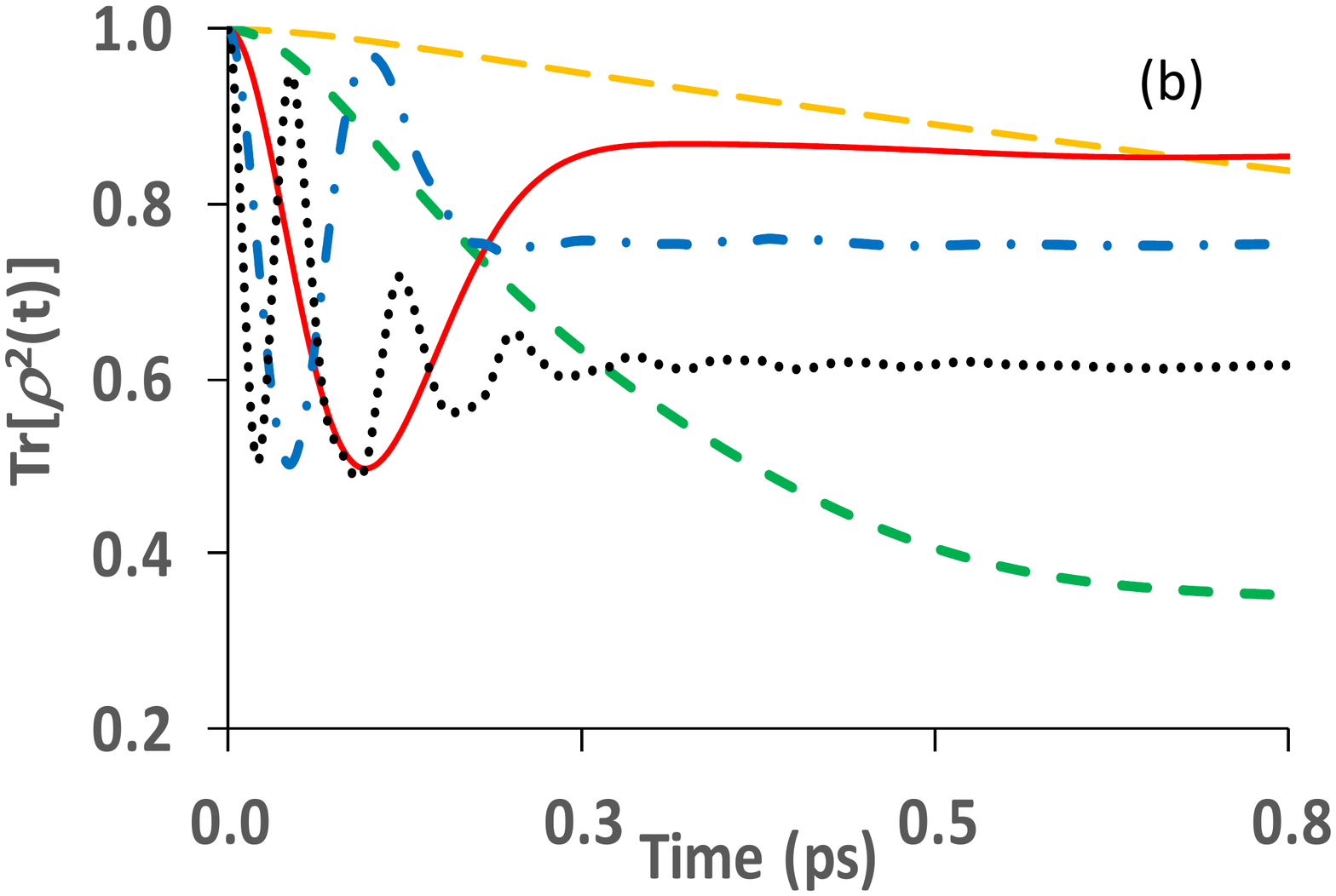}
\caption{Panel (a) : Modulus of the coherence between the doublet states ${\rho _{{D_+}{D_-}}}(t)$ for different $\eta$ parameters. Panel (b) : Purity of the system matrix density $Tr\left[ {{\rho_S^2}(t)} \right]$ for the same $\eta$. Dots : $\eta$ = 0.16; dashes-dots: $\eta$ = 0.04; full line : $\eta$ = 0.01, short dashes : $\eta$ = $10^{-3}$, long dashes : $\eta$ = $10^{-4}$.  }
\label{fig:purity}
\end{figure}

The critical observable, the modulus of the coherence $\vert {{\rho _{{D_+}{D_-}}}(t)} \vert$ between the doublet states, may be seen in figure~\ref{fig:cohe} as contourplots in a time and   $\eta$ parameter map or in figure~\ref{fig:purity}a for some selected couplings. The amplitude of the created coherence shows a clear dependence on the system-bath coupling. In the optimal situation corresponding to values near $\eta $ = 0.015, as predicted by the Redfield analysis, the coherence modulus reaches very close to the maximum possible value of 0.5 in about 100 fs and remains stable for more that 1 ps. For very weak coupling, a coherence is observed but its amplitude remains below 0.1. For the strong coupling, the early dynamics leads to a high amplitude around 0.5 but due to the bath interaction the asymptotic value stabilizes below the optimal coherence.

The purity of the system density matrix  $Tr\left[ {{\rho_S^2}(t)} \right]$ is given in figure \ref{fig:purity}b. In the weak coupling cases ( $\eta = 10^{-4}$ or $10^{-3}$), the purity is mainly determined by the mixed state with the initial state which is not yet relaxed. The purity confirms that the most favorable situation is the moderate coupling case around $\eta $ = 0.01, where the asymptotic purity is well above the purity of a Boltzmann mixture at room temperature ($\approx 0.5$ ) and shows that incoherent relaxation produces a superposition state in the doublet with relatively little entropy generation. 

\begin{figure}
 \includegraphics[width =1.\columnwidth]{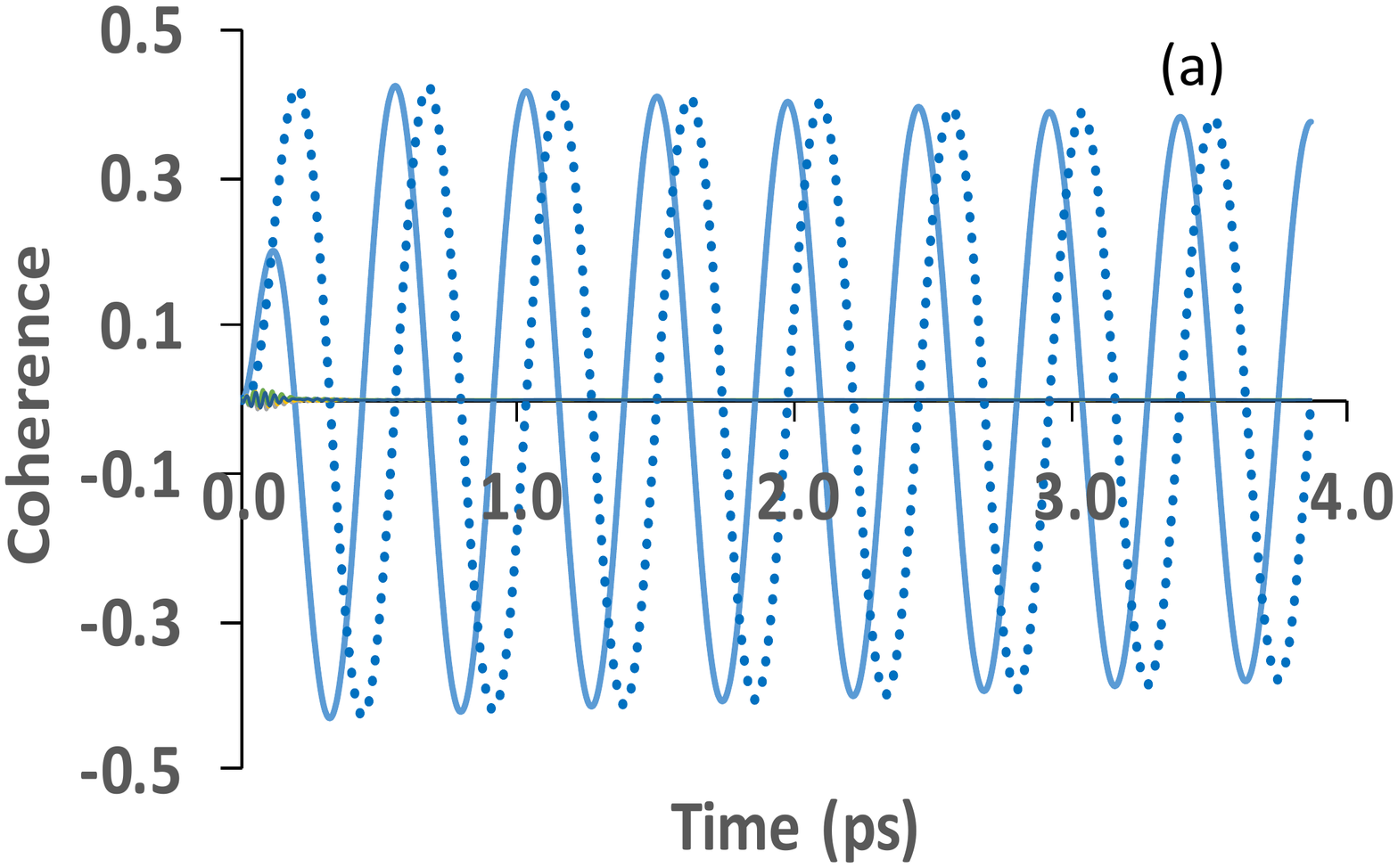}
  \includegraphics[width =1.\columnwidth]{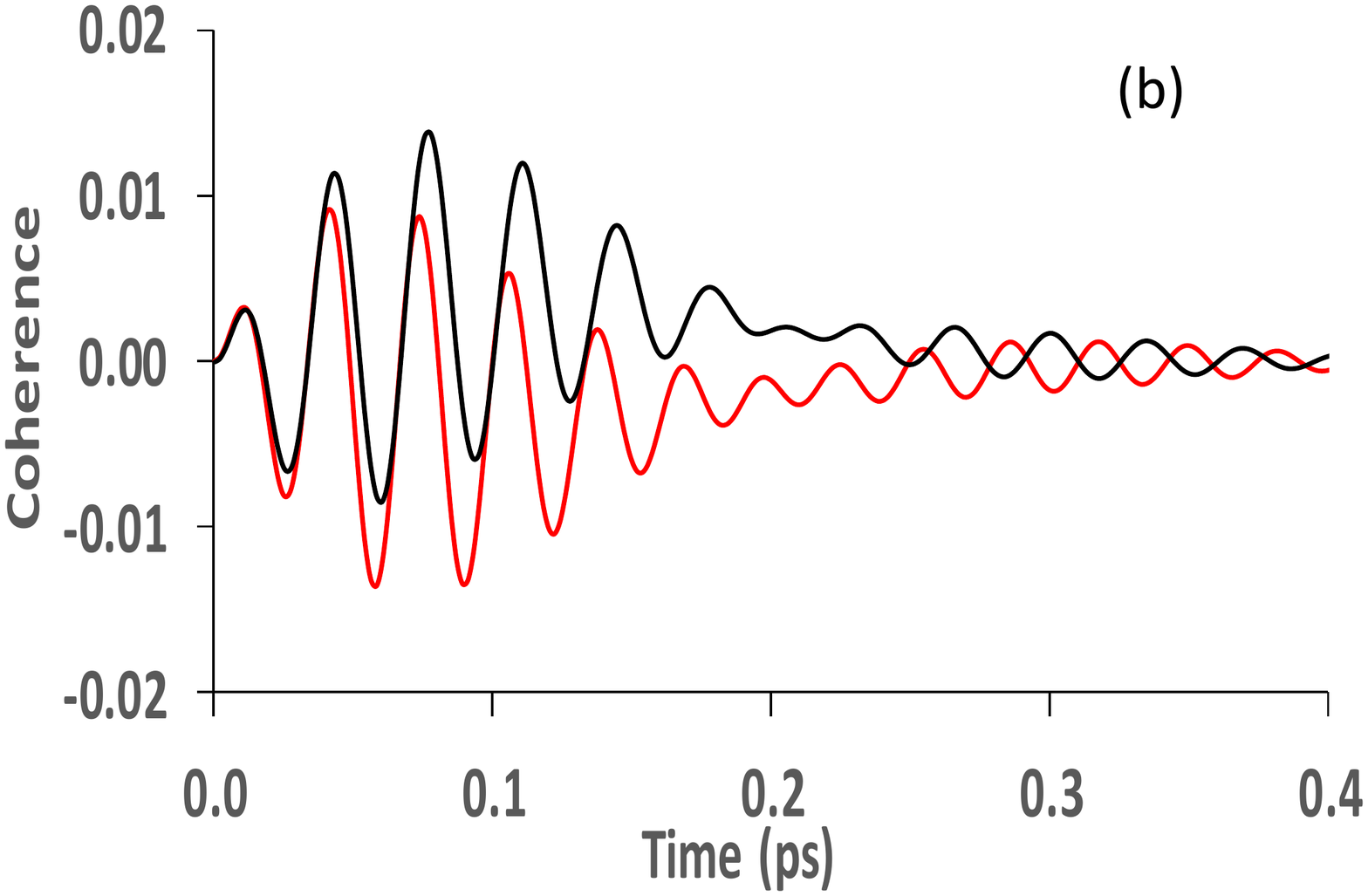}
\caption{ Panel (a) : Coherence ${\rho _{{D_+}{D_-}}}(t)$ (full blue line : real part, blue dashes : imaginary part ) for the  $\eta$ = 0.01. All the other coherences are in black lines. Panel (b): zoom of the real part of the coherences ${\rho _{B{D_-}}}(t)$ (red line) and ${\rho _{B{D_+}}}(t)$ (black line).}
\label{fig:allcohe}
\end{figure}

Figure~\ref{fig:allcohe}a illustrates the stability of the created coherence ${\rho _{{D_+}{D_-}}}(t)$ for a favourable case with $\eta $ = 0.01 and the difference with the other coherences ${\rho _{B{D_-}}}(t)$ or ${\rho _{B{D_+}}}(t)$. As shown in figure~\ref{fig:allcohe}b, the latter completely disappears after 500 fs and their amplitudes never exceed $0.015$ so they are smaller by more than one order of magnitude. As previously discussed, this establishes the coherence generation arises from an \emph{incoherent} decay. 

\begin{figure}
\includegraphics[width =1.0\columnwidth]{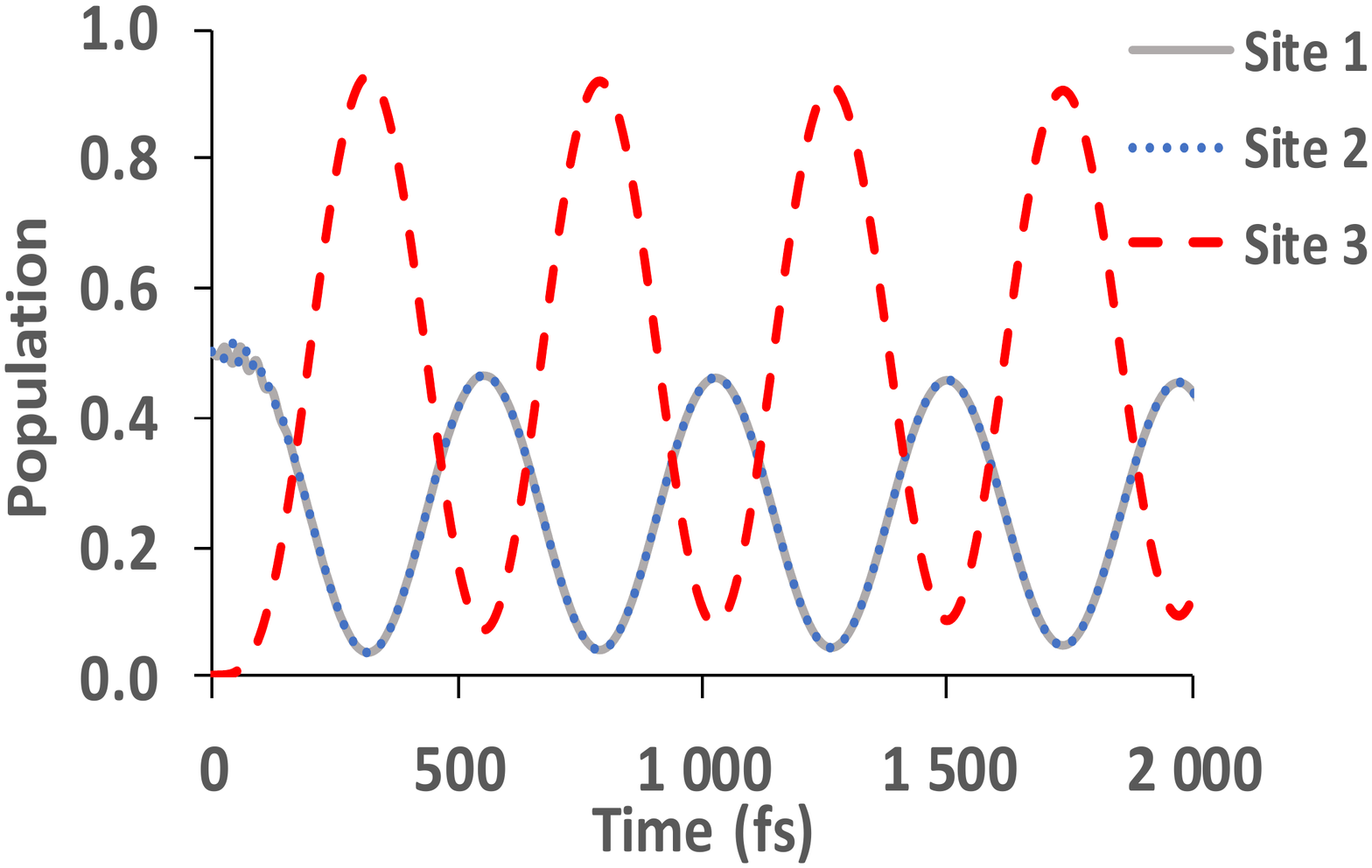}
\caption{Populations in the real-space site basis of the coupled network for coupling to quantum noise ($\eta=0.01$).}
\label{quantumsites}
\end{figure}

These beats could potentially be detected in an experimental set up similar to Ref. \cite{potovcnik2018studying}. The rapid relaxation effectively prepares a nearly pure superposition state that coherently evolves over a subsequent time $t$ as $|\psi (t)\rangle\approx\frac{1}{\sqrt{2}}(e^{-iE_{D_{+}}t}|D_{+}\rangle+e^{-iE_{D_{-}}t}|D_{-}\rangle)$. Expanding this state in the site basis and noting that $|D_{+}\rangle+|D_{-}\rangle=|1\rangle-|2\rangle$ and $|D_{+}\rangle-|D_{-}\rangle=|3\rangle$, it can be easily seen that the evolution of the wave function phases leads to oscillatory real-space motion of the excitation between sites $1\&2$ and $3$. This is illustrated in Fig. \ref{quantumsites}, where a periodic and near-unity population of site three with frequency $E_{D_{+}D_{-}}$ can be seen. As the resonator emission only arises from population of site $3$, oscillations in its population should be observable as a periodic modulation in the resonator signal at a frequency $E_{D_{+}D_{-}}$.

\begin{figure}
\centering
  \includegraphics[width =1.\columnwidth]{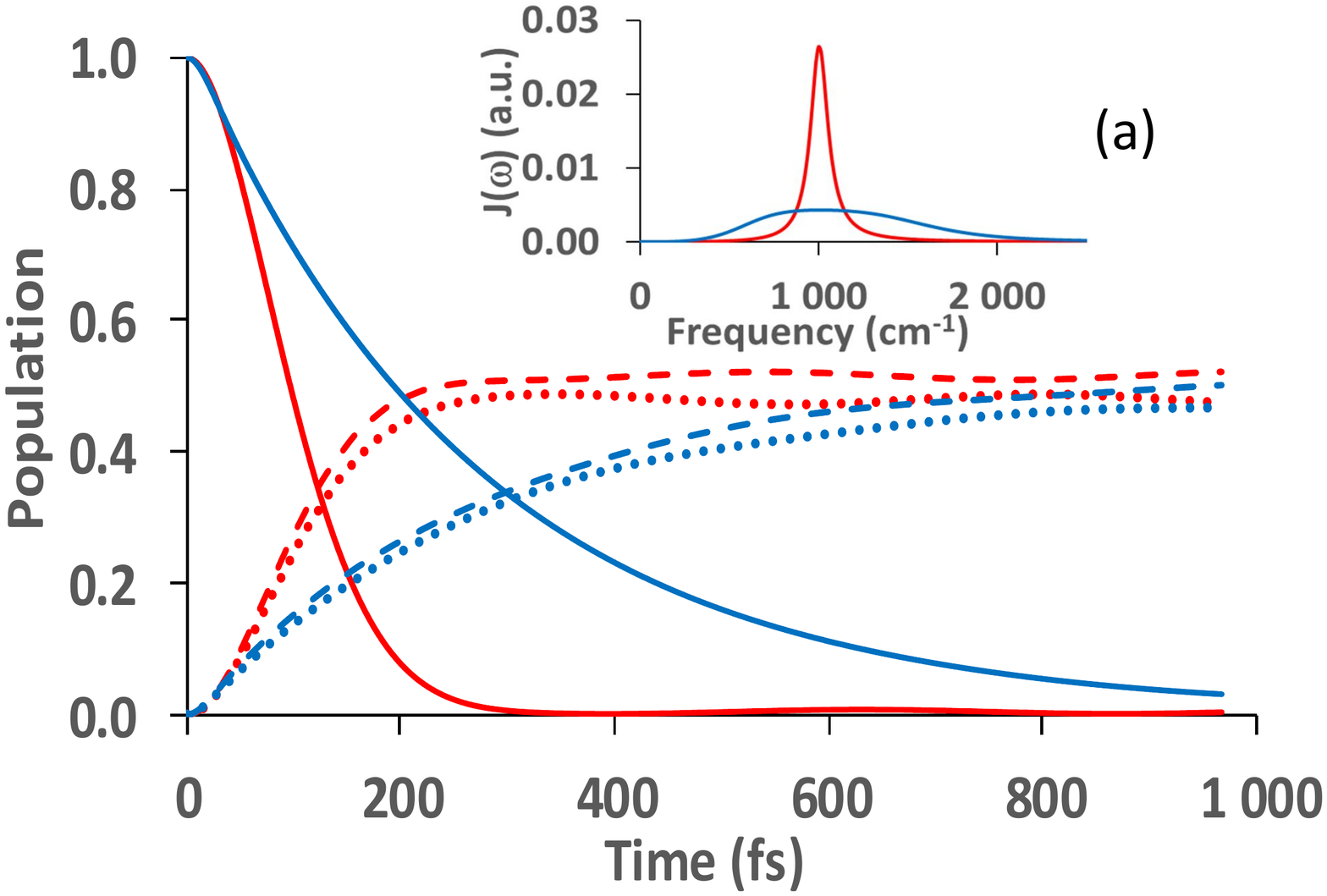} 
  \includegraphics[width =1.\columnwidth]{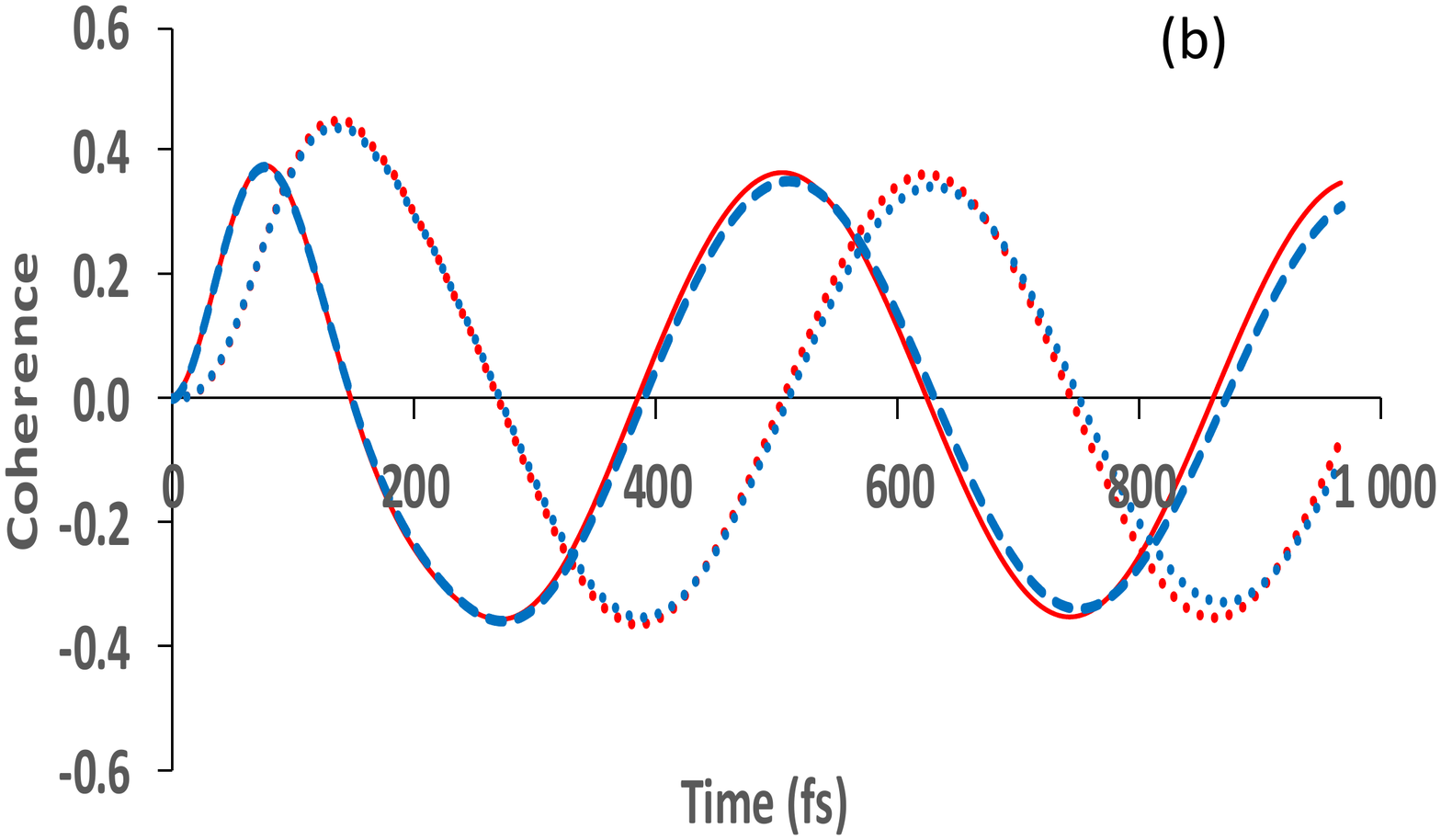} 
\caption{Comparison of the population evolution and of the coherence ${\rho _{{D_-}{D_+}}}(t)$ in the doublet states for the two spectral density shown in the inset. The renormalization energies correpond to $\eta$ = 0.01. Thin peak : red curves and broad peak : blue curves. Upper panel: population in the bright state $\vert B \rangle $ (full lines), dark states $\vert D_- \rangle $ (dots) and $\vert D_+ \rangle $ (dashes). Lower panel: real part of the coherence ${\rho _{{D_-}{D_+}}}(t)$ (thin peak : full lines, broad peak : dashes) and imaginary parts (thin peak : red dots and broad peak : blue dots).}
\label{fig:peak}
\end{figure}

The stability of the coherence creation via the incoherent decay was checked with respect to the shape of the spectral density. We compare the sharp spectral density with a broader one (the parameters are given in the appendix). They are schematized in the inset of Figure~\ref{fig:peak}. The renormalisation energy is calibrated to be nearly equal in both cases. Figure~\ref{fig:peak}a presents the population evolution in the $\vert B \rangle $ and $\vert D_{\pm} \rangle $ doublet for $\eta$ = 0.01. The decay is slower in the broad peak case but the population in the two doublet states is still growing simultaneously. As shown in figure~\ref{fig:peak}b, the coherence ${\rho _{{D_-}{D_+}}}(t)$ presents a similar profile in both coupling schemes. We note that this example serves to show that the coherence generation is a result of the coupling matrix elements and transitions rates (thus appearing at the level of the master equation), and do not arise from vibronic mixing effects that require a strong coupling to a resonant and sharp (underdamped) vibrational mode at the Hamitlonian level. This suggests that the conditions for noise engineering, both in quantum simulators and physical realisations, that are required for noise-driven coherence are, in fact, rather lenient.

While we have shown in this section that it is possible to find a parameter regime where quantum noise can lead to coherence generation via incoherent relaxation, the longevity of the resulting, near-perfect superposition states is perhaps not so surprising. Due to the absence of incoherent transitions back to the high energy bright state, the only mechanisms of dephasing in the doublet are due to intra-doublet relaxation and/or pure dephasing. For both the peaked and broad spectral densities we have used, the spectral weight at the energy gap $E_{D_{+}D_{-}}$ is extremely small and pure dephasing vanishes at long times \cite{chin2013role,Kreisbeck2012}, making the doublet state effectively \emph{decoupled} from the environment. As we shall show, this situation changes dramatically in the experimentally relevant case of classical noise.     
 
\subsection{CLASSICAL STOCHASTIC NOISE}
\label{classical}
At the limit of very high temperature, the bath correlation function becomes real and therefore corresponds to a coloured classically stochastic noise (see Fig.~\ref{fig:corre}). This suggests that it should be possible to simulate the effects of stocastic noise, including any non-markovianity, by making the simulation `temperature' much larger than the other intrinsic energy scales of the system and bath, while rescaling the coupling to the bath to maintain physically reasonable transition rates. In the Golden Rule approximation, the decay rates depend on $J\left( \omega  \right)\left( n(\omega ) + 1 \right)$ which becomes $ J\left( \omega  \right){k_B}T/\omega $ at the high temperature limit. In order to keep the transition rates at similar values to those in Section \ref{sec:quantumnoise} (which also maintains the significance of our parameter $\eta$), we set an artifically high temperature ($10^3 K$) whilst simultaneously dividing the renormalization energy by a factor of $k_{B}T$. We note that this procedure captures the essential `infinite temperature' property of classically stochastic noise: the up and down transition rates are now effectively equal so that any coherence created by relaxation is now subject to potentially strong dephasing noise arising from the rapid transitions that drive the system towards the maximally mixed Boltzmann state. 

\begin{figure}
\centering
 \includegraphics[width =1.\columnwidth]{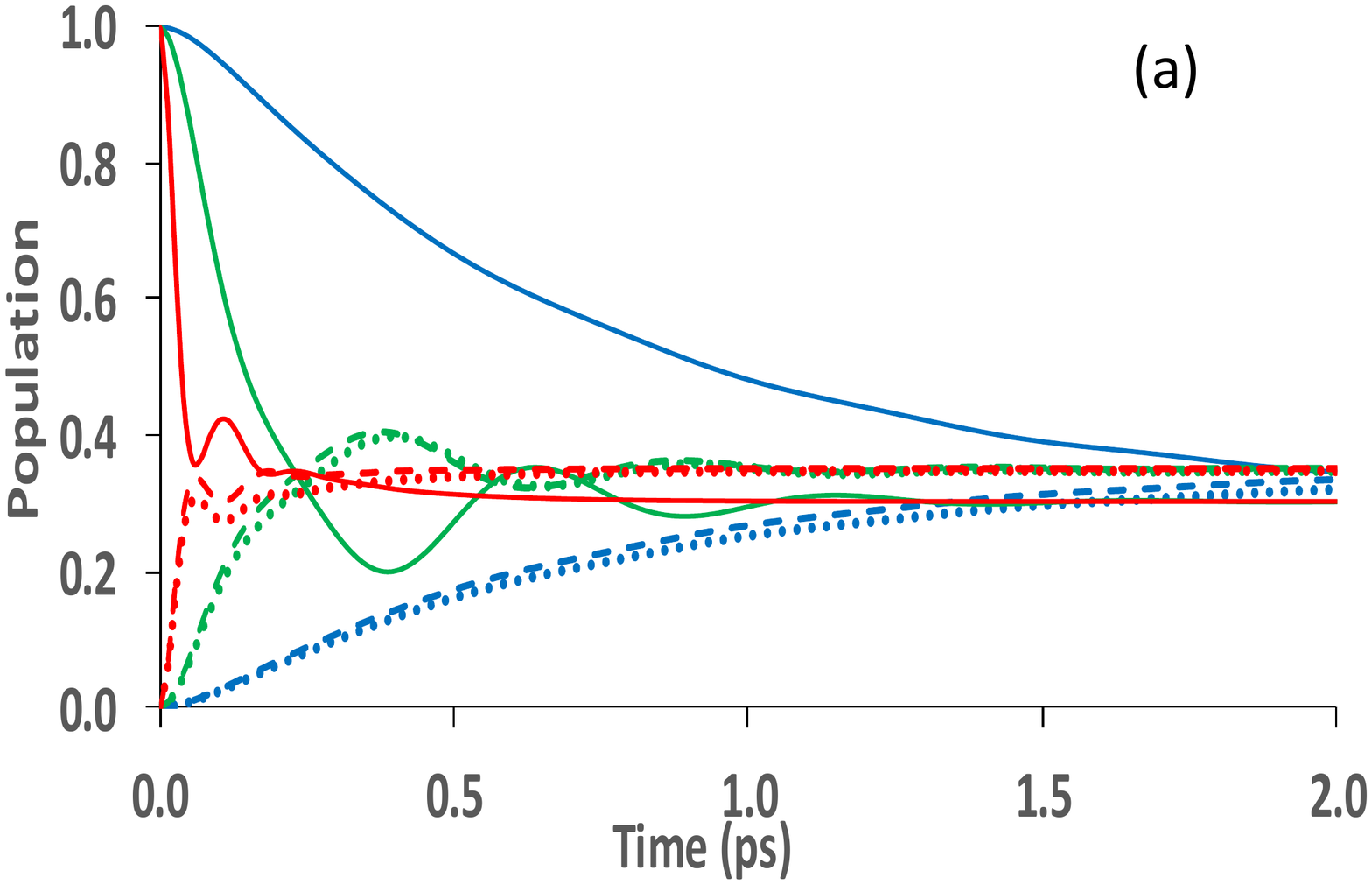}
 \includegraphics[width =1.\columnwidth]{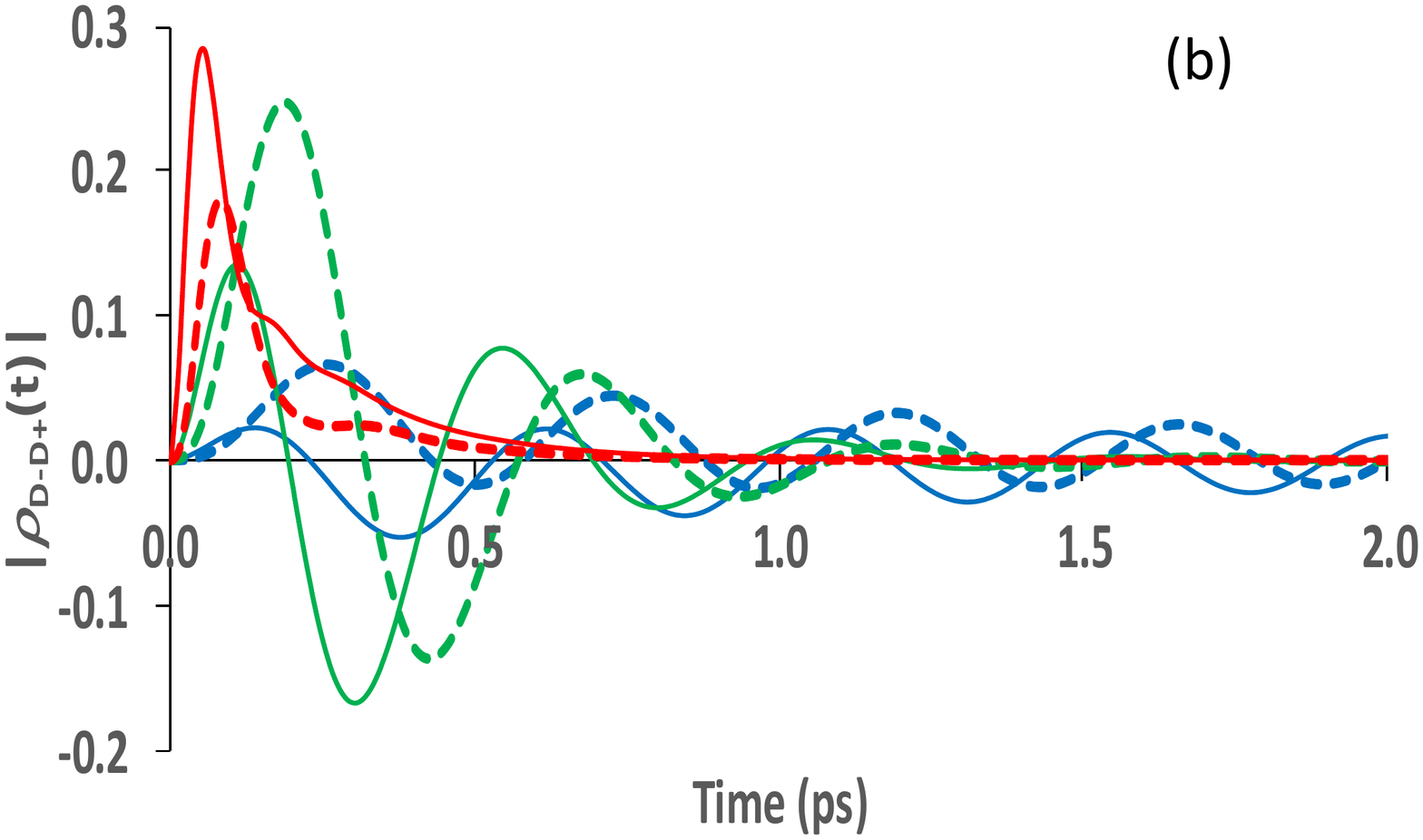}  
\caption{
 Upper panel:  Population evolution of the bright state $\vert B \rangle $ (full lines) and in the dark doublet $\vert {{D_-}} \rangle $ (dots) and $\vert {{D_+}} \rangle $ (dashes) for three coupling parameters : $\eta $ = 10$^{-2}$ (red), $\eta $ = 10$^{-3}$ (green), and $\eta $ = 10$^{-4}$ (blue). Lower panel: real part (full lines) and imaginary part (dots) of the coherence between the doublet state $\rho _{{D_-}{D_5}}(t)$ for the different parameters $\eta$.}
 \label{fig:pop10000}
 \end{figure}

Figure~\ref{fig:pop10000}a presents the population evolution for different $\eta$ parameters. The case $\eta $ = 0.01 (red curves) may be compared with the quantum noise case (see Fig.~\ref{fig:pop298}) where this coupling range gave optimal generation of the doublet state. With a classical noise case, the superposition is still created on ultrafast timescales but the peak coherence amplitude is smaller ($0.3$) and coherence is completely destroyed after just 0.1ps, as seen in figure ~\ref{fig:pop10000}b (red curves). This dephasing time is much faster than the period of coherent oscillations in the dark doublet, so no beating can be resolved in the time domain. Reducing the coupling by an order of magnitude leads to large amplitude coherence generation within $\sim 200 $ fs, but the dephasing rate  is now slower than the beating period, allowing about $2-3$ cycles of beating to be observed over about $1$ ps. For very weak coupling $\eta $ = 10$^{-4}$ (blue curves) the population decay is very slow but a coherence of weak amplitude ($<0.1$) is created and maintained \emph{during} the entire decay (about 2ps) of the system to thermal equilibrium.

We therefore confirm that the noise-induced generation of coherences survives in the case of classically stochastic noise and, although it is much more fragile, as seen in Fig. \ref{fig:pop10000}, there is a parameter regime where it is possible to resolve the quantum oscillations in the temporal domain. 

\begin{figure}
\includegraphics[width =1.0\columnwidth]{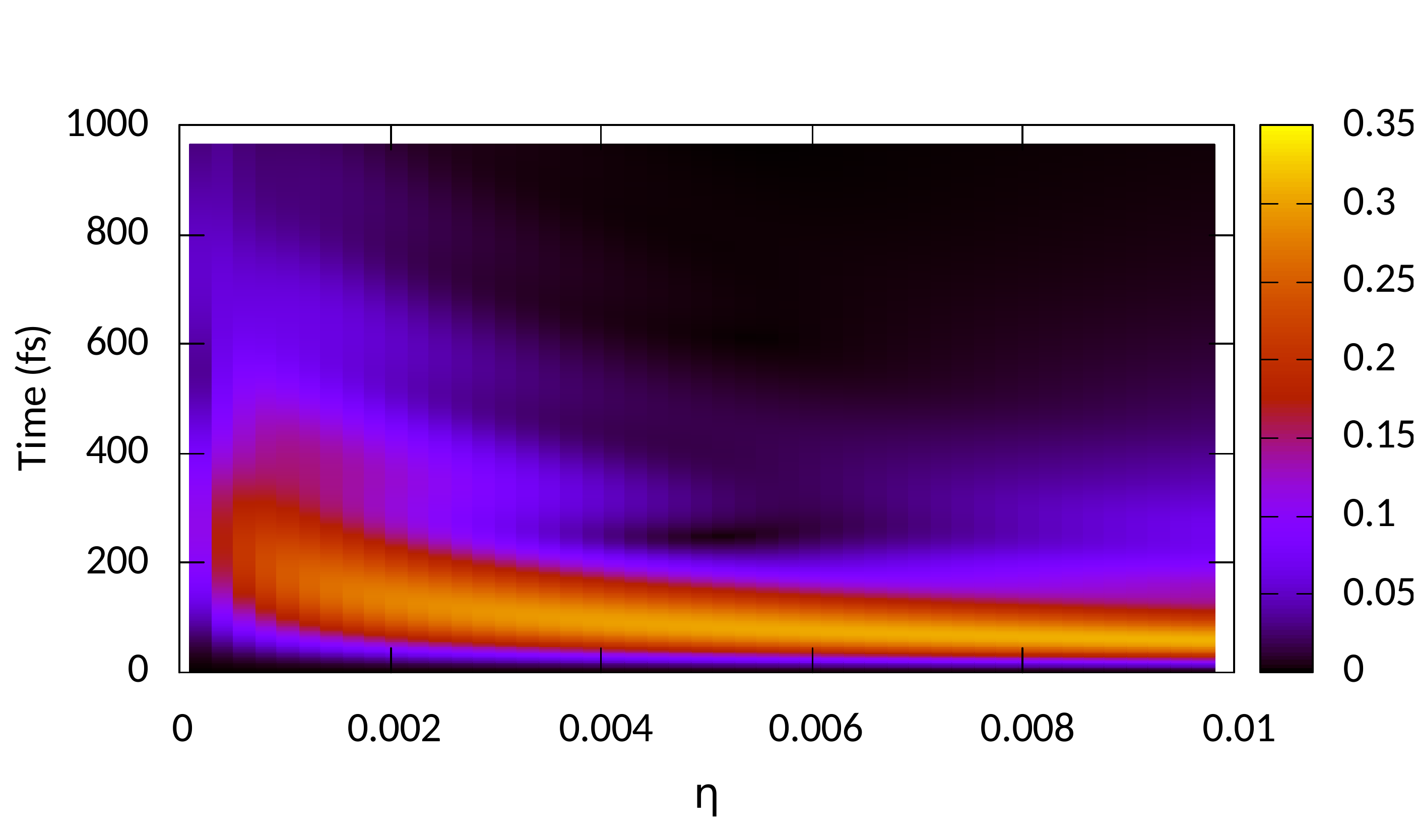}
\caption{ Iso-value contours in the modulus of the coherence between the doublet states ${\rho _{{D_+}{D_-}}}(t)$ for different $\eta$ parameters (Eq. \ref{eta}) at $T =$10000$K$. }
\label{fig:cohe10000}
\end{figure}

Compared to the quantum (cold) case, we note the following differences in coherence generation that may be relevant for future experiments in superconducting quantum cicuits and/or molecular array architectures. Firstly, the near equality of upwards and downward transition rates creates a competition between the fast relaxation needed to generate coherence and the dephasing that arises from the uphill transitions. As shown above, in order to resolve the beats, a compromise must be struck between the maximum possible amplitude of the coherence and the lifetime of the oscillations. The optimum point will depend on the method of detection and the leveraging between acceptable signal-to-noise (favouring large amplitude) and the available time/frequency resolution. Secondly, at larger coupling stengths, oscillations are also seen in the populations of the eigenstates which are damped on the same timescale as the corresponding coherences. Unlike the quantum case, these oscillations are not due to reversible energy exchange with the environment, but result from the coherent \emph{real-space} motion of the excitation in the doublet states. Again, an approximate but intuitive understanding can be obtained from the structure of the Redfield equations. 

Due to the choice of coupling to the environment (local coupling to site $2$), the rapid initial non-secular relaxation in the regime of coherence generation (relaxing to a superposition state) can also be seen as a relaxation of the bright state into the non-stationary dark state $|D\rangle=\frac{1}{\sqrt{2}}(|D_{+}\rangle+|D_{-}\rangle)$. As all the uphill rates are also the same, subsequent relaxation back to the bright state only arises when this non-stationary state is populated. However, due to the coherent evolution of the $|D\rangle$ state, excitations move in real-space to site $3$, which is not coupled to the environment, so that the population in the doublet states is temporarily unable to make any uphil transitions. As the oscillatory quantum beats return population to the $|D\rangle$ state, the uphill transitions become allowed again, depopulating the dark doublet states in a periodic way. These motion-induced modulations of the uphill transition rate are the origin of the eigenstate population oscillations seen in Fig.~\ref{fig:pop10000}a. 

\begin{figure}
\includegraphics[width =1.0\columnwidth]{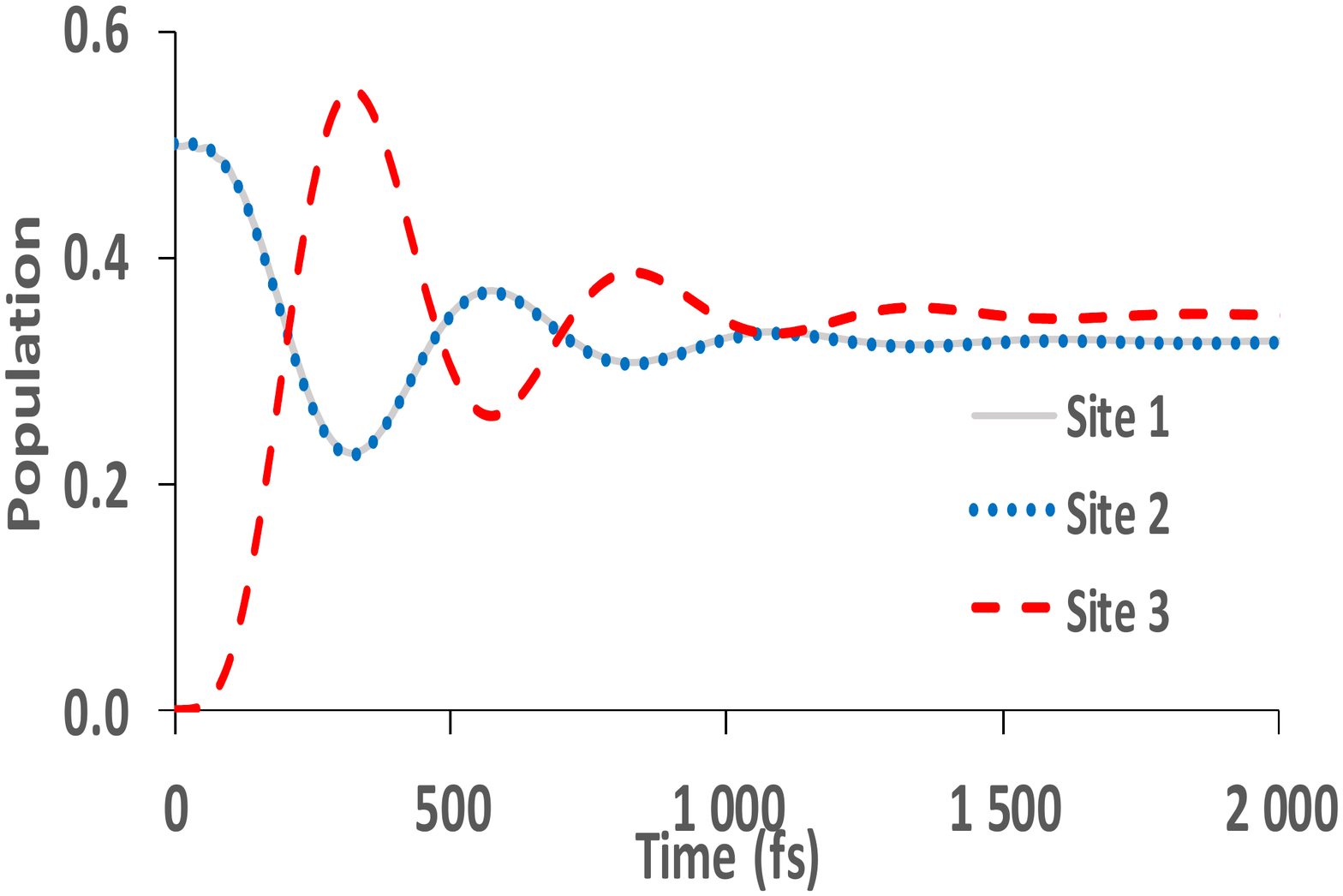}
\caption{Populations in the real-space site basis of the coupled network for coupling to classical noise( $\eta=0.001$). }
\label{classicalsites}
\end{figure}

Interestingly, this novel modulation of the transition rates effectively results in a transient and periodic violation of detailed balance  \cite{oviedo2016phase,Scully2010}, as the suppression of the upwards transitions leads to an `overshoot' of population transfer from the bright state, as if the bath were (temporally) at a much lower temperture. This effect is most prominent for the intermediate coupling ($\eta=0.01$), where the coherence is both large enough and long-lasting enough to allow a few near-complete oscillations of the excitation between sites $1\&2$ and $3$. This interpretation is confirmed by looking at the populations in the site basis for this coupling, as shown in Fig. \ref{classicalsites}. Comparing to Fig.~\ref{fig:pop10000}a, we see that the eigenstate oscillations occur at precisely the times when site $3$ is maximally populated and uphill transitions are suppressed. Experimentally, these coherent modulations of the bright state decay would be detectable through the emission in the transmission/excitation waveguide and would be in anti-phase with modulations in the resonator signal. Indeed, because these coherent dynamics directly effect the eigenstates populations, it is likely that they will be even easier to detect that the oscillations in the eigenstate coherences that appear in the regime quantum dissipation.  We note that this effect is also responsible for the oscillations in the modulus of the coherence that be seen in Fig.~ \ref{fig:cohe10000}. 

For completeness, we finally comment on the nature of the population oscillations in the quantum case of Fig. \ref{fig:pop298}. These arise from a completely different physics, and the oscillations only occur between the doublet states at strong coupling. This is due to the increasing relevance of the reorganisation energy, compared to the electronic couplings and detunings between the sites of the network. Again, due to the coupling of the bath to just site $2$, this has the effect of detuning that site, so that the effective eigenstates in the lower doublet no longer correspond to the fully delocalised states $|D_{+}\rangle \& |D_{-}\rangle$. The increasing amplitude of the oscillations we observe with stronger coupling are essentially related to the increasing misalignment of the measurement and effective eigenstate basis. It is clear that a non-stationary state of these eigenstates is still prepared by the rapid relaxation of the bright state, but the `quality' of the super positions that are formed degrades with increasing coupling strength. This degradation is already suggested by the trends in the Redfield rates shown in Fig. \ref{fig:rate298}, and confirmed by the decreasing purity of the system after relaxation in Fig. \ref{fig:purity}. This strong coupling effect does not occur in the classical case, as the rescaling of the system-bath coupling needed to take the high-temperature limit means that the reorganisation energy is always negligible compared to the system parameters.

\section{Conclusions}
In this article we have demonstrated that the system of bright and dark excitons realised in the qubit architecture of Ref. \cite{potovcnik2018studying} possesses near-ideal properties w.r.t. the non-secular processes that allow \emph{incoherent} dynamics to generate \emph{coherent} wave-like motion. By using the non-perturbative HEOM technique, we have verified that under conditions of classical noise, it should be possible to experimentally resolve these coherent dynamics through the related oscillatory motion across the network in real-space, although the effectively high (`infinite') temperature of the classical stochastic bath leads to strong dephasing via incoherent `uphill' population transitions from the dark manifold. Nevertheless, a previously unanticipated prediction arises from this fact and is seen in our simulations: the periodic violation of detailed balance caused by the suppression of uphill transitions as the excitation moves coherently away from the site-local source of the noise. At the same time, the existence of resolved quantum oscillations also appears to be sensitive to the correlation time of the environment, with non-perturbative theories showing stronger coherent dynamics at the same coupling stength as a simple Redfield approach. The differences persist over the correlation time of the environment, and it may be possible that this is related to the finite 'switch on' time of the rapid transitions that are present from $t=0$ in the Markovian theory. Although in general these effects do not correlate with any obvious feature in the measure of non-markovianity, it is clear that `memory' or temporally non-local effects constitute another handle by which coherent energy transfer might be manipulated. Indeed, the real-time switch from a weaker to stronger dissipative coupling might be more generally important for coherent real-space motion, as suggested for quasi-coherent charge separation in organic bulk heterojunctions \cite{Bredas2016,gelinas2014ultrafast,smith2014ultrafast,smith2015phonon}.  

In the case of quantum noise, which might be realised by using the multi-level nature of superconducting qubits to simulate a quantum harmonic oscillator \cite{mostame2012quantum}, the longevity of the superposition  states is only limited by the form of the spectral density and the strength of interactions at the small energy gap between the dark doublet of states, or at zero frequency (pure dephasing). Spectral functions that vanish rapidly at low frequencies while having large amplitudes at the much larger bright-dark energy gap would therefore be advantageous for coherence generation. However, the coherent dynamics are suppressed at very strong coupling - regardless of the  shape of the spectral density - by the growing reorganisation energy of the environment, which detunes and localises the low-lying excitations.   Finally, our physical understanding of the numerical results has often relied on predictions from Redfield theory that arise from the site-local noise in our model. Given the capabilities of present simulators to apply site-specific noises, it would be very interesting in the future to consider how applying different spatial and spectral correlations to noises impacts coherent dynamics in quantum energy transfer networks.

\appendix

\section{Non-Markovianity analysis}
\label{appendix:NM}

\begin{figure}
\centering
  \includegraphics[width =1.\columnwidth]{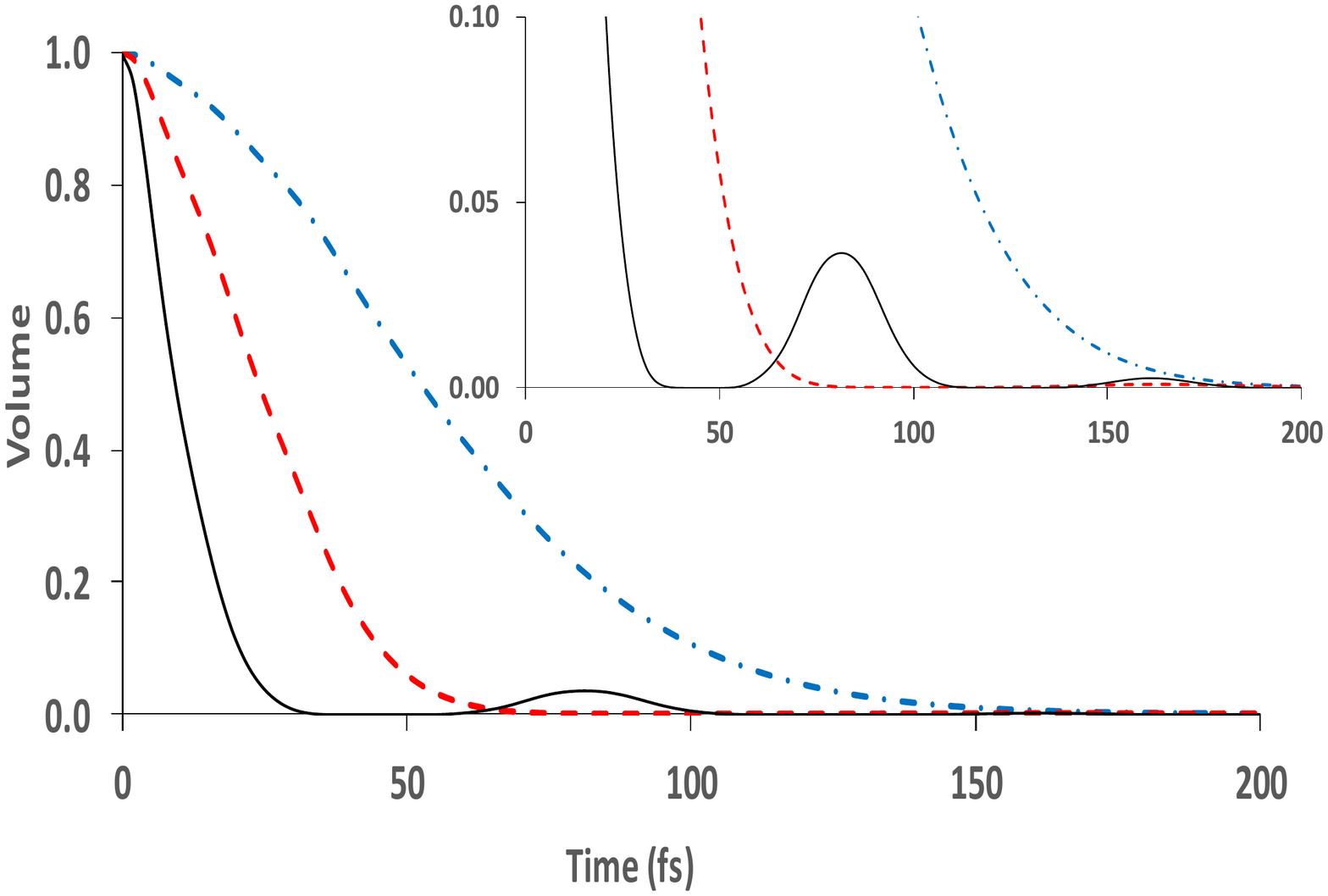}
\caption{Volume of the accessible state $V(t)$. Full line: $\eta $ = 0.16, dashes: $\eta $ = 0.04  , dashes-dots; $\eta $ = 0.01.}
\label{fig:volume}
\end{figure}

We briefly discuss the non-Markovianity of the dynamics for some coupling ranges.  Numerous non-Markovianity witnesses have been proposed in the literature \cite{Laine_2009,Laine_2010, Anderson_2014}  but we consider here only the volume of accessible states in the generalized Bloch sphere \cite{Paternostro_2011}.  Eqs.(\ref{heom}) define a time local dynamical map $\rho_S (t) = {\phi _t}\left[ {\rho_S (0)} \right]$ which may be expressed in an operator basis set $\left\{ {{G_m}} \right\}$ for the Liouville space of dimension ${d^2}$ by a generalization of the Pauli matrices for $d = 3$. The expansion of the matrix density in this basis leads to the Bloch representation of the system. In matrix form the map reads ${F_{m,n}}(t) = Tr\left( {{G_m}{\phi _t}\left[ {{G_n}} \right]} \right)$ and the volume of accessible states in the Bloch sphere can be obtained from the determinant of this matrix
\begin{equation}
V(t) = \det ({\bf{F}}).
\label{volume}
\end{equation}
A non-monotonous decrease of this volume is a signature on a non-Markovian back flow from the bath to the system. This is illustrated in figure ~\ref{fig:volume} by the volume of the accessible states in the Bloch sphere for differnt $\eta$ parameters. For instance, bumps in the volume are obtained for the strong coupling case $\eta $ = 0.16 (full line). This justifies that dynamics must be treated beyond the Redfield approximation and probably beyond second order regime. The perturbative regime for $\eta < 0.02$ leads to a smooth evolution of the population (see Figs.~\ref{fig:pop298} or ~\ref{fig:pop10000}). On the contrary, oscillations are observed during the decay of the bright state for stronger coupling. This behavior may be related to some non Markovian effects characterized by back-flow from the environment to the system.
 
\begin{figure}
 \centering
  \includegraphics[width =1.\columnwidth]{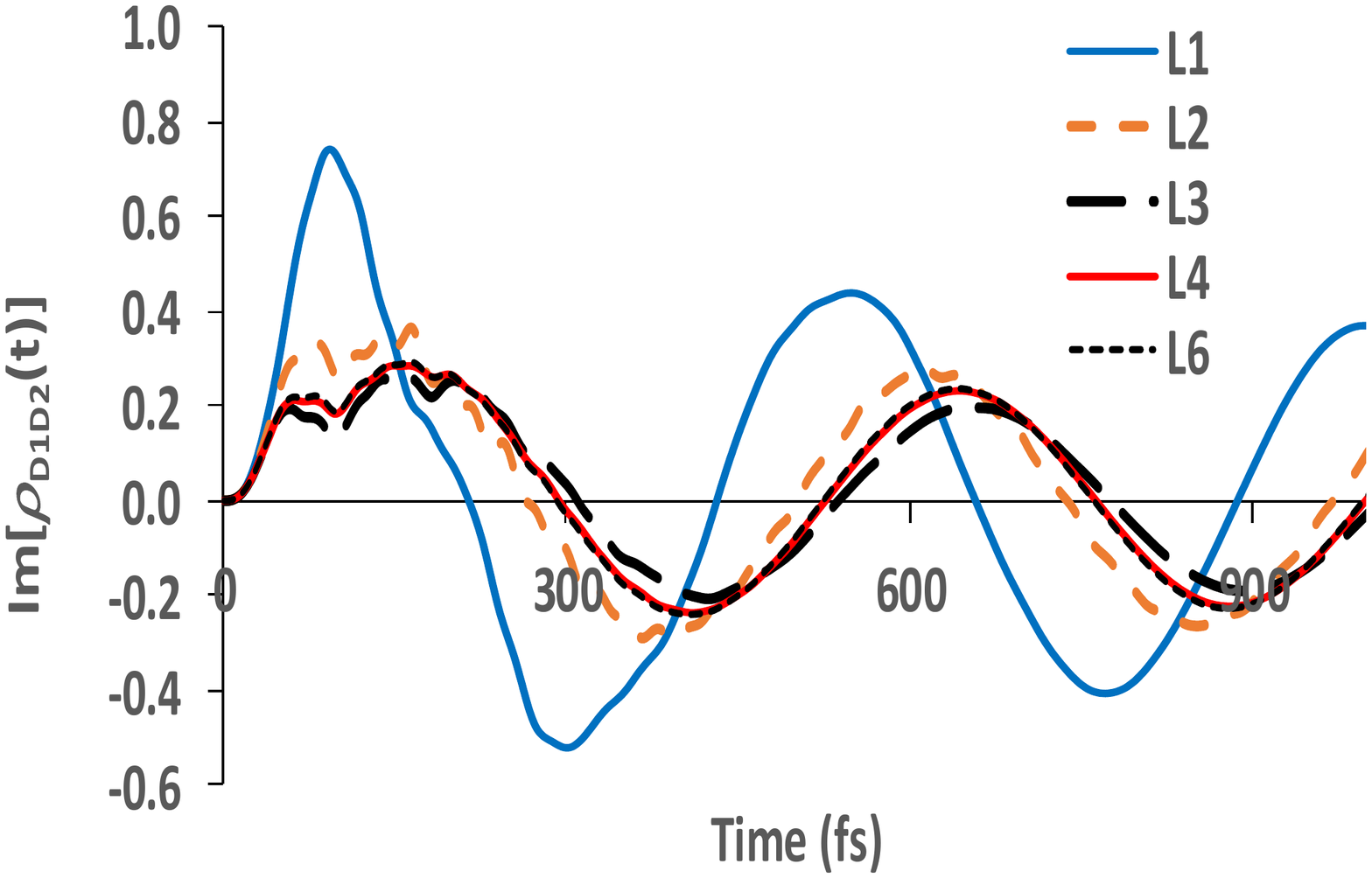}
\caption{Convergence of the real part of the coherence $\rho _{{D_-}{D_+}}(t)$ with respect to the level $L$ of the hierarchy  corresponding to order $2L$ in perturbation for the case $\eta $ = 0.16. }
\label{fig:heom}
\end{figure}

The level of HEOM hierarchy ensuring convergence of the simulation depends on the strength of the system-bath coupling. Level $L$ in the hierarchy corresponds to order $2L$ in perturbation approach. Convergence is checked in figure~\ref{fig:heom} by analyzing the coherence that always converges more slowly than the populations. The cases with $\eta \le $ 0.02 remain in the perturbative regime. On the contrary, for the case $\eta$ = 0.16, the regime is obviously non-perturbative and level $L$ = 4 is required.
 
\begin{figure}
\includegraphics[width =1.0\columnwidth]{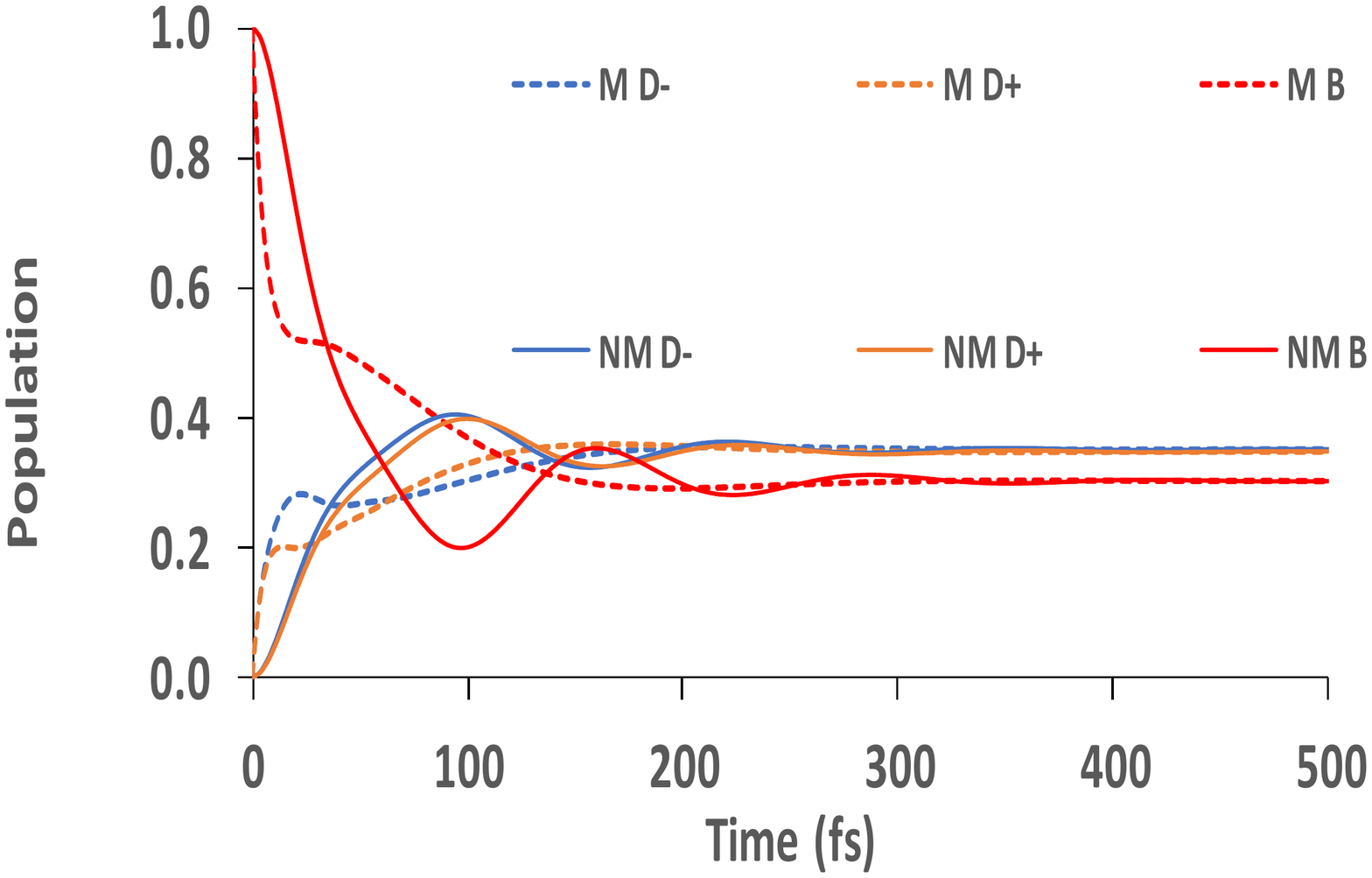}
\caption{Comparison of the eigenstate population dynamics computed with the Markovian (M) Redfield equations (dashed lines) and the numerically non-Markovian (NM) exact results obtained by HEOM (full lines) at relatively weak coupling $\eta=0.01$.  }
\label{nonmarkov}
\end{figure}

We do find that although the Markovian 2nd-order perturbative theory (Bloch-Redfield) allows an intuitive, qualitative understanding of the dynamics, it fails to describe the dynamics with quantitative accuracy (c.f. HEOM results). Even at relatively weak coupling, this can lead to significant differences, as shown in Fig.~\ref{nonmarkov}. Given the relatively weak coupling involved, we believe that the orgin of these differences are related to the Markov (time local) approximation in Redfield theory, which does not take into account the long correlation time of the environmental spectral functions that we consider. This is supported by the results in Fig.~\ref{nonmarkov}, which show that the differences in dynamics become negligible for times longer than the bath correlation time ($\approx 200$ fs), as shown in Fig.~\ref{fig:corre}. As the oscillatory coherent dynamics induced by the relaxation are much more prominent in the HEOM results, this indicates that a proper treatment of extended bath correlation times can also be an important factor for noise-induced coherence generation in the classical case, further emphasising the need for methods such as HEOM.    

\section{Parameters of the spectral density}
\label{appendix:spectraldensity}

The parameters of the superohmic expression (Eq.\ref{J}) for the thin and broad spectral density used in the HEOM simulations are gathered in the following table. The $p$ parameter is taken as $1.95\times10^{-14}\times f $ where $f$ is adjusted to obtain the different renormalization energies. 

\begin{table}[ht]
	
	{\begin{tabular}{cccc} \toprule
			 $\Omega_l$ (a.u.)    & $\Gamma_1$ (a.u.)   & $\Omega_2$ (a.u.) & $\Gamma_1$ (a.u.) \\[0.5ex]
			\hline \noalign{\vskip 1.0ex}
			$9.562\times10^{-4} $  & $6.3537\times10^{-3} $  & $4.5639\times10^{-3} $  & $2.7188\times10^{-4}$ \\[0.5ex]
			$2.762\times10^{-3} $  & $1.6554\times10^{-3} $  & $6.4639\times10^{-3} $  & $2.5319\times10^{-3}$  \\[0.5ex]
		\end{tabular}}
		\label{table:SD_values}
	\end{table}

\section*{Acknowledgment}
		
We are grateful to Anton Poto{\v{c}}nik for helpful discussions and suggestions. Alex Chin acknowledges the Jean D'Alembert Chaire from IDEX Paris-Saclay, contrat CNRS 157819. Etienne Mangaud acknowledges support from the ANR-DFG COQS, under Grant No. \mbox{ANR-15-CE30-0023-01.} This work has been performed within the French GDR 3575 THEMS and we thank Olivier Dulieu for his support.


\begin{thebibliography}{87}%
\makeatletter
\providecommand \@ifxundefined [1]{%
 \@ifx{#1\undefined}
}%
\providecommand \@ifnum [1]{%
 \ifnum #1\expandafter \@firstoftwo
 \else \expandafter \@secondoftwo
 \fi
}%
\providecommand \@ifx [1]{%
 \ifx #1\expandafter \@firstoftwo
 \else \expandafter \@secondoftwo
 \fi
}%
\providecommand \natexlab [1]{#1}%
\providecommand \enquote  [1]{``#1''}%
\providecommand \bibnamefont  [1]{#1}%
\providecommand \bibfnamefont [1]{#1}%
\providecommand \citenamefont [1]{#1}%
\providecommand \href@noop [0]{\@secondoftwo}%
\providecommand \href [0]{\begingroup \@sanitize@url \@href}%
\providecommand \@href[1]{\@@startlink{#1}\@@href}%
\providecommand \@@href[1]{\endgroup#1\@@endlink}%
\providecommand \@sanitize@url [0]{\catcode `\\12\catcode `\$12\catcode
  `\&12\catcode `\#12\catcode `\^12\catcode `\_12\catcode `\%12\relax}%
\providecommand \@@startlink[1]{}%
\providecommand \@@endlink[0]{}%
\providecommand \url  [0]{\begingroup\@sanitize@url \@url }%
\providecommand \@url [1]{\endgroup\@href {#1}{\urlprefix }}%
\providecommand \urlprefix  [0]{URL }%
\providecommand \Eprint [0]{\href }%
\providecommand \doibase [0]{http://dx.doi.org/}%
\providecommand \selectlanguage [0]{\@gobble}%
\providecommand \bibinfo  [0]{\@secondoftwo}%
\providecommand \bibfield  [0]{\@secondoftwo}%
\providecommand \translation [1]{[#1]}%
\providecommand \BibitemOpen [0]{}%
\providecommand \bibitemStop [0]{}%
\providecommand \bibitemNoStop [0]{.\EOS\space}%
\providecommand \EOS [0]{\spacefactor3000\relax}%
\providecommand \BibitemShut  [1]{\csname bibitem#1\endcsname}%
\let\auto@bib@innerbib\@empty
\bibitem [{\citenamefont {Poto{\v{c}}nik}\ \emph {et~al.}(2018)\citenamefont
  {Poto{\v{c}}nik}, \citenamefont {Bargerbos}, \citenamefont {Schr{\"o}der},
  \citenamefont {Khan}, \citenamefont {Collodo}, \citenamefont {Gasparinetti},
  \citenamefont {Salath{\'e}}, \citenamefont {Creatore}, \citenamefont
  {Eichler}, \citenamefont {T{\"u}reci} \emph
  {et~al.}}]{potovcnik2018studying}%
  \BibitemOpen
  \bibfield  {author} {\bibinfo {author} {\bibfnamefont {A.}~\bibnamefont
  {Poto{\v{c}}nik}}, \bibinfo {author} {\bibfnamefont {A.}~\bibnamefont
  {Bargerbos}}, \bibinfo {author} {\bibfnamefont {F.~A.}\ \bibnamefont
  {Schr{\"o}der}}, \bibinfo {author} {\bibfnamefont {S.~A.}\ \bibnamefont
  {Khan}}, \bibinfo {author} {\bibfnamefont {M.~C.}\ \bibnamefont {Collodo}},
  \bibinfo {author} {\bibfnamefont {S.}~\bibnamefont {Gasparinetti}}, \bibinfo
  {author} {\bibfnamefont {Y.}~\bibnamefont {Salath{\'e}}}, \bibinfo {author}
  {\bibfnamefont {C.}~\bibnamefont {Creatore}}, \bibinfo {author}
  {\bibfnamefont {C.}~\bibnamefont {Eichler}}, \bibinfo {author} {\bibfnamefont
  {H.~E.}\ \bibnamefont {T{\"u}reci}},  \emph {et~al.},\ }\href@noop {}
  {\bibfield  {journal} {\bibinfo  {journal} {Nature Communications}\ }\textbf
  {\bibinfo {volume} {9}},\ \bibinfo {pages} {904} (\bibinfo {year}
  {2018})}\BibitemShut {NoStop}%
\bibitem [{\citenamefont {Br{\'e}das}\ \emph {et~al.}(2017)\citenamefont
  {Br{\'e}das}, \citenamefont {Sargent},\ and\ \citenamefont
  {Scholes}}]{bredas2017photovoltaic}%
  \BibitemOpen
  \bibfield  {author} {\bibinfo {author} {\bibfnamefont {J.-L.}\ \bibnamefont
  {Br{\'e}das}}, \bibinfo {author} {\bibfnamefont {E.~H.}\ \bibnamefont
  {Sargent}}, \ and\ \bibinfo {author} {\bibfnamefont {G.~D.}\ \bibnamefont
  {Scholes}},\ }\href@noop {} {\bibfield  {journal} {\bibinfo  {journal}
  {Nature materials}\ }\textbf {\bibinfo {volume} {16}},\ \bibinfo {pages} {35}
  (\bibinfo {year} {2017})}\BibitemShut {NoStop}%
\bibitem [{\citenamefont {Scholes}\ \emph {et~al.}(2017)\citenamefont
  {Scholes}, \citenamefont {Fleming}, \citenamefont {Chen}, \citenamefont
  {Aspuru-Guzik}, \citenamefont {Buchleitner}, \citenamefont {Coker},
  \citenamefont {Engel}, \citenamefont {van Grondelle}, \citenamefont
  {Ishizaki}, \citenamefont {Jonas}, \citenamefont {Lundeen}, \citenamefont
  {McCusker}, \citenamefont {Mukamel}, \citenamefont {Ogilvie}, \citenamefont
  {Olaya-Castro}, \citenamefont {Ratner}, \citenamefont {Spano}, \citenamefont
  {Whaley},\ and\ \citenamefont {Zhu}}]{Scholes2017}%
  \BibitemOpen
  \bibfield  {author} {\bibinfo {author} {\bibfnamefont {G.~D.}\ \bibnamefont
  {Scholes}}, \bibinfo {author} {\bibfnamefont {G.~R.}\ \bibnamefont
  {Fleming}}, \bibinfo {author} {\bibfnamefont {L.~X.}\ \bibnamefont {Chen}},
  \bibinfo {author} {\bibfnamefont {A.}~\bibnamefont {Aspuru-Guzik}}, \bibinfo
  {author} {\bibfnamefont {A.}~\bibnamefont {Buchleitner}}, \bibinfo {author}
  {\bibfnamefont {D.~F.}\ \bibnamefont {Coker}}, \bibinfo {author}
  {\bibfnamefont {G.~S.}\ \bibnamefont {Engel}}, \bibinfo {author}
  {\bibfnamefont {R.}~\bibnamefont {van Grondelle}}, \bibinfo {author}
  {\bibfnamefont {A.}~\bibnamefont {Ishizaki}}, \bibinfo {author}
  {\bibfnamefont {D.~M.}\ \bibnamefont {Jonas}}, \bibinfo {author}
  {\bibfnamefont {J.~S.}\ \bibnamefont {Lundeen}}, \bibinfo {author}
  {\bibfnamefont {J.~K.}\ \bibnamefont {McCusker}}, \bibinfo {author}
  {\bibfnamefont {S.}~\bibnamefont {Mukamel}}, \bibinfo {author} {\bibfnamefont
  {J.~P.}\ \bibnamefont {Ogilvie}}, \bibinfo {author} {\bibfnamefont
  {A.}~\bibnamefont {Olaya-Castro}}, \bibinfo {author} {\bibfnamefont {M.~A.}\
  \bibnamefont {Ratner}}, \bibinfo {author} {\bibfnamefont {F.~C.}\
  \bibnamefont {Spano}}, \bibinfo {author} {\bibfnamefont {K.~B.}\ \bibnamefont
  {Whaley}}, \ and\ \bibinfo {author} {\bibfnamefont {X.}~\bibnamefont {Zhu}},\
  }\href {\doibase 10.1038/nature21425} {\bibfield  {journal} {\bibinfo
  {journal} {Nature}\ }\textbf {\bibinfo {volume} {543}},\ \bibinfo {pages}
  {647} (\bibinfo {year} {2017})}\BibitemShut {NoStop}%
\bibitem [{Cas(2014)}]{Castro2014}%
  \BibitemOpen
  \href {\doibase doi:10.1038/ncomms4012} {\bibfield  {journal} {\bibinfo
  {journal} {Nature Communications}\ }\textbf {\bibinfo {volume} {5}} (\bibinfo
  {year} {2014}),\ doi:10.1038/ncomms4012}\BibitemShut {NoStop}%
\bibitem [{\citenamefont {Chin}\ \emph
  {et~al.}(2013{\natexlab{a}})\citenamefont {Chin}, \citenamefont {Prior},
  \citenamefont {Rosenbach}, \citenamefont {Caycedo-Soler}, \citenamefont
  {Huelga},\ and\ \citenamefont {Plenio}}]{chin2013role}%
  \BibitemOpen
  \bibfield  {author} {\bibinfo {author} {\bibfnamefont {A.}~\bibnamefont
  {Chin}}, \bibinfo {author} {\bibfnamefont {J.}~\bibnamefont {Prior}},
  \bibinfo {author} {\bibfnamefont {R.}~\bibnamefont {Rosenbach}}, \bibinfo
  {author} {\bibfnamefont {F.}~\bibnamefont {Caycedo-Soler}}, \bibinfo {author}
  {\bibfnamefont {S.}~\bibnamefont {Huelga}}, \ and\ \bibinfo {author}
  {\bibfnamefont {M.}~\bibnamefont {Plenio}},\ }\href@noop {} {\bibfield
  {journal} {\bibinfo  {journal} {Nature Physics}\ }\textbf {\bibinfo {volume}
  {9}},\ \bibinfo {pages} {113} (\bibinfo {year}
  {2013}{\natexlab{a}})}\BibitemShut {NoStop}%
\bibitem [{\citenamefont {Collini}\ \emph {et~al.}(2010)\citenamefont
  {Collini}, \citenamefont {Wong}, \citenamefont {Wilk}, \citenamefont {Curmi},
  \citenamefont {Brumer},\ and\ \citenamefont {Scholes}}]{Collini2010}%
  \BibitemOpen
  \bibfield  {author} {\bibinfo {author} {\bibfnamefont {E.}~\bibnamefont
  {Collini}}, \bibinfo {author} {\bibfnamefont {C.~Y.}\ \bibnamefont {Wong}},
  \bibinfo {author} {\bibfnamefont {K.~E.}\ \bibnamefont {Wilk}}, \bibinfo
  {author} {\bibfnamefont {P.~M.~G.}\ \bibnamefont {Curmi}}, \bibinfo {author}
  {\bibfnamefont {P.}~\bibnamefont {Brumer}}, \ and\ \bibinfo {author}
  {\bibfnamefont {G.~D.}\ \bibnamefont {Scholes}},\ }\href {\doibase
  10.1038/nature08811} {\bibfield  {journal} {\bibinfo  {journal} {Nature}\ }
  (\bibinfo {year} {2010}),\ 10.1038/nature08811}\BibitemShut {NoStop}%
\bibitem [{\citenamefont {Fuller}\ \emph {et~al.}(2014)\citenamefont {Fuller},
  \citenamefont {Pan}, \citenamefont {Gelzinis}, \citenamefont {Butkus},
  \citenamefont {Senlik}, \citenamefont {Wilcox}, \citenamefont {Yocum},
  \citenamefont {Valkunas}, \citenamefont {Abramavicius},\ and\ \citenamefont
  {Ogilvie}}]{fuller2014vibronic}%
  \BibitemOpen
  \bibfield  {author} {\bibinfo {author} {\bibfnamefont {F.~D.}\ \bibnamefont
  {Fuller}}, \bibinfo {author} {\bibfnamefont {J.}~\bibnamefont {Pan}},
  \bibinfo {author} {\bibfnamefont {A.}~\bibnamefont {Gelzinis}}, \bibinfo
  {author} {\bibfnamefont {V.}~\bibnamefont {Butkus}}, \bibinfo {author}
  {\bibfnamefont {S.~S.}\ \bibnamefont {Senlik}}, \bibinfo {author}
  {\bibfnamefont {D.~E.}\ \bibnamefont {Wilcox}}, \bibinfo {author}
  {\bibfnamefont {C.~F.}\ \bibnamefont {Yocum}}, \bibinfo {author}
  {\bibfnamefont {L.}~\bibnamefont {Valkunas}}, \bibinfo {author}
  {\bibfnamefont {D.}~\bibnamefont {Abramavicius}}, \ and\ \bibinfo {author}
  {\bibfnamefont {J.~P.}\ \bibnamefont {Ogilvie}},\ }\href@noop {} {\bibfield
  {journal} {\bibinfo  {journal} {Nature chemistry}\ }\textbf {\bibinfo
  {volume} {6}},\ \bibinfo {pages} {706} (\bibinfo {year} {2014})}\BibitemShut
  {NoStop}%
\bibitem [{\citenamefont {Kreisbeck}\ and\ \citenamefont
  {Kramer}(2012)}]{Kreisbeck2012}%
  \BibitemOpen
  \bibfield  {author} {\bibinfo {author} {\bibfnamefont {C.}~\bibnamefont
  {Kreisbeck}}\ and\ \bibinfo {author} {\bibfnamefont {T.}~\bibnamefont
  {Kramer}},\ }\href {\doibase 10.1021/jz3012029} {\bibfield  {journal}
  {\bibinfo  {journal} {J. Phys. Chem. Lett.}\ }\textbf {\bibinfo {volume}
  {3}},\ \bibinfo {pages} {2828} (\bibinfo {year} {2012})},\ \Eprint
  {http://arxiv.org/abs/1203.1485} {arXiv:1203.1485} \BibitemShut {NoStop}%
\bibitem [{\citenamefont {Lambert}\ \emph {et~al.}(2013)\citenamefont
  {Lambert}, \citenamefont {Chen}, \citenamefont {Cheng}, \citenamefont {Li},
  \citenamefont {Chen},\ and\ \citenamefont {Nori}}]{Lambert_2013}%
  \BibitemOpen
  \bibfield  {author} {\bibinfo {author} {\bibfnamefont {N.}~\bibnamefont
  {Lambert}}, \bibinfo {author} {\bibfnamefont {Y.~N.}\ \bibnamefont {Chen}},
  \bibinfo {author} {\bibfnamefont {Y.~C.}\ \bibnamefont {Cheng}}, \bibinfo
  {author} {\bibfnamefont {C.~M.}\ \bibnamefont {Li}}, \bibinfo {author}
  {\bibfnamefont {G.~Y.}\ \bibnamefont {Chen}}, \ and\ \bibinfo {author}
  {\bibfnamefont {F.}~\bibnamefont {Nori}},\ }\href {\doibase
  10.1038/nphys2474} {\bibfield  {journal} {\bibinfo  {journal} {Nat. Phys.}\
  }\textbf {\bibinfo {volume} {9}},\ \bibinfo {pages} {10} (\bibinfo {year}
  {2013})}\BibitemShut {NoStop}%
\bibitem [{\citenamefont {Lee}\ \emph {et~al.}(2007)\citenamefont {Lee},
  \citenamefont {Cheng},\ and\ \citenamefont {Fleming}}]{Lee1462}%
  \BibitemOpen
  \bibfield  {author} {\bibinfo {author} {\bibfnamefont {H.}~\bibnamefont
  {Lee}}, \bibinfo {author} {\bibfnamefont {Y.-C.}\ \bibnamefont {Cheng}}, \
  and\ \bibinfo {author} {\bibfnamefont {G.~R.}\ \bibnamefont {Fleming}},\
  }\href {\doibase 10.1126/science.1142188} {\bibfield  {journal} {\bibinfo
  {journal} {Science}\ }\textbf {\bibinfo {volume} {316}},\ \bibinfo {pages}
  {1462} (\bibinfo {year} {2007})},\ \Eprint
  {http://arxiv.org/abs/http://science.sciencemag.org/content/316/5830/1462.full.pdf}
  {http://science.sciencemag.org/content/316/5830/1462.full.pdf} \BibitemShut
  {NoStop}%
\bibitem [{\citenamefont {Panitchayangkoon}\ \emph {et~al.}(2010)\citenamefont
  {Panitchayangkoon}, \citenamefont {Hayes}, \citenamefont {Fransted},
  \citenamefont {Caram}, \citenamefont {Harel}, \citenamefont {Wen},
  \citenamefont {Blankenship},\ and\ \citenamefont
  {Engel}}]{Panitchayangkoon2010}%
  \BibitemOpen
  \bibfield  {author} {\bibinfo {author} {\bibfnamefont {G.}~\bibnamefont
  {Panitchayangkoon}}, \bibinfo {author} {\bibfnamefont {D.}~\bibnamefont
  {Hayes}}, \bibinfo {author} {\bibfnamefont {K.~A.}\ \bibnamefont {Fransted}},
  \bibinfo {author} {\bibfnamefont {J.~R.}\ \bibnamefont {Caram}}, \bibinfo
  {author} {\bibfnamefont {E.}~\bibnamefont {Harel}}, \bibinfo {author}
  {\bibfnamefont {J.}~\bibnamefont {Wen}}, \bibinfo {author} {\bibfnamefont
  {R.~E.}\ \bibnamefont {Blankenship}}, \ and\ \bibinfo {author} {\bibfnamefont
  {G.~S.}\ \bibnamefont {Engel}},\ }\href {\doibase 10.1073/pnas.1005484107}
  {\bibfield  {journal} {\bibinfo  {journal} {Proceedings of the National
  Academy of Sciences}\ }\textbf {\bibinfo {volume} {107}},\ \bibinfo {pages}
  {12766} (\bibinfo {year} {2010})},\ \Eprint
  {http://arxiv.org/abs/http://www.pnas.org/content/107/29/12766.full.pdf}
  {http://www.pnas.org/content/107/29/12766.full.pdf} \BibitemShut {NoStop}%
\bibitem [{\citenamefont {Romero}\ \emph {et~al.}(2014)\citenamefont {Romero},
  \citenamefont {Augulis}, \citenamefont {Novoderezhkin}, \citenamefont
  {Ferretti}, \citenamefont {Thieme}, \citenamefont {Zigmantas},\ and\
  \citenamefont {Van~Grondelle}}]{romero2014quantum}%
  \BibitemOpen
  \bibfield  {author} {\bibinfo {author} {\bibfnamefont {E.}~\bibnamefont
  {Romero}}, \bibinfo {author} {\bibfnamefont {R.}~\bibnamefont {Augulis}},
  \bibinfo {author} {\bibfnamefont {V.~I.}\ \bibnamefont {Novoderezhkin}},
  \bibinfo {author} {\bibfnamefont {M.}~\bibnamefont {Ferretti}}, \bibinfo
  {author} {\bibfnamefont {J.}~\bibnamefont {Thieme}}, \bibinfo {author}
  {\bibfnamefont {D.}~\bibnamefont {Zigmantas}}, \ and\ \bibinfo {author}
  {\bibfnamefont {R.}~\bibnamefont {Van~Grondelle}},\ }\href@noop {} {\bibfield
   {journal} {\bibinfo  {journal} {Nature physics}\ }\textbf {\bibinfo {volume}
  {10}},\ \bibinfo {pages} {676} (\bibinfo {year} {2014})}\BibitemShut
  {NoStop}%
\bibitem [{\citenamefont {Ishizaki}\ and\ \citenamefont
  {Fleming}(2009{\natexlab{a}})}]{Akihito09}%
  \BibitemOpen
  \bibfield  {author} {\bibinfo {author} {\bibfnamefont {A.}~\bibnamefont
  {Ishizaki}}\ and\ \bibinfo {author} {\bibfnamefont {G.~R.}\ \bibnamefont
  {Fleming}},\ }\href {\doibase 10.1073/pnas.0908989106} {\bibfield  {journal}
  {\bibinfo  {journal} {Proc. Natl. Acad. Sci. USA}\ }\textbf {\bibinfo
  {volume} {106}},\ \bibinfo {pages} {17255} (\bibinfo {year}
  {2009}{\natexlab{a}})}\BibitemShut {NoStop}%
\bibitem [{\citenamefont {Chen}\ \emph {et~al.}(2015)\citenamefont {Chen},
  \citenamefont {Lambert}, \citenamefont {Cheng}, \citenamefont {Chen},\ and\
  \citenamefont {Nori}}]{Chen2015}%
  \BibitemOpen
  \bibfield  {author} {\bibinfo {author} {\bibfnamefont {H.-B.}\ \bibnamefont
  {Chen}}, \bibinfo {author} {\bibfnamefont {N.}~\bibnamefont {Lambert}},
  \bibinfo {author} {\bibfnamefont {Y.-C.}\ \bibnamefont {Cheng}}, \bibinfo
  {author} {\bibfnamefont {Y.-N.}\ \bibnamefont {Chen}}, \ and\ \bibinfo
  {author} {\bibfnamefont {F.}~\bibnamefont {Nori}},\ }\href {\doibase
  10.1038/srep12753} {\bibfield  {journal} {\bibinfo  {journal} {Sci. Rep.}\
  }\textbf {\bibinfo {volume} {5}},\ \bibinfo {pages} {12753} (\bibinfo {year}
  {2015})},\ \Eprint {http://arxiv.org/abs/arXiv:1503.02412v1}
  {arXiv:arXiv:1503.02412v1} \BibitemShut {NoStop}%
\bibitem [{\citenamefont {Chin}\ \emph {et~al.}(2012)\citenamefont {Chin},
  \citenamefont {Huelga},\ and\ \citenamefont {Plenio}}]{Chin2012}%
  \BibitemOpen
  \bibfield  {author} {\bibinfo {author} {\bibfnamefont {A.~W.}\ \bibnamefont
  {Chin}}, \bibinfo {author} {\bibfnamefont {S.~F.}\ \bibnamefont {Huelga}}, \
  and\ \bibinfo {author} {\bibfnamefont {M.~B.}\ \bibnamefont {Plenio}},\
  }\href {\doibase 10.1098/rsta.2011.0224} {\bibfield  {journal} {\bibinfo
  {journal} {Phil. Trans. R. Soc. A}\ }\textbf {\bibinfo {volume} {370}},\
  \bibinfo {pages} {3638} (\bibinfo {year} {2012})},\ \Eprint
  {http://arxiv.org/abs/1203.5072v1} {1203.5072v1} \BibitemShut {NoStop}%
\bibitem [{\citenamefont {Chin}\ \emph
  {et~al.}(2013{\natexlab{b}})\citenamefont {Chin}, \citenamefont {Prior},
  \citenamefont {Rosenbach}, \citenamefont {Caycedo-Soler}, \citenamefont
  {Huelga},\ and\ \citenamefont {Plenio}}]{Chin2013}%
  \BibitemOpen
  \bibfield  {author} {\bibinfo {author} {\bibfnamefont {A.~W.}\ \bibnamefont
  {Chin}}, \bibinfo {author} {\bibfnamefont {J.}~\bibnamefont {Prior}},
  \bibinfo {author} {\bibfnamefont {R.}~\bibnamefont {Rosenbach}}, \bibinfo
  {author} {\bibfnamefont {F.}~\bibnamefont {Caycedo-Soler}}, \bibinfo {author}
  {\bibfnamefont {S.~F.}\ \bibnamefont {Huelga}}, \ and\ \bibinfo {author}
  {\bibfnamefont {M.~B.}\ \bibnamefont {Plenio}},\ }\href
  {http://dx.doi.org/10.1038/nphys2515} {\bibfield  {journal} {\bibinfo
  {journal} {Nat Phys}\ }\textbf {\bibinfo {volume} {9}},\ \bibinfo {pages}
  {113} (\bibinfo {year} {2013}{\natexlab{b}})}\BibitemShut {NoStop}%
\bibitem [{\citenamefont {Dijkstra}\ and\ \citenamefont
  {Tanimura}(2010)}]{Dijkstra_2010}%
  \BibitemOpen
  \bibfield  {author} {\bibinfo {author} {\bibfnamefont {A.~G.}\ \bibnamefont
  {Dijkstra}}\ and\ \bibinfo {author} {\bibfnamefont {Y.}~\bibnamefont
  {Tanimura}},\ }\href {\doibase 10.1103/physrevlett.104.250401} {\bibfield
  {journal} {\bibinfo  {journal} {Phys. Rev. Lett.}\ }\textbf {\bibinfo
  {volume} {104}} (\bibinfo {year} {2010}),\
  10.1103/physrevlett.104.250401}\BibitemShut {NoStop}%
\bibitem [{\citenamefont {Irish}\ \emph {et~al.}(2014)\citenamefont {Irish},
  \citenamefont {G\'omez-Bombarelli},\ and\ \citenamefont {Lovett}}]{elinor1}%
  \BibitemOpen
  \bibfield  {author} {\bibinfo {author} {\bibfnamefont {E.~K.}\ \bibnamefont
  {Irish}}, \bibinfo {author} {\bibfnamefont {R.}~\bibnamefont
  {G\'omez-Bombarelli}}, \ and\ \bibinfo {author} {\bibfnamefont {B.~W.}\
  \bibnamefont {Lovett}},\ }\href {\doibase 10.1103/PhysRevA.90.012510}
  {\bibfield  {journal} {\bibinfo  {journal} {Phys. Rev. A}\ }\textbf {\bibinfo
  {volume} {90}},\ \bibinfo {pages} {012510} (\bibinfo {year}
  {2014})}\BibitemShut {NoStop}%
\bibitem [{\citenamefont {Iles-Smith}\ \emph {et~al.}(2016)\citenamefont
  {Iles-Smith}, \citenamefont {Dijkstra}, \citenamefont {Lambert},\ and\
  \citenamefont {Nazir}}]{Iles_Smith_2015}%
  \BibitemOpen
  \bibfield  {author} {\bibinfo {author} {\bibfnamefont {J.}~\bibnamefont
  {Iles-Smith}}, \bibinfo {author} {\bibfnamefont {A.~G.}\ \bibnamefont
  {Dijkstra}}, \bibinfo {author} {\bibfnamefont {N.}~\bibnamefont {Lambert}}, \
  and\ \bibinfo {author} {\bibfnamefont {A.}~\bibnamefont {Nazir}},\ }\href
  {\doibase 10.1063/1.4940218} {\bibfield  {journal} {\bibinfo  {journal} {J.
  Chem. Phys.}\ }\textbf {\bibinfo {volume} {144}},\ \bibinfo {pages} {044110}
  (\bibinfo {year} {2016})}\BibitemShut {NoStop}%
\bibitem [{\citenamefont {Killoran}\ \emph {et~al.}(2015)\citenamefont
  {Killoran}, \citenamefont {Huelga},\ and\ \citenamefont
  {Plenio}}]{killoran2015}%
  \BibitemOpen
  \bibfield  {author} {\bibinfo {author} {\bibfnamefont {N.}~\bibnamefont
  {Killoran}}, \bibinfo {author} {\bibfnamefont {S.~F.}\ \bibnamefont
  {Huelga}}, \ and\ \bibinfo {author} {\bibfnamefont {M.~B.}\ \bibnamefont
  {Plenio}},\ }\href {\doibase 10.1063/1.4932307} {\bibfield  {journal}
  {\bibinfo  {journal} {The Journal of Chemical Physics}\ }\textbf {\bibinfo
  {volume} {143}},\ \bibinfo {pages} {155102} (\bibinfo {year} {2015})},\
  \Eprint {http://arxiv.org/abs/http://dx.doi.org/10.1063/1.4932307}
  {http://dx.doi.org/10.1063/1.4932307} \BibitemShut {NoStop}%
\bibitem [{\citenamefont {Mal{\'{y}}}\ \emph {et~al.}(2016)\citenamefont
  {Mal{\'{y}}}, \citenamefont {Somsen}, \citenamefont {Novoderezhkin},
  \citenamefont {Man{\v{c}}al},\ and\ \citenamefont
  {van{\hspace{0.25em}}Grondelle}}]{Mal__2016}%
  \BibitemOpen
  \bibfield  {author} {\bibinfo {author} {\bibfnamefont {P.}~\bibnamefont
  {Mal{\'{y}}}}, \bibinfo {author} {\bibfnamefont {O.~J.~G.}\ \bibnamefont
  {Somsen}}, \bibinfo {author} {\bibfnamefont {V.~I.}\ \bibnamefont
  {Novoderezhkin}}, \bibinfo {author} {\bibfnamefont {T.}~\bibnamefont
  {Man{\v{c}}al}}, \ and\ \bibinfo {author} {\bibfnamefont {R.}~\bibnamefont
  {van{\hspace{0.25em}}Grondelle}},\ }\href {\doibase 10.1002/cphc.201500965}
  {\bibfield  {journal} {\bibinfo  {journal} {{ChemPhysChem}}\ }\textbf
  {\bibinfo {volume} {17}},\ \bibinfo {pages} {1356} (\bibinfo {year}
  {2016})}\BibitemShut {NoStop}%
\bibitem [{\citenamefont {Qin}\ \emph {et~al.}(2017)\citenamefont {Qin},
  \citenamefont {Shen}, \citenamefont {Zhao},\ and\ \citenamefont
  {Yi}}]{Qin2017}%
  \BibitemOpen
  \bibfield  {author} {\bibinfo {author} {\bibfnamefont {M.}~\bibnamefont
  {Qin}}, \bibinfo {author} {\bibfnamefont {H.~Z.}\ \bibnamefont {Shen}},
  \bibinfo {author} {\bibfnamefont {X.~L.}\ \bibnamefont {Zhao}}, \ and\
  \bibinfo {author} {\bibfnamefont {X.~X.}\ \bibnamefont {Yi}},\ }\href
  {\doibase 10.1103/PhysRevA.96.012125} {\bibfield  {journal} {\bibinfo
  {journal} {Phys. Rev. A}\ }\textbf {\bibinfo {volume} {96}},\ \bibinfo
  {pages} {012125} (\bibinfo {year} {2017})}\BibitemShut {NoStop}%
\bibitem [{\citenamefont {Santamore}\ \emph {et~al.}(2013)\citenamefont
  {Santamore}, \citenamefont {Lambert},\ and\ \citenamefont
  {Nori}}]{Santamore2013}%
  \BibitemOpen
  \bibfield  {author} {\bibinfo {author} {\bibfnamefont {D.~H.}\ \bibnamefont
  {Santamore}}, \bibinfo {author} {\bibfnamefont {N.}~\bibnamefont {Lambert}},
  \ and\ \bibinfo {author} {\bibfnamefont {F.}~\bibnamefont {Nori}},\ }\href
  {\doibase 10.1103/PhysRevB.87.075422} {\bibfield  {journal} {\bibinfo
  {journal} {Phys. Rev. B}\ }\textbf {\bibinfo {volume} {87}},\ \bibinfo
  {pages} {075422} (\bibinfo {year} {2013})},\ \Eprint
  {http://arxiv.org/abs/1210.7098} {arXiv:1210.7098} \BibitemShut {NoStop}%
\bibitem [{\citenamefont {Stones}\ and\ \citenamefont
  {Olaya-Castro}(2016)}]{Stones2016}%
  \BibitemOpen
  \bibfield  {author} {\bibinfo {author} {\bibfnamefont {R.}~\bibnamefont
  {Stones}}\ and\ \bibinfo {author} {\bibfnamefont {A.}~\bibnamefont
  {Olaya-Castro}},\ }\bibfield  {booktitle} {\emph {\bibinfo {booktitle}
  {Chem}},\ }\href {\doibase 10.1016/j.chempr.2016.11.014} {\bibfield
  {journal} {\bibinfo  {journal} {Chem}\ }\textbf {\bibinfo {volume} {1}},\
  \bibinfo {pages} {822} (\bibinfo {year} {2016})}\BibitemShut {NoStop}%
\bibitem [{\citenamefont {Bakulin}\ \emph {et~al.}(2016)\citenamefont
  {Bakulin}, \citenamefont {Morgan}, \citenamefont {Kehoe}, \citenamefont
  {Wilson}, \citenamefont {Chin}, \citenamefont {Zigmantas}, \citenamefont
  {Egorova},\ and\ \citenamefont {Rao}}]{bakulin2016real}%
  \BibitemOpen
  \bibfield  {author} {\bibinfo {author} {\bibfnamefont {A.~A.}\ \bibnamefont
  {Bakulin}}, \bibinfo {author} {\bibfnamefont {S.~E.}\ \bibnamefont {Morgan}},
  \bibinfo {author} {\bibfnamefont {T.~B.}\ \bibnamefont {Kehoe}}, \bibinfo
  {author} {\bibfnamefont {M.~W.}\ \bibnamefont {Wilson}}, \bibinfo {author}
  {\bibfnamefont {A.~W.}\ \bibnamefont {Chin}}, \bibinfo {author}
  {\bibfnamefont {D.}~\bibnamefont {Zigmantas}}, \bibinfo {author}
  {\bibfnamefont {D.}~\bibnamefont {Egorova}}, \ and\ \bibinfo {author}
  {\bibfnamefont {A.}~\bibnamefont {Rao}},\ }\href@noop {} {\bibfield
  {journal} {\bibinfo  {journal} {Nature chemistry}\ }\textbf {\bibinfo
  {volume} {8}},\ \bibinfo {pages} {16} (\bibinfo {year} {2016})}\BibitemShut
  {NoStop}%
\bibitem [{\citenamefont {Falke}\ \emph {et~al.}(2014)\citenamefont {Falke},
  \citenamefont {Rozzi}, \citenamefont {Brida}, \citenamefont {Maiuri},
  \citenamefont {Amato}, \citenamefont {Sommer}, \citenamefont {De~Sio},
  \citenamefont {Rubio}, \citenamefont {Cerullo}, \citenamefont {Molinari}
  \emph {et~al.}}]{falke2014coherent}%
  \BibitemOpen
  \bibfield  {author} {\bibinfo {author} {\bibfnamefont {S.~M.}\ \bibnamefont
  {Falke}}, \bibinfo {author} {\bibfnamefont {C.~A.}\ \bibnamefont {Rozzi}},
  \bibinfo {author} {\bibfnamefont {D.}~\bibnamefont {Brida}}, \bibinfo
  {author} {\bibfnamefont {M.}~\bibnamefont {Maiuri}}, \bibinfo {author}
  {\bibfnamefont {M.}~\bibnamefont {Amato}}, \bibinfo {author} {\bibfnamefont
  {E.}~\bibnamefont {Sommer}}, \bibinfo {author} {\bibfnamefont
  {A.}~\bibnamefont {De~Sio}}, \bibinfo {author} {\bibfnamefont
  {A.}~\bibnamefont {Rubio}}, \bibinfo {author} {\bibfnamefont
  {G.}~\bibnamefont {Cerullo}}, \bibinfo {author} {\bibfnamefont
  {E.}~\bibnamefont {Molinari}},  \emph {et~al.},\ }\href@noop {} {\bibfield
  {journal} {\bibinfo  {journal} {Science}\ }\textbf {\bibinfo {volume}
  {344}},\ \bibinfo {pages} {1001} (\bibinfo {year} {2014})}\BibitemShut
  {NoStop}%
\bibitem [{\citenamefont {Lim}\ \emph {et~al.}(2015)\citenamefont {Lim},
  \citenamefont {Pale{\v{c}}ek}, \citenamefont {Caycedo-Soler}, \citenamefont
  {Lincoln}, \citenamefont {Prior}, \citenamefont {Von~Berlepsch},
  \citenamefont {Huelga}, \citenamefont {Plenio}, \citenamefont {Zigmantas},\
  and\ \citenamefont {Hauer}}]{lim2015vibronic}%
  \BibitemOpen
  \bibfield  {author} {\bibinfo {author} {\bibfnamefont {J.}~\bibnamefont
  {Lim}}, \bibinfo {author} {\bibfnamefont {D.}~\bibnamefont {Pale{\v{c}}ek}},
  \bibinfo {author} {\bibfnamefont {F.}~\bibnamefont {Caycedo-Soler}}, \bibinfo
  {author} {\bibfnamefont {C.~N.}\ \bibnamefont {Lincoln}}, \bibinfo {author}
  {\bibfnamefont {J.}~\bibnamefont {Prior}}, \bibinfo {author} {\bibfnamefont
  {H.}~\bibnamefont {Von~Berlepsch}}, \bibinfo {author} {\bibfnamefont {S.~F.}\
  \bibnamefont {Huelga}}, \bibinfo {author} {\bibfnamefont {M.~B.}\
  \bibnamefont {Plenio}}, \bibinfo {author} {\bibfnamefont {D.}~\bibnamefont
  {Zigmantas}}, \ and\ \bibinfo {author} {\bibfnamefont {J.}~\bibnamefont
  {Hauer}},\ }\href@noop {} {\bibfield  {journal} {\bibinfo  {journal} {Nature
  communications}\ }\textbf {\bibinfo {volume} {6}},\ \bibinfo {pages}
  {ncomms8755} (\bibinfo {year} {2015})}\BibitemShut {NoStop}%
\bibitem [{\citenamefont {Novelli}\ \emph {et~al.}(2015)\citenamefont
  {Novelli}, \citenamefont {Nazir}, \citenamefont {Richards}, \citenamefont
  {Roozbeh}, \citenamefont {Wilk}, \citenamefont {Curmi},\ and\ \citenamefont
  {Davis}}]{novelli2015vibronic}%
  \BibitemOpen
  \bibfield  {author} {\bibinfo {author} {\bibfnamefont {F.}~\bibnamefont
  {Novelli}}, \bibinfo {author} {\bibfnamefont {A.}~\bibnamefont {Nazir}},
  \bibinfo {author} {\bibfnamefont {G.~H.}\ \bibnamefont {Richards}}, \bibinfo
  {author} {\bibfnamefont {A.}~\bibnamefont {Roozbeh}}, \bibinfo {author}
  {\bibfnamefont {K.~E.}\ \bibnamefont {Wilk}}, \bibinfo {author}
  {\bibfnamefont {P.~M.}\ \bibnamefont {Curmi}}, \ and\ \bibinfo {author}
  {\bibfnamefont {J.~A.}\ \bibnamefont {Davis}},\ }\href@noop {} {\bibfield
  {journal} {\bibinfo  {journal} {The journal of physical chemistry letters}\
  }\textbf {\bibinfo {volume} {6}},\ \bibinfo {pages} {4573} (\bibinfo {year}
  {2015})}\BibitemShut {NoStop}%
\bibitem [{\citenamefont {Boulais}\ \emph {et~al.}(2018)\citenamefont
  {Boulais}, \citenamefont {Sawaya}, \citenamefont {Veneziano}, \citenamefont
  {Andreoni}, \citenamefont {Banal}, \citenamefont {Kondo}, \citenamefont
  {Mandal}, \citenamefont {Lin}, \citenamefont {Schlau-Cohen}, \citenamefont
  {Woodbury} \emph {et~al.}}]{boulais2018programmed}%
  \BibitemOpen
  \bibfield  {author} {\bibinfo {author} {\bibfnamefont {{\'E}.}~\bibnamefont
  {Boulais}}, \bibinfo {author} {\bibfnamefont {N.~P.}\ \bibnamefont {Sawaya}},
  \bibinfo {author} {\bibfnamefont {R.}~\bibnamefont {Veneziano}}, \bibinfo
  {author} {\bibfnamefont {A.}~\bibnamefont {Andreoni}}, \bibinfo {author}
  {\bibfnamefont {J.~L.}\ \bibnamefont {Banal}}, \bibinfo {author}
  {\bibfnamefont {T.}~\bibnamefont {Kondo}}, \bibinfo {author} {\bibfnamefont
  {S.}~\bibnamefont {Mandal}}, \bibinfo {author} {\bibfnamefont
  {S.}~\bibnamefont {Lin}}, \bibinfo {author} {\bibfnamefont {G.~S.}\
  \bibnamefont {Schlau-Cohen}}, \bibinfo {author} {\bibfnamefont {N.~W.}\
  \bibnamefont {Woodbury}},  \emph {et~al.},\ }\href@noop {} {\bibfield
  {journal} {\bibinfo  {journal} {Nature materials}\ }\textbf {\bibinfo
  {volume} {17}},\ \bibinfo {pages} {159} (\bibinfo {year} {2018})}\BibitemShut
  {NoStop}%
\bibitem [{\citenamefont {G{\'e}linas}\ \emph {et~al.}(2014)\citenamefont
  {G{\'e}linas}, \citenamefont {Rao}, \citenamefont {Kumar}, \citenamefont
  {Smith}, \citenamefont {Chin}, \citenamefont {Clark}, \citenamefont {van~der
  Poll}, \citenamefont {Bazan},\ and\ \citenamefont
  {Friend}}]{gelinas2014ultrafast}%
  \BibitemOpen
  \bibfield  {author} {\bibinfo {author} {\bibfnamefont {S.}~\bibnamefont
  {G{\'e}linas}}, \bibinfo {author} {\bibfnamefont {A.}~\bibnamefont {Rao}},
  \bibinfo {author} {\bibfnamefont {A.}~\bibnamefont {Kumar}}, \bibinfo
  {author} {\bibfnamefont {S.~L.}\ \bibnamefont {Smith}}, \bibinfo {author}
  {\bibfnamefont {A.~W.}\ \bibnamefont {Chin}}, \bibinfo {author}
  {\bibfnamefont {J.}~\bibnamefont {Clark}}, \bibinfo {author} {\bibfnamefont
  {T.~S.}\ \bibnamefont {van~der Poll}}, \bibinfo {author} {\bibfnamefont
  {G.~C.}\ \bibnamefont {Bazan}}, \ and\ \bibinfo {author} {\bibfnamefont
  {R.~H.}\ \bibnamefont {Friend}},\ }\href@noop {} {\bibfield  {journal}
  {\bibinfo  {journal} {Science}\ }\textbf {\bibinfo {volume} {343}},\ \bibinfo
  {pages} {512} (\bibinfo {year} {2014})}\BibitemShut {NoStop}%
\bibitem [{\citenamefont {Hemmig}\ \emph {et~al.}(2016)\citenamefont {Hemmig},
  \citenamefont {Creatore}, \citenamefont {Wu~nsch}, \citenamefont {Hecker},
  \citenamefont {Mair}, \citenamefont {Parker}, \citenamefont {Emmott},
  \citenamefont {Tinnefeld}, \citenamefont {Keyser},\ and\ \citenamefont
  {Chin}}]{hemmig2016programming}%
  \BibitemOpen
  \bibfield  {author} {\bibinfo {author} {\bibfnamefont {E.~A.}\ \bibnamefont
  {Hemmig}}, \bibinfo {author} {\bibfnamefont {C.}~\bibnamefont {Creatore}},
  \bibinfo {author} {\bibfnamefont {B.}~\bibnamefont {Wu~nsch}}, \bibinfo
  {author} {\bibfnamefont {L.}~\bibnamefont {Hecker}}, \bibinfo {author}
  {\bibfnamefont {P.}~\bibnamefont {Mair}}, \bibinfo {author} {\bibfnamefont
  {M.~A.}\ \bibnamefont {Parker}}, \bibinfo {author} {\bibfnamefont
  {S.}~\bibnamefont {Emmott}}, \bibinfo {author} {\bibfnamefont
  {P.}~\bibnamefont {Tinnefeld}}, \bibinfo {author} {\bibfnamefont {U.~F.}\
  \bibnamefont {Keyser}}, \ and\ \bibinfo {author} {\bibfnamefont {A.~W.}\
  \bibnamefont {Chin}},\ }\href@noop {} {\bibfield  {journal} {\bibinfo
  {journal} {Nano letters}\ }\textbf {\bibinfo {volume} {16}},\ \bibinfo
  {pages} {2369} (\bibinfo {year} {2016})}\BibitemShut {NoStop}%
\bibitem [{\citenamefont {Duan}\ \emph {et~al.}(2017)\citenamefont {Duan},
  \citenamefont {Prokhorenko}, \citenamefont {Cogdell}, \citenamefont {Ashraf},
  \citenamefont {Stevens}, \citenamefont {Thorwart},\ and\ \citenamefont
  {Miller}}]{duan2017nature}%
  \BibitemOpen
  \bibfield  {author} {\bibinfo {author} {\bibfnamefont {H.-G.}\ \bibnamefont
  {Duan}}, \bibinfo {author} {\bibfnamefont {V.~I.}\ \bibnamefont
  {Prokhorenko}}, \bibinfo {author} {\bibfnamefont {R.~J.}\ \bibnamefont
  {Cogdell}}, \bibinfo {author} {\bibfnamefont {K.}~\bibnamefont {Ashraf}},
  \bibinfo {author} {\bibfnamefont {A.~L.}\ \bibnamefont {Stevens}}, \bibinfo
  {author} {\bibfnamefont {M.}~\bibnamefont {Thorwart}}, \ and\ \bibinfo
  {author} {\bibfnamefont {R.~D.}\ \bibnamefont {Miller}},\ }\href@noop {}
  {\bibfield  {journal} {\bibinfo  {journal} {Proceedings of the National
  Academy of Sciences}\ }\textbf {\bibinfo {volume} {114}},\ \bibinfo {pages}
  {8493} (\bibinfo {year} {2017})}\BibitemShut {NoStop}%
\bibitem [{\citenamefont {May}\ \emph {et~al.}(2008)\citenamefont {May} \emph
  {et~al.}}]{may2008charge}%
  \BibitemOpen
  \bibfield  {author} {\bibinfo {author} {\bibfnamefont {V.}~\bibnamefont
  {May}} \emph {et~al.},\ }\href@noop {} {\emph {\bibinfo {title} {Charge and
  energy transfer dynamics in molecular systems}}}\ (\bibinfo  {publisher}
  {John Wiley \& Sons},\ \bibinfo {year} {2008})\BibitemShut {NoStop}%
\bibitem [{\citenamefont {Renger}\ \emph {et~al.}(2001)\citenamefont {Renger},
  \citenamefont {May},\ and\ \citenamefont {K{\"u}hn}}]{renger2001ultrafast}%
  \BibitemOpen
  \bibfield  {author} {\bibinfo {author} {\bibfnamefont {T.}~\bibnamefont
  {Renger}}, \bibinfo {author} {\bibfnamefont {V.}~\bibnamefont {May}}, \ and\
  \bibinfo {author} {\bibfnamefont {O.}~\bibnamefont {K{\"u}hn}},\ }\href@noop
  {} {\bibfield  {journal} {\bibinfo  {journal} {Physics Reports}\ }\textbf
  {\bibinfo {volume} {343}},\ \bibinfo {pages} {137} (\bibinfo {year}
  {2001})}\BibitemShut {NoStop}%
\bibitem [{\citenamefont {Scholes}\ \emph {et~al.}(2011)\citenamefont
  {Scholes}, \citenamefont {Fleming}, \citenamefont {Olaya-Castro},\ and\
  \citenamefont {Van~Grondelle}}]{scholes2011lessons}%
  \BibitemOpen
  \bibfield  {author} {\bibinfo {author} {\bibfnamefont {G.~D.}\ \bibnamefont
  {Scholes}}, \bibinfo {author} {\bibfnamefont {G.~R.}\ \bibnamefont
  {Fleming}}, \bibinfo {author} {\bibfnamefont {A.}~\bibnamefont
  {Olaya-Castro}}, \ and\ \bibinfo {author} {\bibfnamefont {R.}~\bibnamefont
  {Van~Grondelle}},\ }\href@noop {} {\bibfield  {journal} {\bibinfo  {journal}
  {Nature chemistry}\ }\textbf {\bibinfo {volume} {3}},\ \bibinfo {pages} {763}
  (\bibinfo {year} {2011})}\BibitemShut {NoStop}%
\bibitem [{\citenamefont {{P. Rebentrost}}\ \emph {et~al.}(2009)\citenamefont
  {{P. Rebentrost}}, \citenamefont {{M. Mohseni}}, \citenamefont {{I. Kassal}},
  \citenamefont {{S. Lloyd}},\ and\ \citenamefont
  {Aspuru-Guzik}}]{RebentrostEtAlNJP2009}%
  \BibitemOpen
  \bibfield  {author} {\bibinfo {author} {\bibnamefont {{P. Rebentrost}}},
  \bibinfo {author} {\bibnamefont {{M. Mohseni}}}, \bibinfo {author}
  {\bibnamefont {{I. Kassal}}}, \bibinfo {author} {\bibnamefont {{S. Lloyd}}},
  \ and\ \bibinfo {author} {\bibfnamefont {A.}~\bibnamefont {Aspuru-Guzik}},\
  }\href {\doibase 10.1088/1367-2630/11/3/033003} {\bibfield  {journal}
  {\bibinfo  {journal} {New J. Phys.}\ }\textbf {\bibinfo {volume} {11}},\
  \bibinfo {pages} {33003} (\bibinfo {year} {2009})}\BibitemShut {NoStop}%
\bibitem [{\citenamefont {Caruso}\ \emph {et~al.}(2009)\citenamefont {Caruso},
  \citenamefont {Chin}, \citenamefont {Datta}, \citenamefont {Huelga},\ and\
  \citenamefont {Plenio}}]{Caruso09}%
  \BibitemOpen
  \bibfield  {author} {\bibinfo {author} {\bibfnamefont {F.}~\bibnamefont
  {Caruso}}, \bibinfo {author} {\bibfnamefont {A.~W.}\ \bibnamefont {Chin}},
  \bibinfo {author} {\bibfnamefont {A.}~\bibnamefont {Datta}}, \bibinfo
  {author} {\bibfnamefont {S.~F.}\ \bibnamefont {Huelga}}, \ and\ \bibinfo
  {author} {\bibfnamefont {M.~B.}\ \bibnamefont {Plenio}},\ }\href {\doibase
  10.1063/1.3223548} {\bibfield  {journal} {\bibinfo  {journal} {J. Chem.
  Phys.}\ }\textbf {\bibinfo {volume} {131}},\ \bibinfo {pages} {105106}
  (\bibinfo {year} {2009})}\BibitemShut {NoStop}%
\bibitem [{\citenamefont {Plenio}\ and\ \citenamefont
  {Huelga}(2008)}]{Plenio08}%
  \BibitemOpen
  \bibfield  {author} {\bibinfo {author} {\bibfnamefont {M.~B.}\ \bibnamefont
  {Plenio}}\ and\ \bibinfo {author} {\bibfnamefont {S.~F.}\ \bibnamefont
  {Huelga}},\ }\href {\doibase 10.1088/1367-2630/10/11/113019} {\bibfield
  {journal} {\bibinfo  {journal} {New J. Phys.}\ }\textbf {\bibinfo {volume}
  {10}},\ \bibinfo {pages} {113019} (\bibinfo {year} {2008})}\BibitemShut
  {NoStop}%
\bibitem [{\citenamefont {Chin}\ \emph
  {et~al.}(2010{\natexlab{a}})\citenamefont {Chin}, \citenamefont {Datta},
  \citenamefont {Caruso}, \citenamefont {Plenio},\ and\ \citenamefont
  {Huelga}}]{Plenio10}%
  \BibitemOpen
  \bibfield  {author} {\bibinfo {author} {\bibfnamefont {A.~W.}\ \bibnamefont
  {Chin}}, \bibinfo {author} {\bibfnamefont {A.}~\bibnamefont {Datta}},
  \bibinfo {author} {\bibfnamefont {F.}~\bibnamefont {Caruso}}, \bibinfo
  {author} {\bibfnamefont {M.~B.}\ \bibnamefont {Plenio}}, \ and\ \bibinfo
  {author} {\bibfnamefont {S.~F.}\ \bibnamefont {Huelga}},\ }\href {\doibase
  10.1088/1367-2630/12/6/065002} {\bibfield  {journal} {\bibinfo  {journal}
  {New J. Phys.}\ }\textbf {\bibinfo {volume} {12}},\ \bibinfo {pages} {065002}
  (\bibinfo {year} {2010}{\natexlab{a}})}\BibitemShut {NoStop}%
\bibitem [{\citenamefont {Chen}\ \emph {et~al.}(2016)\citenamefont {Chen},
  \citenamefont {Chiu},\ and\ \citenamefont {Chen}}]{Chen2016}%
  \BibitemOpen
  \bibfield  {author} {\bibinfo {author} {\bibfnamefont {H.-B.}\ \bibnamefont
  {Chen}}, \bibinfo {author} {\bibfnamefont {P.-Y.}\ \bibnamefont {Chiu}}, \
  and\ \bibinfo {author} {\bibfnamefont {Y.-N.}\ \bibnamefont {Chen}},\ }\href
  {\doibase 10.1103/PhysRevE.94.052101} {\bibfield  {journal} {\bibinfo
  {journal} {Phys. Rev. E}\ }\textbf {\bibinfo {volume} {94}},\ \bibinfo
  {pages} {052101} (\bibinfo {year} {2016})}\BibitemShut {NoStop}%
\bibitem [{\citenamefont {Huelga}\ and\ \citenamefont
  {Plenio}(2013)}]{Huelga2013}%
  \BibitemOpen
  \bibfield  {author} {\bibinfo {author} {\bibfnamefont {S.~F.}\ \bibnamefont
  {Huelga}}\ and\ \bibinfo {author} {\bibfnamefont {M.~B.}\ \bibnamefont
  {Plenio}},\ }\href {\doibase 10.1080/00405000.2013.829687} {\bibfield
  {journal} {\bibinfo  {journal} {Contemp. Phys.}\ }\textbf {\bibinfo {volume}
  {54}},\ \bibinfo {pages} {181} (\bibinfo {year} {2013})},\ \Eprint
  {http://arxiv.org/abs/1307.3530v1} {1307.3530v1} \BibitemShut {NoStop}%
\bibitem [{\citenamefont {Schulze}\ and\ \citenamefont
  {Kuhn}(2015)}]{schulze2015explicit}%
  \BibitemOpen
  \bibfield  {author} {\bibinfo {author} {\bibfnamefont {J.}~\bibnamefont
  {Schulze}}\ and\ \bibinfo {author} {\bibfnamefont {O.}~\bibnamefont {Kuhn}},\
  }\href@noop {} {\bibfield  {journal} {\bibinfo  {journal} {The Journal of
  Physical Chemistry B}\ }\textbf {\bibinfo {volume} {119}},\ \bibinfo {pages}
  {6211} (\bibinfo {year} {2015})}\BibitemShut {NoStop}%
\bibitem [{\citenamefont {de~Vega}\ and\ \citenamefont
  {Alonso}(2017)}]{deVega2017}%
  \BibitemOpen
  \bibfield  {author} {\bibinfo {author} {\bibfnamefont {I.}~\bibnamefont
  {de~Vega}}\ and\ \bibinfo {author} {\bibfnamefont {D.}~\bibnamefont
  {Alonso}},\ }\href {\doibase 10.1103/RevModPhys.89.015001} {\bibfield
  {journal} {\bibinfo  {journal} {Rev. Mod. Phys.}\ }\textbf {\bibinfo {volume}
  {89}},\ \bibinfo {pages} {015001} (\bibinfo {year} {2017})}\BibitemShut
  {NoStop}%
\bibitem [{\citenamefont {Fruchtman}\ \emph {et~al.}(2016)\citenamefont
  {Fruchtman}, \citenamefont {Lambert},\ and\ \citenamefont
  {Gauger}}]{Fruchtman2016}%
  \BibitemOpen
  \bibfield  {author} {\bibinfo {author} {\bibfnamefont {A.}~\bibnamefont
  {Fruchtman}}, \bibinfo {author} {\bibfnamefont {N.}~\bibnamefont {Lambert}},
  \ and\ \bibinfo {author} {\bibfnamefont {E.~M.}\ \bibnamefont {Gauger}},\
  }\href {\doibase 10.1038/srep28204} {\bibfield  {journal} {\bibinfo
  {journal} {Sci. Rep.}\ }\textbf {\bibinfo {volume} {6}},\ \bibinfo {pages}
  {28204} (\bibinfo {year} {2016})},\ \Eprint {http://arxiv.org/abs/1512.09086}
  {arXiv:1512.09086} \BibitemShut {NoStop}%
\bibitem [{\citenamefont {Hughes}\ \emph {et~al.}(2009)\citenamefont {Hughes},
  \citenamefont {Christ},\ and\ \citenamefont {Burghardt}}]{Hughes_2009}%
  \BibitemOpen
  \bibfield  {author} {\bibinfo {author} {\bibfnamefont {K.~H.}\ \bibnamefont
  {Hughes}}, \bibinfo {author} {\bibfnamefont {C.~D.}\ \bibnamefont {Christ}},
  \ and\ \bibinfo {author} {\bibfnamefont {I.}~\bibnamefont {Burghardt}},\
  }\href {\doibase 10.1063/1.3159671} {\bibfield  {journal} {\bibinfo
  {journal} {The Journal of Chemical Physics}\ }\textbf {\bibinfo {volume}
  {131}},\ \bibinfo {pages} {024109} (\bibinfo {year} {2009})}\BibitemShut
  {NoStop}%
\bibitem [{\citenamefont {Chin}\ \emph
  {et~al.}(2010{\natexlab{b}})\citenamefont {Chin}, \citenamefont {Rivas},
  \citenamefont {Huelga},\ and\ \citenamefont {Plenio}}]{chin2010exact}%
  \BibitemOpen
  \bibfield  {author} {\bibinfo {author} {\bibfnamefont {A.~W.}\ \bibnamefont
  {Chin}}, \bibinfo {author} {\bibfnamefont {{\'A}.}~\bibnamefont {Rivas}},
  \bibinfo {author} {\bibfnamefont {S.~F.}\ \bibnamefont {Huelga}}, \ and\
  \bibinfo {author} {\bibfnamefont {M.~B.}\ \bibnamefont {Plenio}},\
  }\href@noop {} {\bibfield  {journal} {\bibinfo  {journal} {Journal of
  Mathematical Physics}\ }\textbf {\bibinfo {volume} {51}},\ \bibinfo {pages}
  {092109} (\bibinfo {year} {2010}{\natexlab{b}})}\BibitemShut {NoStop}%
\bibitem [{\citenamefont {Prior}\ \emph {et~al.}(2010)\citenamefont {Prior},
  \citenamefont {Chin}, \citenamefont {Huelga},\ and\ \citenamefont
  {Plenio}}]{prior2010efficient}%
  \BibitemOpen
  \bibfield  {author} {\bibinfo {author} {\bibfnamefont {J.}~\bibnamefont
  {Prior}}, \bibinfo {author} {\bibfnamefont {A.~W.}\ \bibnamefont {Chin}},
  \bibinfo {author} {\bibfnamefont {S.~F.}\ \bibnamefont {Huelga}}, \ and\
  \bibinfo {author} {\bibfnamefont {M.~B.}\ \bibnamefont {Plenio}},\
  }\href@noop {} {\bibfield  {journal} {\bibinfo  {journal} {Physical review
  letters}\ }\textbf {\bibinfo {volume} {105}},\ \bibinfo {pages} {050404}
  (\bibinfo {year} {2010})}\BibitemShut {NoStop}%
\bibitem [{\citenamefont {Makri}(1998)}]{Makri_1998}%
  \BibitemOpen
  \bibfield  {author} {\bibinfo {author} {\bibfnamefont {N.}~\bibnamefont
  {Makri}},\ }\href {\doibase 10.1021/jp980359y} {\bibfield  {journal}
  {\bibinfo  {journal} {The Journal of Physical Chemistry A}\ }\textbf
  {\bibinfo {volume} {102}},\ \bibinfo {pages} {4414} (\bibinfo {year}
  {1998})}\BibitemShut {NoStop}%
\bibitem [{\citenamefont {Manthe}(2008)}]{manthe2008multilayer}%
  \BibitemOpen
  \bibfield  {author} {\bibinfo {author} {\bibfnamefont {U.}~\bibnamefont
  {Manthe}},\ }\href@noop {} {\bibfield  {journal} {\bibinfo  {journal} {The
  Journal of chemical physics}\ }\textbf {\bibinfo {volume} {128}},\ \bibinfo
  {pages} {164116} (\bibinfo {year} {2008})}\BibitemShut {NoStop}%
\bibitem [{\citenamefont {Martinazzo}\ \emph {et~al.}(2011)\citenamefont
  {Martinazzo}, \citenamefont {Vacchini}, \citenamefont {Hughes},\ and\
  \citenamefont {Burghardt}}]{Martinazzo_2011}%
  \BibitemOpen
  \bibfield  {author} {\bibinfo {author} {\bibfnamefont {R.}~\bibnamefont
  {Martinazzo}}, \bibinfo {author} {\bibfnamefont {B.}~\bibnamefont
  {Vacchini}}, \bibinfo {author} {\bibfnamefont {K.~H.}\ \bibnamefont
  {Hughes}}, \ and\ \bibinfo {author} {\bibfnamefont {I.}~\bibnamefont
  {Burghardt}},\ }\href {\doibase 10.1063/1.3532408} {\bibfield  {journal}
  {\bibinfo  {journal} {The Journal of Chemical Physics}\ }\textbf {\bibinfo
  {volume} {134}},\ \bibinfo {pages} {011101} (\bibinfo {year}
  {2011})}\BibitemShut {NoStop}%
\bibitem [{\citenamefont {Nalbach}\ \emph {et~al.}(2011)\citenamefont
  {Nalbach}, \citenamefont {Ishizaki}, \citenamefont {Fleming},\ and\
  \citenamefont {Thorwart}}]{Peter11}%
  \BibitemOpen
  \bibfield  {author} {\bibinfo {author} {\bibfnamefont {P.}~\bibnamefont
  {Nalbach}}, \bibinfo {author} {\bibfnamefont {A.}~\bibnamefont {Ishizaki}},
  \bibinfo {author} {\bibfnamefont {G.~R.}\ \bibnamefont {Fleming}}, \ and\
  \bibinfo {author} {\bibfnamefont {M.}~\bibnamefont {Thorwart}},\ }\href
  {\doibase 10.1088/1367-2630/13/6/063040} {\bibfield  {journal} {\bibinfo
  {journal} {New J. Phys.}\ }\textbf {\bibinfo {volume} {13}},\ \bibinfo
  {pages} {063040} (\bibinfo {year} {2011})}\BibitemShut {NoStop}%
\bibitem [{\citenamefont {Prior}\ \emph {et~al.}(2013)\citenamefont {Prior},
  \citenamefont {de~Vega}, \citenamefont {Chin}, \citenamefont {Huelga},\ and\
  \citenamefont {Plenio}}]{prior2013quantum}%
  \BibitemOpen
  \bibfield  {author} {\bibinfo {author} {\bibfnamefont {J.}~\bibnamefont
  {Prior}}, \bibinfo {author} {\bibfnamefont {I.}~\bibnamefont {de~Vega}},
  \bibinfo {author} {\bibfnamefont {A.~W.}\ \bibnamefont {Chin}}, \bibinfo
  {author} {\bibfnamefont {S.~F.}\ \bibnamefont {Huelga}}, \ and\ \bibinfo
  {author} {\bibfnamefont {M.~B.}\ \bibnamefont {Plenio}},\ }\href@noop {}
  {\bibfield  {journal} {\bibinfo  {journal} {Physical Review A}\ }\textbf
  {\bibinfo {volume} {87}},\ \bibinfo {pages} {013428} (\bibinfo {year}
  {2013})}\BibitemShut {NoStop}%
\bibitem [{\citenamefont {Schr{\"o}der}\ \emph {et~al.}(2017)\citenamefont
  {Schr{\"o}der}, \citenamefont {Turban}, \citenamefont {Musser}, \citenamefont
  {Hine},\ and\ \citenamefont {Chin}}]{schroder2017multi}%
  \BibitemOpen
  \bibfield  {author} {\bibinfo {author} {\bibfnamefont {F.~A.}\ \bibnamefont
  {Schr{\"o}der}}, \bibinfo {author} {\bibfnamefont {D.~H.}\ \bibnamefont
  {Turban}}, \bibinfo {author} {\bibfnamefont {A.~J.}\ \bibnamefont {Musser}},
  \bibinfo {author} {\bibfnamefont {N.~D.}\ \bibnamefont {Hine}}, \ and\
  \bibinfo {author} {\bibfnamefont {A.~W.}\ \bibnamefont {Chin}},\ }\href@noop
  {} {\bibfield  {journal} {\bibinfo  {journal} {arXiv preprint
  arXiv:1710.01362}\ } (\bibinfo {year} {2017})}\BibitemShut {NoStop}%
\bibitem [{\citenamefont {Strasberg}\ \emph {et~al.}(2016)\citenamefont
  {Strasberg}, \citenamefont {Schaller}, \citenamefont {Lambert},\ and\
  \citenamefont {Brandes}}]{Strasberg_2016}%
  \BibitemOpen
  \bibfield  {author} {\bibinfo {author} {\bibfnamefont {P.}~\bibnamefont
  {Strasberg}}, \bibinfo {author} {\bibfnamefont {G.}~\bibnamefont {Schaller}},
  \bibinfo {author} {\bibfnamefont {N.}~\bibnamefont {Lambert}}, \ and\
  \bibinfo {author} {\bibfnamefont {T.}~\bibnamefont {Brandes}},\ }\href
  {\doibase 10.1088/1367-2630/18/7/073007} {\bibfield  {journal} {\bibinfo
  {journal} {New Journal of Physics}\ }\textbf {\bibinfo {volume} {18}},\
  \bibinfo {pages} {073007} (\bibinfo {year} {2016})}\BibitemShut {NoStop}%
\bibitem [{\citenamefont {Tanimura}\ and\ \citenamefont
  {Kubo}(1989)}]{Tanimura_1989}%
  \BibitemOpen
  \bibfield  {author} {\bibinfo {author} {\bibfnamefont {Y.}~\bibnamefont
  {Tanimura}}\ and\ \bibinfo {author} {\bibfnamefont {R.}~\bibnamefont
  {Kubo}},\ }\href {\doibase 10.1143/jpsj.58.101} {\bibfield  {journal}
  {\bibinfo  {journal} {Journal of the Physical Society of Japan}\ }\textbf
  {\bibinfo {volume} {58}},\ \bibinfo {pages} {101} (\bibinfo {year}
  {1989})}\BibitemShut {NoStop}%
\bibitem [{\citenamefont {Thoss}\ \emph {et~al.}(2001)\citenamefont {Thoss},
  \citenamefont {Wang},\ and\ \citenamefont {Miller}}]{Thoss_2001}%
  \BibitemOpen
  \bibfield  {author} {\bibinfo {author} {\bibfnamefont {M.}~\bibnamefont
  {Thoss}}, \bibinfo {author} {\bibfnamefont {H.}~\bibnamefont {Wang}}, \ and\
  \bibinfo {author} {\bibfnamefont {W.~H.}\ \bibnamefont {Miller}},\ }\href
  {\doibase 10.1063/1.1385562} {\bibfield  {journal} {\bibinfo  {journal} {The
  Journal of Chemical Physics}\ }\textbf {\bibinfo {volume} {115}},\ \bibinfo
  {pages} {2991} (\bibinfo {year} {2001})}\BibitemShut {NoStop}%
\bibitem [{\citenamefont {Ishizaki}\ and\ \citenamefont
  {Tanimura}(2005)}]{Ishizaki_2005}%
  \BibitemOpen
  \bibfield  {author} {\bibinfo {author} {\bibfnamefont {A.}~\bibnamefont
  {Ishizaki}}\ and\ \bibinfo {author} {\bibfnamefont {Y.}~\bibnamefont
  {Tanimura}},\ }\href {\doibase 10.1143/JPSJ.74.3131} {\bibfield  {journal}
  {\bibinfo  {journal} {Journal of the Physical Society of Japan}\ }\textbf
  {\bibinfo {volume} {74}},\ \bibinfo {pages} {3131} (\bibinfo {year}
  {2005})},\ \Eprint
  {http://arxiv.org/abs/https://doi.org/10.1143/JPSJ.74.3131}
  {https://doi.org/10.1143/JPSJ.74.3131} \BibitemShut {NoStop}%
\bibitem [{\citenamefont {Tanimura}(2006)}]{Tanimura_2006}%
  \BibitemOpen
  \bibfield  {author} {\bibinfo {author} {\bibfnamefont {Y.}~\bibnamefont
  {Tanimura}},\ }\href {\doibase 10.1143/JPSJ.75.082001} {\bibfield  {journal}
  {\bibinfo  {journal} {Journal of the Physical Society of Japan}\ }\textbf
  {\bibinfo {volume} {75}},\ \bibinfo {pages} {082001} (\bibinfo {year}
  {2006})},\ \Eprint
  {http://arxiv.org/abs/https://doi.org/10.1143/JPSJ.75.082001}
  {https://doi.org/10.1143/JPSJ.75.082001} \BibitemShut {NoStop}%
\bibitem [{\citenamefont {Xu}\ and\ \citenamefont {Yan}(2007)}]{Xu_2007}%
  \BibitemOpen
  \bibfield  {author} {\bibinfo {author} {\bibfnamefont {R.-X.}\ \bibnamefont
  {Xu}}\ and\ \bibinfo {author} {\bibfnamefont {Y.}~\bibnamefont {Yan}},\
  }\href {\doibase 10.1103/PhysRevE.75.031107} {\bibfield  {journal} {\bibinfo
  {journal} {Phys. Rev. E}\ }\textbf {\bibinfo {volume} {75}},\ \bibinfo
  {pages} {031107} (\bibinfo {year} {2007})}\BibitemShut {NoStop}%
\bibitem [{\citenamefont {Str{\"u}mpfer}\ and\ \citenamefont
  {Schulten}(2012)}]{Schulten_2012}%
  \BibitemOpen
  \bibfield  {author} {\bibinfo {author} {\bibfnamefont {J.}~\bibnamefont
  {Str{\"u}mpfer}}\ and\ \bibinfo {author} {\bibfnamefont {K.}~\bibnamefont
  {Schulten}},\ }\href {\doibase 10.1021/ct3003833} {\bibfield  {journal}
  {\bibinfo  {journal} {Journal of Chemical Theory and Computation}\ }\textbf
  {\bibinfo {volume} {8}},\ \bibinfo {pages} {2808} (\bibinfo {year} {2012})},\
  \bibinfo {note} {pMID: 23105920},\ \Eprint
  {http://arxiv.org/abs/https://doi.org/10.1021/ct3003833}
  {https://doi.org/10.1021/ct3003833} \BibitemShut {NoStop}%
\bibitem [{\citenamefont {Ishizaki}\ and\ \citenamefont
  {Fleming}(2009{\natexlab{b}})}]{Ishizaki_2009}%
  \BibitemOpen
  \bibfield  {author} {\bibinfo {author} {\bibfnamefont {A.}~\bibnamefont
  {Ishizaki}}\ and\ \bibinfo {author} {\bibfnamefont {G.~R.}\ \bibnamefont
  {Fleming}},\ }\href {\doibase 10.1063/1.3155372} {\bibfield  {journal}
  {\bibinfo  {journal} {The Journal of Chemical Physics}\ }\textbf {\bibinfo
  {volume} {130}},\ \bibinfo {pages} {234111} (\bibinfo {year}
  {2009}{\natexlab{b}})},\ \Eprint
  {http://arxiv.org/abs/https://doi.org/10.1063/1.3155372}
  {https://doi.org/10.1063/1.3155372} \BibitemShut {NoStop}%
\bibitem [{\citenamefont {Shi}\ \emph {et~al.}(2009)\citenamefont {Shi},
  \citenamefont {Chen}, \citenamefont {Nan}, \citenamefont {Xu},\ and\
  \citenamefont {Yan}}]{Shi_2009}%
  \BibitemOpen
  \bibfield  {author} {\bibinfo {author} {\bibfnamefont {Q.}~\bibnamefont
  {Shi}}, \bibinfo {author} {\bibfnamefont {L.}~\bibnamefont {Chen}}, \bibinfo
  {author} {\bibfnamefont {G.}~\bibnamefont {Nan}}, \bibinfo {author}
  {\bibfnamefont {R.-X.}\ \bibnamefont {Xu}}, \ and\ \bibinfo {author}
  {\bibfnamefont {Y.}~\bibnamefont {Yan}},\ }\href {\doibase 10.1063/1.3077918}
  {\bibfield  {journal} {\bibinfo  {journal} {The Journal of Chemical Physics}\
  }\textbf {\bibinfo {volume} {130}},\ \bibinfo {pages} {084105} (\bibinfo
  {year} {2009})},\ \Eprint
  {http://arxiv.org/abs/https://doi.org/10.1063/1.3077918}
  {https://doi.org/10.1063/1.3077918} \BibitemShut {NoStop}%
\bibitem [{\citenamefont {Dodin}\ \emph {et~al.}(2016)\citenamefont {Dodin},
  \citenamefont {Tscherbul},\ and\ \citenamefont {Brumer}}]{brumer2016}%
  \BibitemOpen
  \bibfield  {author} {\bibinfo {author} {\bibfnamefont {A.}~\bibnamefont
  {Dodin}}, \bibinfo {author} {\bibfnamefont {T.~V.}\ \bibnamefont
  {Tscherbul}}, \ and\ \bibinfo {author} {\bibfnamefont {P.}~\bibnamefont
  {Brumer}},\ }\href {\doibase 10.1063/1.4954243} {\bibfield  {journal}
  {\bibinfo  {journal} {The Journal of Chemical Physics}\ }\textbf {\bibinfo
  {volume} {144}},\ \bibinfo {pages} {244108} (\bibinfo {year} {2016})},\
  \Eprint {http://arxiv.org/abs/http://dx.doi.org/10.1063/1.4954243}
  {http://dx.doi.org/10.1063/1.4954243} \BibitemShut {NoStop}%
\bibitem [{\citenamefont {Han}\ \emph {et~al.}(2013)\citenamefont {Han},
  \citenamefont {Shapiro},\ and\ \citenamefont {Brumer}}]{han2013nature}%
  \BibitemOpen
  \bibfield  {author} {\bibinfo {author} {\bibfnamefont {A.~C.}\ \bibnamefont
  {Han}}, \bibinfo {author} {\bibfnamefont {M.}~\bibnamefont {Shapiro}}, \ and\
  \bibinfo {author} {\bibfnamefont {P.}~\bibnamefont {Brumer}},\ }\href@noop {}
  {\bibfield  {journal} {\bibinfo  {journal} {The Journal of Physical Chemistry
  A}\ }\textbf {\bibinfo {volume} {117}},\ \bibinfo {pages} {8199} (\bibinfo
  {year} {2013})}\BibitemShut {NoStop}%
\bibitem [{\citenamefont {Gorman}\ \emph {et~al.}(2018)\citenamefont {Gorman},
  \citenamefont {Hemmerling}, \citenamefont {Megidish}, \citenamefont
  {Moeller}, \citenamefont {Schindler}, \citenamefont {Sarovar},\ and\
  \citenamefont {Haeffner}}]{gorman2018engineering}%
  \BibitemOpen
  \bibfield  {author} {\bibinfo {author} {\bibfnamefont {D.~J.}\ \bibnamefont
  {Gorman}}, \bibinfo {author} {\bibfnamefont {B.}~\bibnamefont {Hemmerling}},
  \bibinfo {author} {\bibfnamefont {E.}~\bibnamefont {Megidish}}, \bibinfo
  {author} {\bibfnamefont {S.~A.}\ \bibnamefont {Moeller}}, \bibinfo {author}
  {\bibfnamefont {P.}~\bibnamefont {Schindler}}, \bibinfo {author}
  {\bibfnamefont {M.}~\bibnamefont {Sarovar}}, \ and\ \bibinfo {author}
  {\bibfnamefont {H.}~\bibnamefont {Haeffner}},\ }\href@noop {} {\bibfield
  {journal} {\bibinfo  {journal} {Physical Review X}\ }\textbf {\bibinfo
  {volume} {8}},\ \bibinfo {pages} {011038} (\bibinfo {year}
  {2018})}\BibitemShut {NoStop}%
\bibitem [{\citenamefont {Creatore}\ \emph {et~al.}(2013)\citenamefont
  {Creatore}, \citenamefont {Parker}, \citenamefont {Emmott},\ and\
  \citenamefont {Chin}}]{Creatore_2013}%
  \BibitemOpen
  \bibfield  {author} {\bibinfo {author} {\bibfnamefont {C.}~\bibnamefont
  {Creatore}}, \bibinfo {author} {\bibfnamefont {M.~A.}\ \bibnamefont
  {Parker}}, \bibinfo {author} {\bibfnamefont {S.}~\bibnamefont {Emmott}}, \
  and\ \bibinfo {author} {\bibfnamefont {A.~W.}\ \bibnamefont {Chin}},\ }\href
  {\doibase 10.1103/physrevlett.111.253601} {\bibfield  {journal} {\bibinfo
  {journal} {Physical Review Letters}\ }\textbf {\bibinfo {volume} {111}}
  (\bibinfo {year} {2013}),\ 10.1103/physrevlett.111.253601}\BibitemShut
  {NoStop}%
\bibitem [{\citenamefont {Dorfman}\ \emph {et~al.}(2013)\citenamefont
  {Dorfman}, \citenamefont {Voronine}, \citenamefont {Mukamel},\ and\
  \citenamefont {Scully}}]{Dorfman_2013}%
  \BibitemOpen
  \bibfield  {author} {\bibinfo {author} {\bibfnamefont {K.~E.}\ \bibnamefont
  {Dorfman}}, \bibinfo {author} {\bibfnamefont {D.~V.}\ \bibnamefont
  {Voronine}}, \bibinfo {author} {\bibfnamefont {S.}~\bibnamefont {Mukamel}}, \
  and\ \bibinfo {author} {\bibfnamefont {M.~O.}\ \bibnamefont {Scully}},\
  }\href {\doibase 10.1073/pnas.1212666110} {\bibfield  {journal} {\bibinfo
  {journal} {Proceedings of the National Academy of Sciences}\ }\textbf
  {\bibinfo {volume} {110}},\ \bibinfo {pages} {2746} (\bibinfo {year}
  {2013})}\BibitemShut {NoStop}%
\bibitem [{\citenamefont {Gelbwaser-Klimovsky}\ and\ \citenamefont
  {Aspuru-Guzik}(2017)}]{guzik}%
  \BibitemOpen
  \bibfield  {author} {\bibinfo {author} {\bibfnamefont {D.}~\bibnamefont
  {Gelbwaser-Klimovsky}}\ and\ \bibinfo {author} {\bibfnamefont
  {A.}~\bibnamefont {Aspuru-Guzik}},\ }\href {\doibase 10.1039/C6SC04350J}
  {\bibfield  {journal} {\bibinfo  {journal} {Chem. Sci.}\ }\textbf {\bibinfo
  {volume} {8}},\ \bibinfo {pages} {1008} (\bibinfo {year} {2017})}\BibitemShut
  {NoStop}%
\bibitem [{\citenamefont {Newman}\ \emph {et~al.}(2017)\citenamefont {Newman},
  \citenamefont {Mintert},\ and\ \citenamefont {Nazir}}]{Newman_2016}%
  \BibitemOpen
  \bibfield  {author} {\bibinfo {author} {\bibfnamefont {D.}~\bibnamefont
  {Newman}}, \bibinfo {author} {\bibfnamefont {F.}~\bibnamefont {Mintert}}, \
  and\ \bibinfo {author} {\bibfnamefont {A.}~\bibnamefont {Nazir}},\ }\href
  {\doibase 10.1103/PhysRevE.95.032139} {\bibfield  {journal} {\bibinfo
  {journal} {Phys. Rev. E}\ }\textbf {\bibinfo {volume} {95}},\ \bibinfo
  {pages} {032139} (\bibinfo {year} {2017})}\BibitemShut {NoStop}%
\bibitem [{\citenamefont {Scully}(2010)}]{Scully2010}%
  \BibitemOpen
  \bibfield  {author} {\bibinfo {author} {\bibfnamefont {M.~O.}\ \bibnamefont
  {Scully}},\ }\href {\doibase 10.1103/PhysRevLett.104.207701} {\bibfield
  {journal} {\bibinfo  {journal} {Phys. Rev. Lett.}\ }\textbf {\bibinfo
  {volume} {104}},\ \bibinfo {pages} {207701} (\bibinfo {year}
  {2010})}\BibitemShut {NoStop}%
\bibitem [{\citenamefont {Zhang}\ \emph {et~al.}(2015)\citenamefont {Zhang},
  \citenamefont {Oh}, \citenamefont {Alharbi}, \citenamefont {Engel},\ and\
  \citenamefont {Kais}}]{Zhang_2015}%
  \BibitemOpen
  \bibfield  {author} {\bibinfo {author} {\bibfnamefont {Y.}~\bibnamefont
  {Zhang}}, \bibinfo {author} {\bibfnamefont {S.}~\bibnamefont {Oh}}, \bibinfo
  {author} {\bibfnamefont {F.~H.}\ \bibnamefont {Alharbi}}, \bibinfo {author}
  {\bibfnamefont {G.~S.}\ \bibnamefont {Engel}}, \ and\ \bibinfo {author}
  {\bibfnamefont {S.}~\bibnamefont {Kais}},\ }\href {\doibase
  10.1039/c4cp05310a} {\bibfield  {journal} {\bibinfo  {journal} {Phys. Chem.
  Chem. Phys.}\ }\textbf {\bibinfo {volume} {17}},\ \bibinfo {pages} {5743}
  (\bibinfo {year} {2015})}\BibitemShut {NoStop}%
\bibitem [{\citenamefont {Heinz-Peter~Breuer}(2002)}]{Breuer2002}%
  \BibitemOpen
  \bibfield  {author} {\bibinfo {author} {\bibfnamefont {F.~P.}\ \bibnamefont
  {Heinz-Peter~Breuer}},\ }\href@noop {} {\emph {\bibinfo {title} {The Theory
  of Open Quantum Systems}}}\ (\bibinfo  {publisher} {Oxford University
  Press},\ \bibinfo {year} {2002})\BibitemShut {NoStop}%
\bibitem [{\citenamefont {Blum}(2013)}]{blum2013density}%
  \BibitemOpen
  \bibfield  {author} {\bibinfo {author} {\bibfnamefont {K.}~\bibnamefont
  {Blum}},\ }\href@noop {} {\emph {\bibinfo {title} {Density matrix theory and
  applications}}}\ (\bibinfo  {publisher} {Springer Science \& Business
  Media},\ \bibinfo {year} {2013})\BibitemShut {NoStop}%
\bibitem [{\citenamefont {Jeske}\ \emph {et~al.}(2015)\citenamefont {Jeske},
  \citenamefont {Ing}, \citenamefont {Plenio}, \citenamefont {Huelga},\ and\
  \citenamefont {Cole}}]{jeske2015bloch}%
  \BibitemOpen
  \bibfield  {author} {\bibinfo {author} {\bibfnamefont {J.}~\bibnamefont
  {Jeske}}, \bibinfo {author} {\bibfnamefont {D.~J.}\ \bibnamefont {Ing}},
  \bibinfo {author} {\bibfnamefont {M.~B.}\ \bibnamefont {Plenio}}, \bibinfo
  {author} {\bibfnamefont {S.~F.}\ \bibnamefont {Huelga}}, \ and\ \bibinfo
  {author} {\bibfnamefont {J.~H.}\ \bibnamefont {Cole}},\ }\href@noop {}
  {\bibfield  {journal} {\bibinfo  {journal} {The Journal of chemical physics}\
  }\textbf {\bibinfo {volume} {142}},\ \bibinfo {pages} {064104} (\bibinfo
  {year} {2015})}\BibitemShut {NoStop}%
\bibitem [{\citenamefont {Eastham}\ \emph {et~al.}(2016)\citenamefont
  {Eastham}, \citenamefont {Kirton}, \citenamefont {Cammack}, \citenamefont
  {Lovett},\ and\ \citenamefont {Keeling}}]{eastham2016bath}%
  \BibitemOpen
  \bibfield  {author} {\bibinfo {author} {\bibfnamefont {P.}~\bibnamefont
  {Eastham}}, \bibinfo {author} {\bibfnamefont {P.}~\bibnamefont {Kirton}},
  \bibinfo {author} {\bibfnamefont {H.}~\bibnamefont {Cammack}}, \bibinfo
  {author} {\bibfnamefont {B.}~\bibnamefont {Lovett}}, \ and\ \bibinfo {author}
  {\bibfnamefont {J.}~\bibnamefont {Keeling}},\ }\href@noop {} {\bibfield
  {journal} {\bibinfo  {journal} {Physical Review A}\ }\textbf {\bibinfo
  {volume} {94}},\ \bibinfo {pages} {012110} (\bibinfo {year}
  {2016})}\BibitemShut {NoStop}%
\bibitem [{\citenamefont {Liu}\ \emph {et~al.}(2014)\citenamefont {Liu},
  \citenamefont {Zhu}, \citenamefont {Bai},\ and\ \citenamefont
  {Shi}}]{Shi_2014}%
  \BibitemOpen
  \bibfield  {author} {\bibinfo {author} {\bibfnamefont {H.}~\bibnamefont
  {Liu}}, \bibinfo {author} {\bibfnamefont {L.}~\bibnamefont {Zhu}}, \bibinfo
  {author} {\bibfnamefont {S.}~\bibnamefont {Bai}}, \ and\ \bibinfo {author}
  {\bibfnamefont {Q.}~\bibnamefont {Shi}},\ }\href {\doibase 10.1063/1.4870035}
  {\bibfield  {journal} {\bibinfo  {journal} {The Journal of Chemical Physics}\
  }\textbf {\bibinfo {volume} {140}},\ \bibinfo {pages} {134106} (\bibinfo
  {year} {2014})},\ \Eprint
  {http://arxiv.org/abs/https://doi.org/10.1063/1.4870035}
  {https://doi.org/10.1063/1.4870035} \BibitemShut {NoStop}%
\bibitem [{\citenamefont {Mangaud}\ \emph {et~al.}(2017)\citenamefont
  {Mangaud}, \citenamefont {Meier},\ and\ \citenamefont
  {Desouter-Lecomte}}]{Mangaud_2017}%
  \BibitemOpen
  \bibfield  {author} {\bibinfo {author} {\bibfnamefont {E.}~\bibnamefont
  {Mangaud}}, \bibinfo {author} {\bibfnamefont {C.}~\bibnamefont {Meier}}, \
  and\ \bibinfo {author} {\bibfnamefont {M.}~\bibnamefont {Desouter-Lecomte}},\
  }\href {\doibase https://doi.org/10.1016/j.chemphys.2017.07.011} {\bibfield
  {journal} {\bibinfo  {journal} {Chemical Physics}\ }\textbf {\bibinfo
  {volume} {494}},\ \bibinfo {pages} {90 } (\bibinfo {year}
  {2017})}\BibitemShut {NoStop}%
\bibitem [{\citenamefont {Pomyalov}\ \emph {et~al.}(2010)\citenamefont
  {Pomyalov}, \citenamefont {Meier},\ and\ \citenamefont
  {Tannor}}]{Tannor_2010}%
  \BibitemOpen
  \bibfield  {author} {\bibinfo {author} {\bibfnamefont {A.}~\bibnamefont
  {Pomyalov}}, \bibinfo {author} {\bibfnamefont {C.}~\bibnamefont {Meier}}, \
  and\ \bibinfo {author} {\bibfnamefont {D.}~\bibnamefont {Tannor}},\ }\href
  {\doibase https://doi.org/10.1016/j.chemphys.2010.02.017} {\bibfield
  {journal} {\bibinfo  {journal} {Chemical Physics}\ }\textbf {\bibinfo
  {volume} {370}},\ \bibinfo {pages} {98 } (\bibinfo {year} {2010})},\ \bibinfo
  {note} {dynamics of molecular systems: From quantum to classical}\BibitemShut
  {NoStop}%
\bibitem [{\citenamefont {Lorenzo}\ \emph {et~al.}(2011)\citenamefont
  {Lorenzo}, \citenamefont {Plastina},\ and\ \citenamefont
  {Paternostro}}]{Paternostro_2011}%
  \BibitemOpen
  \bibfield  {author} {\bibinfo {author} {\bibfnamefont {S.}~\bibnamefont
  {Lorenzo}}, \bibinfo {author} {\bibfnamefont {F.}~\bibnamefont {Plastina}}, \
  and\ \bibinfo {author} {\bibfnamefont {M.}~\bibnamefont {Paternostro}},\
  }\href {\doibase 10.1103/PhysRevA.84.032124} {\bibfield  {journal} {\bibinfo
  {journal} {Phys. Rev. A}\ }\textbf {\bibinfo {volume} {84}},\ \bibinfo
  {pages} {032124} (\bibinfo {year} {2011})}\BibitemShut {NoStop}%
\bibitem [{\citenamefont {Oviedo-Casado}\ \emph {et~al.}(2016)\citenamefont
  {Oviedo-Casado}, \citenamefont {Prior}, \citenamefont {Chin}, \citenamefont
  {Rosenbach}, \citenamefont {Huelga},\ and\ \citenamefont
  {Plenio}}]{oviedo2016phase}%
  \BibitemOpen
  \bibfield  {author} {\bibinfo {author} {\bibfnamefont {S.}~\bibnamefont
  {Oviedo-Casado}}, \bibinfo {author} {\bibfnamefont {J.}~\bibnamefont
  {Prior}}, \bibinfo {author} {\bibfnamefont {A.}~\bibnamefont {Chin}},
  \bibinfo {author} {\bibfnamefont {R.}~\bibnamefont {Rosenbach}}, \bibinfo
  {author} {\bibfnamefont {S.}~\bibnamefont {Huelga}}, \ and\ \bibinfo {author}
  {\bibfnamefont {M.}~\bibnamefont {Plenio}},\ }\href@noop {} {\bibfield
  {journal} {\bibinfo  {journal} {Physical Review A}\ }\textbf {\bibinfo
  {volume} {93}},\ \bibinfo {pages} {020102} (\bibinfo {year}
  {2016})}\BibitemShut {NoStop}%
\bibitem [{\citenamefont {Jean-Luc~Bredas}(2016)}]{Bredas2016}%
  \BibitemOpen
  \bibfield  {author} {\bibinfo {author} {\bibfnamefont {G.~D.~S.}\
  \bibnamefont {Jean-Luc~Bredas}, \bibfnamefont {Edward H.~Sargent}},\ }\href
  {\doibase 10.1038/nmat4767} {\bibfield  {journal} {\bibinfo  {journal}
  {Nature Materials}\ } (\bibinfo {year} {2016}),\
  10.1038/nmat4767}\BibitemShut {NoStop}%
\bibitem [{\citenamefont {Smith}\ and\ \citenamefont
  {Chin}(2014)}]{smith2014ultrafast}%
  \BibitemOpen
  \bibfield  {author} {\bibinfo {author} {\bibfnamefont {S.~L.}\ \bibnamefont
  {Smith}}\ and\ \bibinfo {author} {\bibfnamefont {A.~W.}\ \bibnamefont
  {Chin}},\ }\href@noop {} {\bibfield  {journal} {\bibinfo  {journal} {Physical
  Chemistry Chemical Physics}\ }\textbf {\bibinfo {volume} {16}},\ \bibinfo
  {pages} {20305} (\bibinfo {year} {2014})}\BibitemShut {NoStop}%
\bibitem [{\citenamefont {Smith}\ and\ \citenamefont
  {Chin}(2015)}]{smith2015phonon}%
  \BibitemOpen
  \bibfield  {author} {\bibinfo {author} {\bibfnamefont {S.~L.}\ \bibnamefont
  {Smith}}\ and\ \bibinfo {author} {\bibfnamefont {A.~W.}\ \bibnamefont
  {Chin}},\ }\href@noop {} {\bibfield  {journal} {\bibinfo  {journal} {Physical
  Review B}\ }\textbf {\bibinfo {volume} {91}},\ \bibinfo {pages} {201302}
  (\bibinfo {year} {2015})}\BibitemShut {NoStop}%
\bibitem [{\citenamefont {Mostame}\ \emph {et~al.}(2012)\citenamefont
  {Mostame}, \citenamefont {Rebentrost}, \citenamefont {Eisfeld}, \citenamefont
  {Kerman}, \citenamefont {Tsomokos},\ and\ \citenamefont
  {Aspuru-Guzik}}]{mostame2012quantum}%
  \BibitemOpen
  \bibfield  {author} {\bibinfo {author} {\bibfnamefont {S.}~\bibnamefont
  {Mostame}}, \bibinfo {author} {\bibfnamefont {P.}~\bibnamefont {Rebentrost}},
  \bibinfo {author} {\bibfnamefont {A.}~\bibnamefont {Eisfeld}}, \bibinfo
  {author} {\bibfnamefont {A.~J.}\ \bibnamefont {Kerman}}, \bibinfo {author}
  {\bibfnamefont {D.~I.}\ \bibnamefont {Tsomokos}}, \ and\ \bibinfo {author}
  {\bibfnamefont {A.}~\bibnamefont {Aspuru-Guzik}},\ }\href@noop {} {\bibfield
  {journal} {\bibinfo  {journal} {New Journal of Physics}\ }\textbf {\bibinfo
  {volume} {14}},\ \bibinfo {pages} {105013} (\bibinfo {year}
  {2012})}\BibitemShut {NoStop}%
\bibitem [{\citenamefont {Breuer}\ \emph {et~al.}(2009)\citenamefont {Breuer},
  \citenamefont {Laine},\ and\ \citenamefont {Piilo}}]{Laine_2009}%
  \BibitemOpen
  \bibfield  {author} {\bibinfo {author} {\bibfnamefont {H.-P.}\ \bibnamefont
  {Breuer}}, \bibinfo {author} {\bibfnamefont {E.-M.}\ \bibnamefont {Laine}}, \
  and\ \bibinfo {author} {\bibfnamefont {J.}~\bibnamefont {Piilo}},\ }\href
  {\doibase 10.1103/PhysRevLett.103.210401} {\bibfield  {journal} {\bibinfo
  {journal} {Phys. Rev. Lett.}\ }\textbf {\bibinfo {volume} {103}},\ \bibinfo
  {pages} {210401} (\bibinfo {year} {2009})}\BibitemShut {NoStop}%
\bibitem [{\citenamefont {Laine}\ \emph {et~al.}(2010)\citenamefont {Laine},
  \citenamefont {Piilo},\ and\ \citenamefont {Breuer}}]{Laine_2010}%
  \BibitemOpen
  \bibfield  {author} {\bibinfo {author} {\bibfnamefont {E.-M.}\ \bibnamefont
  {Laine}}, \bibinfo {author} {\bibfnamefont {J.}~\bibnamefont {Piilo}}, \ and\
  \bibinfo {author} {\bibfnamefont {H.-P.}\ \bibnamefont {Breuer}},\ }\href
  {\doibase 10.1103/PhysRevA.81.062115} {\bibfield  {journal} {\bibinfo
  {journal} {Phys. Rev. A}\ }\textbf {\bibinfo {volume} {81}},\ \bibinfo
  {pages} {062115} (\bibinfo {year} {2010})}\BibitemShut {NoStop}%
\bibitem [{\citenamefont {Hall}\ \emph {et~al.}(2014)\citenamefont {Hall},
  \citenamefont {Cresser}, \citenamefont {Li},\ and\ \citenamefont
  {Andersson}}]{Anderson_2014}%
  \BibitemOpen
  \bibfield  {author} {\bibinfo {author} {\bibfnamefont {M.~J.~W.}\
  \bibnamefont {Hall}}, \bibinfo {author} {\bibfnamefont {J.~D.}\ \bibnamefont
  {Cresser}}, \bibinfo {author} {\bibfnamefont {L.}~\bibnamefont {Li}}, \ and\
  \bibinfo {author} {\bibfnamefont {E.}~\bibnamefont {Andersson}},\ }\href
  {\doibase 10.1103/PhysRevA.89.042120} {\bibfield  {journal} {\bibinfo
  {journal} {Phys. Rev. A}\ }\textbf {\bibinfo {volume} {89}},\ \bibinfo
  {pages} {042120} (\bibinfo {year} {2014})}\BibitemShut {NoStop}%
\end{thebibliography}
\end{document}